\documentclass[12pt,a4paper]{article}

\usepackage{amsmath,amssymb}
\usepackage[nosort]{cite}
\usepackage[hyperref,bulletsep]{collect}

\makeatletter
\let\old@startsection=\@startsection
\renewcommand{\@startsection}[6]{\old@startsection{#1}{#2}{#3}{#4}{#5}{#6\mathversion{bold}}}
\makeatother

\makeatletter \@addtoreset{equation}{section} \makeatother

\makeatletter
\let\old@makecaption=\@makecaption
\def\@makecaption{\small\old@makecaption}
\makeatother

\let\oldPhi=\Phi
\let\oldPsi=\Psi
\let\oldGamma=\Gamma
\let\oldDelta=\Delta
\let\oldSigma=\Sigma
\let\oldTheta=\Theta
\let\oldPi=\Pi
\renewcommand{\Phi}{\mathnormal{\oldPhi}}
\renewcommand{\Psi}{\mathnormal{\oldPsi}}
\renewcommand{\Gamma}{\mathnormal{\oldGamma}}
\renewcommand{\Sigma}{\mathnormal{\oldSigma}}
\renewcommand{\Delta}{\mathnormal{\oldDelta}}
\renewcommand{\Theta}{\mathnormal{\oldTheta}}
\renewcommand{\Pi}{\mathnormal{\oldPi}}

\newcommand{\ham}{\mathcal{H}}
\newcommand{\charge}{\mathcal{Q}}
\newcommand{\gen}[1]{\mathfrak{#1}}

\newcommand{\superN}{\mathcal{N}}
\newcommand{\fldZ}{\mathcal{Z}}
\newcommand{\fldD}{\mathcal{D}}
\newcommand{\fldS}{\mathcal{X}}
\newcommand{\fldF}{\mathcal{U}}
\newcommand{\fldG}{\mathcal{\dot U}}
\newcommand{\fldA}{\mathcal{A}}

\newcommand{\gym}{g\indups{YM}}

\newcommand{\Tr}{\mathop{\mathrm{Tr}}}

\newcommand{\osc}[1]{\mathbf{#1}}

\newcommand{\order}[1]{\mathcal{O}(#1)}

\newcommand{\resolvsl}[1][]{\makebox[0pt][l]{\hspace{0.06em}$/$}#1G}
\newcommand{\resolvHsl}[1][]{\makebox[0pt][l]{\hspace{0.15em}$/$}#1H}
\newcommand{\sheetsl}[1][]{\makebox[0pt][l]{\hspace{0.06em}$/$}#1p}
\newcommand{\contour}{\mathcal{C}}

\newcommand{\sgrad}{\eta}
\newcommand{\sgrada}{{\eta_1}}
\newcommand{\sgradb}{{\eta_2}}

\newcommand{\sheetsign}{\varepsilon}

\newcommand{\scataux}{\theta}
\newcommand{\scatmain}{\psi}
\newcommand{\scatsym}{\Psi}

\newcommand{\MMM}[2]{{\arraycolsep0pt\begin{array}[b]{c}\makebox[0cm]{$\atopfrac{#2}{\downarrow}$}\\#1\end{array}}}

\ifx\genfrac\sdflkaj
\newcommand{\atopfrac}[2]{{{#1}\above0pt{#2}}}
\else
\newcommand{\atopfrac}[2]{\genfrac{}{}{0pt}{}{#1}{#2}}
\fi
\newcommand{\sfrac}[2]{{\textstyle\frac{#1}{#2}}}
\newcommand{\half}{\sfrac{1}{2}}
\newcommand{\quarter}{\sfrac{1}{4}}

\newcommand{\indup}[1]{_{\mathrm{#1}}}
\newcommand{\indups}[1]{_{\mathrm{\scriptscriptstyle #1}}}

\newcommand{\rep}[1]{{\mathbf{#1}}}
\newcommand{\matr}[2]{\left(\begin{array}{#1}#2\end{array}\right)}

\newcommand{\upper}[1]{^{\mathrm{\scriptscriptstyle #1}}}
\newcommand{\lrbrk}[1]{\left(#1\right)}
\newcommand{\bigbrk}[1]{\bigl(#1\bigr)}

\newcommand{\set}[1]{\{#1\}}
\newcommand{\state}[1]{\mathopen{|}#1\mathclose{\rangle}}

\newcommand{\alg}[1]{\mathfrak{#1}}

\newcommand{\nn}{\nonumber}
\newcommand{\nln}{\nonumber\\}
\newcommand{\nl}[1][0pt]{\nonumber\\[#1]&\hspace{-4\arraycolsep}&\mathord{}}
\newcommand{\nlnum}{\\&\hspace{-4\arraycolsep}&\mathord{}}
\newcommand{\earel}[1]{\mathrel{}&\hspace{-2\arraycolsep}#1\hspace{-2\arraycolsep}&\mathrel{}}
\newcommand{\eq}{\earel{=}}

\def\[{\begin{equation}}
\def\]{\end{equation}}
\def\<{\begin{eqnarray}}
\def\>{\end{eqnarray}}

\newenvironment{bulletlist}{\begin{list}{$\bullet$}{\leftmargin1.5em\itemsep0pt}}{\end{list}}
\newcounter{enumlistcnt}
\renewcommand{\theenumlistcnt}{\protect\emph{\roman{enumlistcnt}}}
\newenvironment{enumlist}{\begin{list}{\theenumlistcnt.}
                         {\usecounter{enumlistcnt}\setcounter{enumlistcnt}{0}\labelwidth 2.5em\itemindent0em\leftmargin\labelwidth\itemsep0pt}}
                         {\end{list}}

\makeatletter
\def\mr@ignsp#1 {\ifx\:#1\@empty\else #1\expandafter\mr@ignsp\fi}%
\newcommand{\multiref}[1]{\begingroup
\xdef\mr@no@sparg{\expandafter\mr@ignsp#1 \: }%
\def\mr@comma{}%
\@for\mr@refs:=\mr@no@sparg\do{\mr@comma\def\mr@comma{,}\ref{\mr@refs}}%
\endgroup}
\makeatother

\newcommand{\hypref}[2]{\ifx\href\asklfhas #2\else\href{#1}{#2}\fi}
\newcommand{\Secref}[1]{Section~\multiref{#1}}
\newcommand{\secref}[1]{Sec.~\multiref{#1}}

\newcommand{\appref}[1]{App.~\multiref{#1}}

\newcommand{\tabref}[1]{Tab.~\multiref{#1}}

\newcommand{\figref}[1]{Fig.~\multiref{#1}}
\renewcommand{\eqref}[1]{(\multiref{#1})}

\ifx\href\asklfhas\newcommand{\href}[2]{#2}\fi
\newcommand{\arxivno}[1]{\href{http://arxiv.org/abs/#1}{#1}}

\begin{document}

\thispagestyle{empty}
\begin{flushright}\footnotesize
\texttt{\arxivno{hep-th/0504190}}\\
\texttt{AEI-2005-092}\\
\texttt{PUTP-2159}\\
\vspace{0.5cm}
\end{flushright}
\vspace{0.5cm}

\renewcommand{\thefootnote}{\fnsymbol{footnote}}
\setcounter{footnote}{0}

\begin{center}
{\Large\textbf{\mathversion{bold}
Long-Range $\alg{psu}(2,2|4)$ Bethe Ans\"atze\\for Gauge Theory and Strings
}\par} \vspace{1cm}

\textsc{Niklas Beisert$^{a}$ and Matthias Staudacher$^b$} \vspace{5mm}

\textit{$^{a}$ Joseph Henry Laboratories\\
Princeton University\\
Princeton, NJ 08544, USA} \vspace{3mm}

\textit{$^{b}$ Max-Planck-Institut f\"ur Gravitationsphysik\\
Albert-Einstein-Institut\\
Am M\"uhlenberg 1, D-14476 Golm, Germany}\vspace{3mm}

\texttt{nbeisert@princeton.edu}\\
\texttt{matthias@aei.mpg.de}\\
\par\vspace{1cm}

\vfill

In Honor of Hans Bethe\vspace{5mm}

\textbf{Abstract}\vspace{5mm}

\begin{minipage}{12.7cm}
We generalize various existing higher-loop Bethe ans\"atze for simple sectors
of the integrable long-range dynamic spin chain describing planar $\mathcal{N}=4$ 
Super Yang-Mills Theory to the full $\alg{psu}(2,2|4)$ symmetry and,
asymptotically, to arbitrary loop order. We perform a large number
of tests of our conjectured equations, such as internal consistency,
comparison to direct three-loop diagonalization and expected
thermodynamic behavior. In the special case of the 
$\alg{su}(1|2)$ subsector, corresponding to a long-range $t$-$J$ model,
we are able to derive, up to three loops, the S-matrix and
the associated nested Bethe ansatz from the gauge theory dilatation 
operator. We conjecture novel all-order S-matrices for
the $\alg{su}(1|2)$ and $\alg{su}(1,1|2)$ subsectors, and show
that they satisfy the Yang-Baxter equation. 
Throughout the paper, we muse about
the idea that quantum string theory 
on $AdS_5\times S^5$ is also described by
a $\alg{psu}(2,2|4)$ spin chain. 
We propose asymptotic all-order
Bethe equations for this putative ``string chain{}'', which differ in a 
systematic fashion from the gauge theory equations.
\end{minipage}

\vspace*{\fill}

\end{center}

\newpage
\setcounter{page}{1}
\renewcommand{\thefootnote}{\arabic{footnote}}
\setcounter{footnote}{0}

\section{Introduction and Overview}
\label{sec:Intro}

Recently a powerful new tool for the study of planar
non-abelian gauge theories and strings on curved space-times,
as well as the conjectured dualities linking the two, has become
available.
\emph{Integrability} has made its appearance in $\mathcal{N}=4$
Super Yang-Mills theory and in IIB string theory on the
$AdS_5 \times S^5$ background. It is beginning to shed 
entirely new light on the AdS/CFT duality. Proving or 
disproving part of the gauge/string correspondence suddenly
seems to be within reach.
The central new tool 
is a technique widely known as the \emph{Bethe ansatz}.
It dates back to the year 1931 when Hans Bethe
solved the Heisenberg spin chain 
in his pioneering work \cite{Bethe:1931hc}.
Its impact on condensed matter theory and mathematical physics 
cannot be underestimated.

The first, crucial observation 
in the context of the gauge/string duality 
was made by Minahan and Zarembo~\cite{Minahan:2002ve}. 
They noticed that the conformal quantum operators in the scalar
field sector of $\mathcal{N}=4$ gauge theory are,
at the planar one-loop level, in one-to-one correspondence with the 
translationally invariant eigenstates of an integrable $\alg{so}(6)$
magnetic quantum spin chain. The spin chain Hamiltonian
corresponds to the gauge theoretic planar one-loop dilatation operator,
whose ``energy{}'' eigenvalues yield the scaling weights of the conformal 
operators. This observation turned out to be a first hint at
a very deep structure. The result generalizes to all local operators
of the planar one-loop $\mathcal{N}=4$ theory \cite{Beisert:2003yb}.
What is more, evidence was found that integrability extends
beyond the one-loop approximation \cite{Beisert:2003tq}.

First indications that planar gauge theories may contain hidden
integrable structures were discovered in a QCD context in 
seminal work by Lipatov
\cite{Lipatov:1993qn,Lipatov:1994yb,Lipatov:1994xy,Lipatov:1997vu}.
References to further interesting work on integrability in QCD may
be found in \cite{Belitsky:2004cz}. New aspects of the more recent
developments \cite{Minahan:2002ve,Beisert:2003yb,Beisert:2003tq}
when comparing to these important earlier insights are that
($i$) the integrability links space-time to internal symmetries,
($ii$) the studied spin chains allow for an interesting thermodynamics
with a large number of lattice sites, 
($iii$) the integrability
extends beyond the leading approximation and leads to novel long-range
spin chains,
($iv$) it allows for comparison with similar structures appearing
in the conjectured dual string theory.

And indeed it was argued in \cite{Bena:2003wd} that superstrings on
$AdS_5 \times S^5$ are classically integrable. 
This allows, in many cases, to find explicit solutions of the non-linear 
equations describing the classical motions of strings on
that background \cite{Arutyunov:2003uj,Arutyunov:2003za}.
More generally, classical integrability even permits to describe generic 
classical motions of the string as solutions of algebraic curves 
\cite{Kazakov:2004qf,Kazakov:2004nh,Beisert:2004ag,Beisert:2005bm,
Schafer-Nameki:2004ik,Beisert:2005di}.
A question of primary importance is clearly whether this
classical integrability extends to the quantum theory.
First encouraging evidence was presented in \cite{Arutyunov:2004vx} where
a Bethe ansatz for the string sigma model was found ``ex\-pe\-ri\-men\-tally{}'' 
in a special case. 
Excitingly, this Bethe ansatz also stems from 
a long-range spin chain, as first noticed in \cite{Beisert:2004jw}.

We may therefore hope to gain a much deeper understanding of the
AdS/CFT correspondence by directly comparing the integrable structures
of gauge and string theory as opposed to considering only the
spectrum of energies.
This is the approach proposed and pursued for semiclassical strings
in \cite{Arutyunov:2003rg,Arutyunov:2004xy,Mikhailov:2005wn,Arutyunov:2005nk,
Kazakov:2004qf,Kazakov:2004nh,Beisert:2004ag,Beisert:2005bm,
Schafer-Nameki:2004ik,Beisert:2005di}
and for quantum strings in 
\cite{Arutyunov:2004vx,Beisert:2004jw,Staudacher:2004tk}.
It leads to a deeper probe of earlier proposals for comparing
the string and gauge theory in the plane wave/BMN 
\cite{Berenstein:2002jq} and the semiclassical limit 
\cite{Frolov:2003qc}
(see \cite{Gubser:2002tv,Frolov:2002av} and 
\cite{Russo:2002sr,Minahan:2002rc} for qualitative and
quantitative precursors in particular cases).
Reviews of the work on these proposals are found in 
\cite{Pankiewicz:2003pg,Plefka:2003nb,Kristjansen:2003uy,Sadri:2003pr,Russo:2004kr}
and \cite{Tseytlin:2003ii,Tseytlin:2004xa,Beisert:2004yq,Zarembo:2004hp,Beisert:2004ry}.
The immense usefulness of the Bethe ansatz in the study of these proposals 
was demonstrated in \cite{Minahan:2002ve} and \cite{Beisert:2003xu,Beisert:2003ea}.

In the case of gauge theory one may directly demonstrate the
emergence of an integrable long-range spin chain from the first few orders
of perturbation theory 
\cite{Minahan:2002ve,Beisert:2003yb,Beisert:2003tq,Beisert:2003ys,Serban:2004jf}.
Of course, with current technology it is not possible to 
give an all-orders, let alone non-perturbative, proof.
However, constructing the correct chain under some reasonable 
assumptions does not appear to be entirely out of reach; for 
first steps in this direction see \cite{Beisert:2004hm}.
In the case of string theory the evidence for an underlying
spin chain structure is entirely indirect 
\cite{Arutyunov:2004vx,Beisert:2004jw,Staudacher:2004tk}.
If true, it should emerge from an exact quantization of the
string sigma model. Despite some progress towards
setting up the quantization of the integrable model
\cite{Swanson:2004qa,Arutyunov:2004yx,Alday:2005gi,Berkovits:2004jw,Berkovits:2004xu} 
it is fair to say that this putative ``string chain{}'' is currently
hiding well.

In this paper we will continue the construction of higher-loop Bethe ans\"atze, 
begun in \cite{Serban:2004jf,Beisert:2004hm,Arutyunov:2004vx,Beisert:2004jw,Staudacher:2004tk},
for the integrable long-range spin chains
which appear to describe gauge and (possibly) string theory.
The approach we shall follow is somewhat similar to
the one commonly applied for the solution of a jigsaw puzzle.
We attempt to self-consistently assemble smaller building blocks
(sectors) into an emergent larger picture, until we end up with
a proposal for the full set of asymptotic Bethe equations.
An important restriction is the condition of asymptoticity:
We suspect that our equations diagonalize the underlying spin chains 
only to $\order{g^{2L}}$, where $g$ is the coupling constant and
$L$ the chain length.
While our derivation contains multiple gaps in need
of proof, we feel that we have performed many of the currently possible
consistency checks without finding any manifest contradictions.
This includes checks against direct diagonalization of the
three-loop Hamiltonian (where available), thermodynamic consistency,
symmetry, and the idea \cite{Arutyunov:2004vx,Staudacher:2004tk}, supported by
the structure found in \cite{Beisert:2003ys}, that the
asymptotic S-matrices of string and gauge theory differ by
a global, flavor-independent ``dressing factor{}'':
\[\label{eq:dressing}
S\upper{string}=\hat{S}\upper{dressing}\,S\upper{gauge}\, .
\]
Hopefully this factor will appear as one goes from weak (gauge theory)
to strong (string theory) coupling and will reconcile the notorious
third-order discrepancies noticed in \cite{Callan:2003xr,Serban:2004jf}
between string and gauge theory in, respectively,
the near-BMN and Frolov-Tseytlin limits.
For an interesting alternative proposal to explain the
discrepancy, see \cite{Minahan:2005jq}.

We start out in \Secref{sec:One} by reconsidering the 
S-matrices and associated Bethe equations of the three
two-component sectors $\alg{su}(2)$, $\alg{su}(1|1)$ and $\alg{sl}(2)$.
We first review their Hamiltonians and 
the perturbative asymptotic Bethe ansatz (PABA) of \cite{Staudacher:2004tk}
which is based, to a large extent, on the Bethe's original work \cite{Bethe:1931hc}. 
In the case of gauge theory we extend the three-loop asymptotic S-matrices
of \cite{Staudacher:2004tk} for $\alg{su}(1|1)$ and $\alg{sl}(2)$ to all 
loops, in analogy with the $\alg{su}(2)$ case \cite{Beisert:2004hm}. 
On the string side we improve the approximate S-matrices of 
\cite{Staudacher:2004tk} to (hopefully) asymptotically exact ones.
In particular, the improved Bethe equations appear to be consistent with 
the existence of an integrable spin chain for quantum strings beyond
the $\alg{su}(2)$ sector \cite{Arutyunov:2004vx,Beisert:2004jw}.
Our construction confirms, for string and gauge theory, the following 
relation between the S-matrices of the three sectors 
\cite{Staudacher:2004tk}:
\[\label{eq:grouptheory}
S_{\alg{sl}(2)}=S_{\alg{su}(1|1)}\, S^{-1}_{\alg{su}(2)}\, S_{\alg{su}(1|1)}\, .
\]
Note that \eqref{eq:dressing} is consistent with our claim
that \eqref{eq:grouptheory} holds for both string and gauge theory. 
 
In \Secref{sec:SU12} we consider the unification of the
$\alg{su}(2)$ and $\alg{su}(1|1)$ sectors into the $\alg{su}(1|2)$
sector.%
\footnote{Conventionally, we will always
place the number referring to spacetime
symmetries before the number
referring to internal symmetry.}
The one-loop Hamiltonian of the latter is identical to the one
of an integrable quantum super spin chain important in 
condensed matter theory: the so-called $t$-$J$ model. 
We then extract, using the PABA, the three-loop
S-matrix from the Hamiltonian of \cite{Beisert:2003ys}.
Using the insights from \Secref{sec:One} we generalize this
S-matrix to all loops (asymptotically). 
Excitingly, the resulting S-matrix \eqref{eq:su12.Sx}, 
which we have not been able to find in the literature, 
satisfies the Yang-Baxter equation. 
We end this section by applying the machinery of the nested Bethe ansatz 
to our S-matrix, thereby deriving two equivalent sets of 
nested long-range $\alg{su}(1|2)$ Bethe equations.

In \Secref{sec:PSU112} we consider the further unification
of all three two-component sectors of \Secref{sec:One} into a
super spin chain with non-compact symmetry $\alg{su}(1,1|2)$.
Alternatively we can say that we combine the $\alg{su}(1|2)$ sector
of the previous section with the derivative sector 
$\alg{sl}(2)=\alg{su}(1,1)$. Here, as in \cite{Staudacher:2004tk}, 
the higher loop Hamiltonian, and thus the PABA, is currently not available.
Upon inspection of the one-loop S-matrix we however find 
a very natural all-loop generalization in complete analogy
with the $\alg{su}(1|2)$ case. We then apply the nested Bethe
ansatz and derive the system of Bethe equations for this sector.
There are four forms of the Bethe equations 
corresponding to different Cartan matrices 
for $\alg{su}(1,1|2)$ which we prove to be equivalent 
by dualization as in \cite{Beisert:2005di}.
Incidentally this yields infinitely many novel higher-loop predictions
for gauge theory operators, some of which should be testable,
at least at the two-loop level, against field theory computations.

In the final \Secref{sec:Odyssee2005} we take a big leap and attempt to
extend the previous system of $\alg{su}(1,1|2)$ Bethe equations 
to the full set of excitations with symmetry $\alg{psu}(2,2|4)$.
Here we have not yet found an appropriate S-matrix;
this is not only due to increased complexity, 
but mainly related to the fact that the higher-loop $\superN=4$ spin chain 
is \emph{dynamic} \cite{Beisert:2003ys}.
The nested Bethe ansatz in this case will most likely require some new ideas.
There are two main sources of inspiration which lead to our equations:
The finite gap solution of the string sigma model
\cite{Beisert:2005bm} gives expressions which
we can generalize beyond the thermodynamic limit 
using our experience from the two-component sectors. 
The possibility of dualizing the Bethe equations 
between several equivalent Cartan matrices
seems to imply a very special structure of the equations.
Our final set of equations is displayed in \tabref{tab:ABE}.
It agrees with a list of properties outlined in \cite{Beisert:2004ry}. 
Most importantly, we find an intriguing symmetry of the higher-loop equations
which appears to be related to the dynamic nature of the underlying chain.


\section{Rank-One Sectors}
\label{sec:One}

In $\superN=4$ SYM there are three sectors of local operators 
where the dilatation operator takes a particularly simple form. 
They are all based on a vacuum state 
\[\label{eq:One.Vacuum}
\state{0,L}=\fldZ^L
\]
which is half-BPS and therefore has exactly vanishing anomalous dimension.
Here $\fldZ$ is a complex combination of two real scalars of the theory.
The excitations of the vacuum are obtained by 
changing some of the $\fldZ$'s into other fields. 
In the $\alg{su}(2)$ sector we replace $\fldZ$ by another
complex scalar $\fldS$. The $\alg{su}(1|1)$ sector has fermionic
excitations $\fldF$. In the third $\alg{sl}(2)=\alg{su}(1,1)$ sector
the excitations are covariant derivatives $\fldD\fldZ$. 
Here, unlike the other two cases, 
it is allowed to have 
multiple excitations $(\fldD^n\fldZ)$ residing at a single site.  

\subsection{Review of Hamiltonians}
\label{sec:One.Dil}

The symmetry algebra of conformal $\superN=4$ SYM is $\alg{psu}(2,2|4)$,
it acts linearly on the set of states.
This representation $\gen{J}(g)$ 
depends on the coupling constant $g$ defined by
\[
g^2=\frac{\gym^2N}{8\pi^2}=\frac{\lambda}{8\pi^2}\,.
\]
In this paper we consider a small coupling constant $g$ 
and apply perturbation theory.
Classically, i.e.~at $g=0$, the action of the symmetry
algebra on the set of states is merely the tensor product
of the action on the individual sites. 
This is how the symmetry algebra acts for common quantum spin chains.
When we turn on interactions, the picture changes:
The range of the action of $\gen{J}(g)$ extends;
for each loop order, i.e.~order in $\lambda\sim g^2$, the generators may act 
on one additional neighboring site at the same time. 
Moreover, they may even create or destroy spin chain sites
and one might therefore consider the spin chain as \emph{dynamic}
\cite{Beisert:2003ys}. 

A priori, there is no natural Hamiltonian $\ham$ for the spin chain,
there is only the symmetry algebra. 
However, in perturbation theory one can identify the
anomalous dimension $\delta\gen{D}(g)$ as a $\alg{u}(1)$ generator
which commutes with $\alg{psu}(2,2|4)$; we shall set
\[
\delta \gen{D}(g)=g^2 \ham(g),
\]
because by definition the classical part of the anomalous dimension 
vanishes and for various reasons 
we would like to have finite energies $E$ at $g=0$.
These are consequently related to the anomalous dimension by
\[
\delta D(g)=g^2 E(g).
\]

In all investigated cases it has turned out 
that in the planar limit
there are additional generators $\charge_r$ of a form 
similar to the one of $\ham$. 
All of them commute with each other,
with $\alg{psu}(2,2|4)$ and with the Hamiltonian. 
This is an implication of the apparent 
higher-loop integrability of $\superN=4$ SYM \cite{Beisert:2003tq}.
In fact, the first two charges $\charge_1,\charge_2$ are 
crucial for extracting physical information from the theory. 
The second charge is nothing but the Hamiltonian
and the eigenvalues therefore match
\[
\ham=\charge_2,\qquad E=Q_2.
\]
The first charge is the logarithm of the spin chain
shift operator. The shift operator permutes the spin chain
state cyclically by one step.%
\footnote{The shift operator is graded 
and generates the appropriate signs for commuting fermions.}
As gauge theory states are identified cyclically,
the shift operator must act trivially on all physical states
\[\label{eq:shift}
\exp(i\charge_1)\simeq 1,\qquad Q_1=2\pi m.
\]
In fact this condition is essential
for the consistency of the model \cite{Beisert:2004ry}.
Finally, the higher charges do not seem to contain interesting physical
information. They serve as hidden symmetries leading to integrability.

\subsection{Spectrum}
\label{sec:One.Spectrum}

\begin{table}\centering
$\begin{array}{|c|l|}\hline
L&(E_0,E_2,E\indup{4g}|E\indup{4s})^P\\\hline\hline
3&(6,-12,42|33)^+\\\hline
4&(4,-6,17|13)^-\\\hline
5&(10E-20,-17E+60,\frac{117}{2}E-230|\frac{107}{2}E-210)^+\\\hline
6&(6,-\frac{21}{2},\frac{555}{16}|\frac{483}{16})^-\\
 &(2,-\frac{3}{2},\frac{37}{16}|\frac{29}{16})^-\\\hline
7&(14E^2-56E+56,-23E^2+172E-224,79E^2-695E+966|74E^2-653E+910)^+
\hspace{-0.2cm}\\\hline
8&(4,-5,\frac{49}{4}|\frac{41}{4})^-\\
 &(8E-8,-13E+18,\frac{179}{4}E-61|\frac{167}{4}E-57)^-\\\hline
\end{array}$
\caption{Spectrum of lowest-lying two-excitation states.}
\label{tab:TwoEx}
\end{table}

\begin{table}\centering
$\begin{array}{|c|c|l|}\hline
L&K&(E_0,E_2,E\indup{4g}|E\indup{4s})^P\\\hline\hline
5+1&3&(6,-9,\frac{63}{2}|\frac{63}{2})^-\\\hline
6+1&3&(5,-\frac{15}{2},25|\frac{95}{4})^\pm\\\hline
7+1&3&(4,-5,14|13)^\pm\\
 & &(6,-9,33|33)^-\\
 &4&({\scriptstyle 20E^2-116E+200},
     {\scriptstyle-32E^2+340E-800},
     {\scriptstyle 112E^2-1400E+3600}|
     {\scriptstyle 101E^2-1304E+3400})^+\\\hline
8+1&3&({\scriptstyle 17E^2-90E+147},
     {\scriptstyle -\frac{51}{2}E^2+\frac{525}{2}E-\frac{1239}{2}},
     {\scriptstyle \frac{169}{2}E^2-\frac{2091}{2}E+\frac{5649}{2}}|
     {\scriptstyle 82E^2-\frac{2037}{2}E+\frac{11025}{4}})^\pm\\
 &4&(5,-\frac{15}{2},\frac{55}{2}|\frac{109}{4})^\pm\\
 & & (12E-24,-18E+54,57E-171|48E-147)^-\\\hline
\end{array}$
\caption{Spectrum of lowest-lying states of the $\alg{su}(2)$ sector.}
\label{tab:SU2}
\end{table}

\begin{table}\centering
$\begin{array}{|c|c|l|}\hline
L&K&(E_0,E_2,E\indup{4g}|E\indup{4s})^P\\\hline\hline
5&4&(10,-20,\frac{145}{2}|\frac{125}{2})^-\\\hline
6&3&(8,-14,49|47)^\pm\\
 &4&(8,-14,46|38)^+\\\hline
7&3&(7,-12,\frac{83}{2}|\frac{153}{4})^\pm\\
 &4&({\scriptstyle 28E^2-252E+728},
     {\scriptstyle-51E^2+906E-3864},
     {\scriptstyle 179E^2-3965E+20090}|
     {\scriptstyle 160E^2-3629E+18634})^-\\
 &6&(14,-28,\frac{203}{2}|\frac{175}{2})^+\\\hline
8&3&(6,-\frac{19}{2},\frac{247}{8}|\frac{223}{8})^\pm\\
 & &(8,-\frac{29}{2},\frac{427}{8}|\frac{419}{8})^\pm\\
 &4&(8,-14,\frac{97}{2}|\frac{89}{2})^+\\
 & &(10,-18,64|61)^\pm\\
 & &(16E-56,-26E+170,\frac{165}{2}E-638|\frac{139}{2}E-554)^+\\
 &5&(12,-22,77|69)^\pm\\
 &6&(12,-22,75|63)^-\\\hline
\end{array}$
\caption{Spectrum of lowest-lying states of the $\alg{su}(1|1)$ sector.}
\label{tab:U11}
\end{table}

Our aim is to determine the spectrum of the Hamiltonian
and thus the spectrum of anomalous dimensions.
The energies of the states can be obtained directly
from the Hamiltonian. In practice, this requires that the state
is not too complicated. More importantly, 
we have to get hold of the Hamiltonian in the first place. 
At the one-loop level the complete Hamiltonian was obtained 
in \cite{Beisert:2003jj}. 
The largest piece of the higher-loop Hamiltonian is
known for the $\alg{su}(2|3)$ sector from \cite{Beisert:2003ys}.
We will obtain all our spectral data from this particular Hamiltonian
and display it in a number of tables, e.g.~\tabref{tab:TwoEx}.
A set of states is specified by the length $L$,
the number of excitations $K$ and the leading three
orders of the energy 
\[
E\indup{g}=E_0+g^2 E_2+g^4 E\indup{4g}+\ldots,
\qquad
E\indup{s}=E_0+g^2 E_2+g^4 E\indup{4s}+\ldots
\]
as well as the (charge conjugation/spin chain inversion) parity $P$.
We shall distinguish between two models, gauge theory, with energy
$E\indup{g}$,
and a string chain, with energy $E\indup{s}$, see below.
The parity may be either $+$ or $-$. 
Many states, however, come in exactly degenerate pairs, 
but with opposite parity \cite{Beisert:2003tq}. 
These pairs are indicated by $\pm$ and are a direct
consequence of the existence of a conserved, 
parity-inverting charge $\charge_3$.
When there is only a single state with given $L$ and $K$,
the state cannot mix and the energy always expands in rational
numbers.
When mixing with other states occurs, however, 
the energies are often irrational.%
\footnote{Note however that the energies cannot be transcendental,
as one might have expected from general experience with perturbative
quantum field theory. The reason is that the higher-loop Bethe ansatz 
always leads to algebraic equations for the energies.}
In those cases we prefer to encode the energies 
as the roots of an algebraic equation.
For mixing of $M$ states, the tables state, in the 
form $\left(X_0(E),X_2(E),X\indup{4g}(E)|X\indup{4s}(E)\right)$, a polynomial
$X(E)=X_0(E)+X_2(E)g^2+X_4(E)g^4$ of degree $M-1$. 
The energies of the $M$ states
are obtained as solutions to
\[
E^M=X(E).
\]
If desired, numerical values may be obtained immediately with an
appropriate root finder. E.g.~with \texttt{Mathematica}
one might use, to an accuracy of $k$ digits,
\begin{center}
\verb"Series[e /. NSolve[e^M == X[e], e, k], {g,0,4}] // Normal // Chop".
\end{center}

In this section we concentrate on subsectors of the full theory 
where the main part of the symmetry algebra has rank one. 
These are the $\alg{su}(2)$, $\alg{su}(1|1)$ and $\alg{sl}(2)$ sectors
introduced above. In later sections we will extend the analysis
to larger sectors and eventually to the complete model.
The $\alg{su}(2)$ and $\alg{su}(1|1)$ sectors are contained in 
the $\alg{su}(2|3)$ sector of \cite{Beisert:2003ys}
for which we know the three-loop Hamiltonian. 
We display the energies of the lowest-lying
spin chain states in \tabref{tab:TwoEx,tab:SU2,tab:U11}.
Some of these results have appeared in \cite{Beisert:2003tq,Beisert:2003ys}
and we have supplemented the values of $E\indup{4s}$.
Note that the two sectors intersect on the set of
states with only two excitations \cite{Beisert:2002tn},
we display those states separately in \tabref{tab:TwoEx}.
For the $\alg{sl}(2)$ sector a higher-loop Hamiltonian is currently
not available.

\subsection{Review of Two-Component Bethe Ans\"atze}
\label{sec:One.Mom}

In this section we will review the perturbative asymptotic Bethe ansatz (PABA)
developed in \cite{Staudacher:2004tk}. 
This is a Bethe ansatz for spin chain excitations 
in position space which is very closely related 
to the original ansatz by Bethe \cite{Bethe:1931hc}, 
but adapted to the long-range Hamiltonians introduced in \secref{sec:One.Dil}.
It starts with the assumption that 
cyclic chains of finite length $L$ are infinite chains with 
periodic boundary conditions.
We shall therefore consider an infinitely long chain
and try to determine its eigenstates.
Only later we will restrict to periodic states
which are given by the solutions to the Bethe equations. 

The vacuum state is a tensor product of fields $\fldZ$
much like the ferromagnetic vacuum of a magnetic chain
\[
\state{0}=\state{\ldots\fldZ\fldZ\fldZ\ldots}.
\]
It is a protected state, its energy is exactly zero
\[
\ham\state{0}=0.
\]
We may place a few excitations 
$\fldA=\fldS,\fldF,\fldD\fldZ$ 
(depending on the sector)
into the vacuum and then try to find the eigenstates 
of the Hamiltonian.
Let us start with a single excitation
\[
\state{\ldots \fldZ\MMM{\fldA}{\ell}\fldZ\ldots}
=
\alpha^\dagger_\ell\state{0}.
\]
The latter representation indicates
that we can view the excitation 
as being produced by some creation operator $\alpha^\dagger_\ell$ 
acting on site $\ell$. 
The Hamiltonian is homogeneous and the appropriate
ansatz for an eigenstate is a plane wave
with momentum $p$
\[
\state{p}=\sum_{\ell}e^{ip\ell}\,\alpha^\dagger_\ell\state{0}.
\]
It is automatically an eigenstate $\ham \state{p}=e(p)\state{p}$
and the energy turns out to be
\cite{Serban:2004jf}
\[
e(p)=4\sin^2(\half p)- 
  8g^2\sin^4(\half p)+ 
  32g^4\sin^6(\half p)+\ldots
\]
which is consistent with the all-loop prediction 
\cite{Beisert:2004hm,Berenstein:2002jq}
\[
e(p)=g^{-2}\sqrt{1+8g^2\sin^2(\half p)}-g^{-2}.
\]

Now we can attack the two-excitation problem,
the eigenstates are given by%
\footnote{The excitation generators
$\alpha^\dagger$ can be either bosonic or 
fermionic. According to their grading,
they will automatically generate
some relative signs
in the states $\state{\ldots\fldZ\fldA\fldZ\ldots\fldZ\fldA\fldZ\ldots}$.}
\[
\state{p_1,p_2}=\sum_{\ell_1,\ell_2}
\Psi_{\ell_1,\ell_2}(p_1,p_2)\,
\alpha^\dagger_{\ell_1}\alpha^\dagger_{\ell_2}\state{0}.
\]
with some wave function $\Psi_{\ell_1,\ell_2}(p_1,p_2)$.
At each fixed loop order the range of the
interaction is finite. 
Asymptotically, the wave function should therefore factorize
into one-particle wave functions with
\[
\Psi_{\ell_1,\ell_2}(p_1,p_2)=e^{ip_1 \ell_1+ip_2 \ell_2}\,A\qquad
\mbox{for}\qquad \ell_1\ll \ell_2
\]
and
\[
\Psi_{\ell_1,\ell_2}(p_1,p_2)=e^{ip_1 \ell_1+ip_2 \ell_2}\,A'\qquad
\mbox{for}\qquad \ell_1\gg \ell_2
\]
where $A$ and $A'$ are independent of $\ell_1,\ell_2$.
The elements of the wave function in the interaction range
$\ell_1\approx \ell_2$ are determined by the 
non-diffractive scattering problem
\[
\ham \state{p_1,p_2}
=\bigbrk{e(p_1)+e(p_2)}\state{p_1,p_2}.
\]
It also fixes the ratio of $A$ and $A'$.
The wave function is unphysical,%
\footnote{In any renormalization scheme
there is freedom of applying a linear
transformation on the set of all local operators;
this is part of how divergencies are absorbed in a quantum field theory.
Physical information must not depend on the change of basis. 
For example, anomalous dimensions and correlation functions
of \emph{eigen}operators are invariant quantities while
the wave function depends on the basis.}
the relevant physical information 
is the phase shift $\Phi$ between the wave function 
on both sides of the interaction 
\[\label{eq:One.Ansatz.Phase}
S(p_2,p_1)=\exp\bigbrk{i\Phi(p_2,p_1)}=\frac{A'}{A}\,.
\]
The amplitude $S(p_2,p_1)$ is the two-body \emph{S-matrix}.

For an integrable Hamiltonian $\ham$ the phase shift $\Phi(p_1,p_2)$
is all we need to know to construct the asymptotic state with
arbitrarily many excitations. Here asymptotic means that 
we neglect those contributions to the exact wave function where 
some excitations are sufficiently close to each other to interact.
These contributions are determined from the asymptotic data by the Hamiltonian,
but they are not relevant for finding the spectrum of energies.
An eigenstate is specified by a set of $K$ momenta $p_k$
\[\label{eq:wf}
\state{\set{p_k}}=
\sum_{\ell_k}
A_{\ell_k} 
\prod_{k=1}^K\lrbrk{\exp(ip_k\ell_k)\, \alpha_{\ell_k}^\dagger}\state{0}.
\]
Here the amplitudes $A_{\ell_k}$
depend asymptotically only on the ordering $\sigma(\ell_k)$ of the (well-separated) 
positions $\ell_k$ of the excitations.
The amplitudes are related among each other through 
the phase shifts (up to one overall constant):
If two orderings $\sigma$ and $\sigma'$ are related by the interchange
of two adjacent excitations, the amplitudes $A_\sigma$ and $A_{\sigma'}$ 
must be related by \eqref{eq:One.Ansatz.Phase}.
Then the energy of the (corresponding exact) eigenstate 
is given by 
\[
E=\sum_{k=1}^K e(p_k).
\]

If the chain were really infinitely long this would already be
the end of the analysis,  and arbitrary values of the momenta
$p_k$ would yield valid solutions. For a chain of length
$L$ we however need to take into account the periodicity
conditions. In particular, if we shift the position of any
particular excitation by $L$ lattice sites the wave function
\eqref{eq:wf} should not change. These $K$ constraints lead to
a set of $K$ \emph{Bethe equations}:
\[\label{eq:One.MomBethe}
\exp(iLp_k)=\prod_{\textstyle\atopfrac{j=1}{j\neq k}}^{K}
\exp\bigbrk{i\Phi(p_k,p_j)}
=\prod_{\textstyle\atopfrac{j=1}{j\neq k}}^{K}
S(p_k,p_j)\, .
\]
Finally, we have to take into account that the eigenvalue of the lattice
shift operator should equal 1 for the translationally invariant
states we are interested in. This leads, from \eqref{eq:shift}, to
the momentum constraint
\[\label{eq:One.MomConstr}
\prod_{k=1}^K \exp(ip_k)=1
\qquad
\mbox{i.e.}
\qquad
Q_1=\sum_{k=1}^K p_k=2\pi m.
\]

This completes our general review of the Bethe ansatz technique
for two-component systems. The case of more than two components
is conceptually similar, but significantly more involved
and will be discussed in the following chapters.

Let us now return to the concrete case of the two-component sectors 
relevant to $\mathcal{N}=4$ gauge theory. 
For the $\alg{su}(2)$ sector, the all-loop scattering phase was conjectured
in \cite{Beisert:2004hm} to be given by
\[\label{eq:One.MomPhase}
\exp\bigbrk{i\Phi_{\fldS}(p_k,p_j)}=
\frac{u(p_k)-u(p_j)+i}{u(p_k)-u(p_j)-i}
\]
with the rapidity function
\[\label{eq:One.MomRap}
u(p)=\half\cot(\half p)\sqrt{1+8g^2\sin^2(\half p)}\,.
\]
This may be proven to three loops by embedding the $\alg{su}(2)$ 
Hamiltonian to $\order{g^4}$ into the Inozemtsev spin chain 
\cite{Serban:2004jf}. For this long-range system exact wave functions 
are known, and the correct phase shift may therefore be extracted.
At four and five loops the conjectured gauge theory Hamiltonian%
\footnote{The conjecture is based on assuming integrability, 
proper scaling in the thermodynamic limit 
and certain features of field-theoretic perturbation theory \cite{Beisert:2004hm,Beisert:2004ry}.
No rigorous proof for BMN scaling exists beyond three loops.
This means that, strictly speaking, the Inozemtsev model has not yet
been completely ruled out, even though we strongly suspect that the Hamiltonian
of  \cite{Beisert:2004hm,Beisert:2004ry} is indeed the correct one.
The conjectured phase \eqref{eq:One.MomPhase} may be verified from the
latter to five-loop order by the PABA of
\cite{Staudacher:2004tk} (M.~S., T.~Klose, unpublished).} 
may no longer be embedded into the Inozemtsev model.

The scattering phase for the fermionic $\alg{su}(1|1)$ sector
was distilled from the $\alg{su}(2|3)$ vertex of \cite{Beisert:2003ys}
by the PABA (``perturbative asymptotic Bethe ansatz{}'') technique 
in \cite{Staudacher:2004tk}. It reads
\<\label{eq:One.MomTheta}
\Phi_{\fldF}(p_k,p_j)\eq
4g^2\sin (\half p_k)\sin(\half p_j)\sin(\half p_k-\half p_j)
\nl
+g^4\sin (\half p_k)\sin (\half p_j)
\Bigl(
-7\sin(\sfrac{1}{2}  p_k - \sfrac{1}{2} p_j)
+\sin(\sfrac{1}{2} p_k-\sfrac{3}{2}p_j)
\nlnum\nonumber
\qquad\qquad\qquad\qquad\qquad\quad
+\sin(\sfrac{3}{2} p_k-\sfrac{1}{2}p_j)
+\sin(\sfrac{3}{2} p_k-\sfrac{3}{2}p_j)
\Bigr)+\order{g^6}.
\>
In the derivative $\alg{sl}(2)$ sector the PABA is currently not
applicable since we are lacking the Hamiltonian beyond the one-loop level.
However, in \cite{Staudacher:2004tk} the simple relation 
\eqref{eq:grouptheory} between the S-matrices was discovered from a 
spectroscopic analysis of strings in the near-plane wave background.
It was then assumed that \eqref{eq:grouptheory} should also hold
in gauge theory. In view of \eqref{eq:One.Ansatz.Phase} this led to
the following conjecture for the scattering phase for~$\alg{sl}(2)$
\[\label{eq:One.MomPhaseD}
\Phi_{\fldD}(p_k,p_j)=
2\Phi_{\fldF}(p_k,p_j)-\Phi_{\fldS}(p_k,p_j).
\]
This turned out to be consistent with the anomalous dimensions
of twist-two operators which are rigorously known to two loops
\cite{Kotikov:2003fb} and were conjectured, based on a fully-fledged 
QCD loop calculation \cite{Moch:2004pa}, to three loops in 
\cite{Kotikov:2004er,Kotikov:2005ne}.%
\footnote{Here, the asymptotic S-matrix \eqref{eq:One.MomPhaseD} works
even better than expected \cite{Staudacher:2004tk}:
For two and three loops, the chain is already \emph{shorter} than 
the range of the interaction, 
but the Bethe ansatz still reproduces the correct result. 
This curiosity is explained by the results of \secref{sec:PSU112.PSU11}
which relate the $L=2$ spin chain to a $L=4$ spin chain by 
supersymmetry. Then the interaction is just sufficiently short up to three loops.}
An involved two-loop test, using sophisticated superspace
Feynman rules (see also \cite{Eden:2003sj,Eden:2004ua,Eden:2005ve}),
for the simplest twist-three operator was recently successfully 
performed in \cite{Eden:2005bt}.

\subsection{The Spectral Parameter Plane}
\label{sec:One.Spec}

In \cite{Beisert:2004hm} an alternative parametrization 
of the $\alg{su}(2)$ Bethe ansatz was presented 
which simplified many expressions. 
It is based on the map between the rapidity ($u$) plane
and a spectral parameter ($x$) plane%
\footnote{The map $x(u)$ has two branches. 
For $g\approx 0$ we will pick the branch where $x\approx u$. 
The other branch is given by $x'=g^2/2x$.}
\[\label{eq:One.SpecRap}
x(u)=\half u+\half u\sqrt{1-2g^2/u^2}\,,
\qquad
u(x)=x+\frac{g^2}{2x}\,.
\]
The relation to the momentum ($p$) plane is given by 
\[\label{eq:One.SpecMom}
\exp(ip)=\frac{x(u+\sfrac{i}{2})}{x(u-\sfrac{i}{2})}\,.
\]

Let us for simplicity define several equivalent parametrizations
of Bethe roots. We shall consider the spectral parameter $x_k$ 
as fundamental. The momentum $p_k$, rapidity $u_k$ 
and shifted spectral parameters $x_k^+$ and $x_k^-$ 
are defined as
\[\label{eq:One.SpecRoots}
u_k=u(x_k),\qquad
x_k^\pm=x(u_k\pm \sfrac{i}{2}),\qquad
p_k=-i\log \frac{x^+_k}{x^-_k}.
\]

The local charges $Q_r$ of the integrable model 
can now be conveniently expressed as
\[\label{eq:One.SpecCharges}
Q_r=\sum_{k=1}^K q_r(x_k),
\qquad
q_r(x_k)=\frac{i}{r-1}\lrbrk{\frac{1}{(x_k^+)^{r-1}}-\frac{1}{(x_k^-)^{r-1}}},
\]
with the regularized first charge $q_1(x_k)=-i\log(x_k^+/x_k^-)=p_k$
being the momentum.
Particularly important are the first two charges,
the total momentum $Q_1$ for the momentum constraint
and the energy $Q_2$ for the anomalous dimension $\delta D$
\[\label{eq:One.SpecConstrEnergy}
Q_1=2\pi m,\qquad 
\delta D=g^2 Q_2.
\]
Before we discuss the Bethe equations let us note some useful 
identities relating the $u$- and $x$-plane
\<\label{eq:One.SpecIdent}
u_k-u_j\eq (x_k-x_j)(1-g^2/2x_kx_j)
\nln\eq(x^\pm_k-x^\pm_j)(1-g^2/2x^\pm_kx^\pm_j),
\nln
u_k-u_j\pm \sfrac{i}{2}\eq 
   (x^\pm_k-x_j)(1-g^2/2x^\pm_kx_j)
\nln\eq
(x_k-x^\mp_j)(1-g^2/2x_kx^\mp_j),
\nln
u_k-u_j\pm i\eq 
   (x^\pm_k-x^\mp_j)(1-g^2/2x^\pm_kx^\mp_j).
\>
They are easily confirmed using the definition of $u(x)$ in
\eqref{eq:One.SpecRap}.

\subsection{Bethe Equations for Spins}
\label{sec:One.Bethe}

Let us next attempt to express the Bethe ans\"atze of 
\secref{sec:One.Mom} through the spectral parameters $x^+$, $x^-$.
We shall discover that this allows to find, in analogy with the 
$\alg{su}(2)$ case \cite{Beisert:2004hm}, very natural
all-loop extensions of the three-loop S-matrices for the
$\alg{su}(1|1)$ and $\alg{sl}(2)$ sectors.  

The Bethe equations \eqref{eq:One.MomBethe,eq:One.MomPhase} for 
the $\alg{su}(2)$ sector 
in the $u$-plane read
\[\label{eq:One.BetheSU2U}
\lrbrk{\frac{x(u_k+\sfrac{i}{2})}{x(u_k-\sfrac{i}{2})}}^L=
\prod_{\textstyle\atopfrac{j=1}{j\neq k}}^K
\frac{u_k-u_j+i}{u_k-u_j-i}\,.
\]
Using the identities \eqref{eq:One.SpecIdent} we can translate them to
the $x^\pm$-plane
\[\label{eq:One.BetheSU2}
\lrbrk{\frac{x^+_k}{x^-_k}}^L=
\prod_{\textstyle\atopfrac{j=1}{j\neq k}}^K
\frac{x_k^+-x_j^-}{x_k^--x_j^+}\,
\frac{1-g^2/2x_k^+x_j^-}{1-g^2/2x_k^-x_j^+}\,.
\]
As it stands this result is neither remarkable nor very helpful.
It turns out, however, that the second term agrees 
\emph{precisely} with the function 
$\exp(i\Phi_{\fldF})$ of \eqref{eq:One.MomTheta}
\[\label{eq:One.BetheFermiPhase}
\exp \bigbrk{i\Phi_{\fldF}(x_k,x_j)}
=
\frac{1-g^2/2x_k^+x_j^-}{1-g^2/2x_k^-x_j^+}
\]
at third loop order $\order{g^4}$ up to which the function 
$\Phi_{\fldF}$ is known
from \cite{Staudacher:2004tk,Beisert:2003ys}. 
This form thus appears to be a natural asymptotic
generalization of $\Phi_{\fldF}$. For simplicity of notation,
we shall \emph{assume} that the $\alg{su}(1|1)$ sector of 
gauge theory at higher loops is indeed 
described by this scattering phase.
We do not have as much justification for this point
of view as for the $\alg{su}(2)$ sector,
where some calculations up to $\order{g^{10}}$ exist
\cite{Beisert:2004ry}, 
but below we shall see that it neatly fulfills 
some non-trivial requirements. 
Therefore, the asymptotic generalization of the 
Bethe equations \eqref{eq:One.MomBethe,eq:One.MomTheta} 
for the $\alg{su}(1|1)$ sector of gauge theory
apparently reads 
\[\label{eq:One.BetheU11}
\lrbrk{\frac{x^+_k}{x^-_k}}^L=
\prod_{\textstyle\atopfrac{j=1}{j\neq k}}^K
\frac{1-g^2/2x_k^+x_j^-}{1-g^2/2x_k^-x_j^+}\,.
\]
Finally, assuming again \eqref{eq:grouptheory}, we find 
that the asymptotic%
\footnote{Curiously the Bethe ansatz \eqref{eq:One.BetheSL2} works
even better than one might have expected \cite{Staudacher:2004tk}
as noted above.
While maybe not too likely, it is not excluded that
\eqref{eq:One.BetheSL2} reproduces the anomalous
dimensions of twist-two operators to all orders in perturbation theory! 
We hope that an appropriate four-loop field theory computation will be 
performed in the future.} 
form of the conjectured Bethe equation 
\eqref{eq:One.MomBethe,eq:One.MomPhaseD} 
for the $\alg{sl}(2)$ sector should be given by
\[\label{eq:One.BetheSL2}
\lrbrk{\frac{x^+_k}{x^-_k}}^L=
\prod_{\textstyle\atopfrac{j=1}{j\neq k}}^K
\frac{x_k^--x_j^+}{x_k^+-x_j^-}\,
\frac{1-g^2/2x_k^+x_j^-}{1-g^2/2x_k^-x_j^+}\,.
\]
We can combine the asymptotic Bethe equations for all
three sectors in the concise form
\[\label{eq:One.BetheAll}
\lrbrk{\frac{x^+_k}{x^-_k}}^L=
\prod_{\textstyle\atopfrac{j=1}{j\neq k}}^K
\lrbrk{\frac{x_k^+-x_j^-}{x_k^--x_j^+}}^\sgrad\,
\frac{1-g^2/2x_k^+x_j^-}{1-g^2/2x_k^-x_j^+}\,.
\]
Here the parameter $\sgrad$ specifies the sector: 
$\sgrad=+1$ for $\alg{su}(2)$, 
$\sgrad=0$ for $\alg{su}(1|1)$ or
$\sgrad=-1$ for $\alg{sl}(2)$.
For all three sectors, gauge theory states obey the
momentum constraint
\[\label{eq:One.MomConX}
\prod_{k=1}^{K}\frac{x^+_{k}}{x^-_{k}}=1
\]
and their anomalous dimension is given by
\[\label{eq:One.EngX}
\delta D=g^2\sum_{k=1}^{K}
\lrbrk{\frac{i}{x^+_{k}}-\frac{i}{x^-_{k}}}\,.
\]

As discussed in \cite{Beisert:2004hm,Staudacher:2004tk},
the three-loop spectrum obtained from \eqref{eq:One.BetheAll}
for $\sgrad=1,0$ agrees with a large number of states
obtained by direct diagonalization of the Hamiltonian,
cf.~\tabref{tab:TwoEx,tab:SU2,tab:U11}.

\subsection{Bethe Equations for Strings}
\label{sec:One.Strings}

The quantization of IIB string theory in the curved $AdS_5 \times S^5$
geometry is currently not understood. The AdS/CFT conjecture
proposes that free strings moving on that background should
correspond to \emph{planar} $\mathcal{N}=4$ gauge theory. More
precisely, it holds that the energies and eigenstates of free strings 
should map to, respectively, scaling dimensions and eigenstates
of the gauge theory planar dilatation operator. In turn, the latter appear
to be given by the energies and eigenstates of certain novel
long-range quantum spin chains. Assuming the, at least approximate, 
validity of the correspondence, and assuming that the spin chains
do not spontaneously evaporate as one goes from weak to strong
coupling, we conclude that 
\emph{quantum strings on  $AdS_5 \times S^5$ should also be 
described by an integrable long-range spin chain}.

For the best studied case of the $\alg{su}(2)$ sector,
Bethe equations for this ``string chain{}'' were proposed in
\cite{Arutyunov:2004vx}. These were based on a discretization
of the finite gap equation describing the classical sigma
model in this sector \cite{Kazakov:2004qf}. Shortly thereafter
it was demonstrated that these equations indeed diagonalize,
to at least five orders in the coupling constant, a long-range
$\alg{su}(2)$ spin chain similar to the one describing weakly
coupled gauge theory \cite{Beisert:2004jw}. Using the variables
introduced above, the equations read
\[\label{eq:afsb}
\lrbrk{\frac{x^+_k}{x^-_k}}^L
=
\prod_{\textstyle\atopfrac{j=1}{j\neq k}}^K
\frac{x_k^+-x_j^-}{x_k^--x_j^+}\,
\frac{1-g^2/2x_k^+x_j^-}{1-g^2/2x_k^-x_j^+}\,
\sigma^2(x_k,x_j)
\]
with the ``stringy{}'' scattering term \cite{Arutyunov:2004vx}
\[\label{eq:One.StringsPhase}
\sigma(x_k,x_j)=
\exp\bigbrk{i\scataux(x_k,x_j)},
\]
where the phase is given by
\[\label{eq:One.StringsAux}
\scataux(x_k,x_j)=
\sum_{r=2}^\infty
\bigbrk{
\scataux_{r,r+1}(x_k,x_j)-\scataux_{r+1,r}(x_k,x_j)},
\]
with
\[\label{eq:One.StringsAuxTerm}
\scataux_{r,s}(x_k,x_j)=(\half g^2)^{(r+s-1)/2}\, q_r(x_k)\,q_s(x_j).
\]
This term may also be summed and expressed through the spectral parameters as
\cite{Beisert:2004jw}
\[\label{eq:One.StringsAnalytic}
\sigma(x_k,x_j)=
\frac{\displaystyle 1-\frac{g^2}{2x_{k}^- x_{j}^+}}{\displaystyle 1-\frac{g^2}{2x_{k}^+ x_{j}^-}}\,
\lrbrk{\frac{\displaystyle 1-\frac{g^2}{2x_k^- x_j^+}}{\displaystyle 1-\frac{g^2}{2x_k^+ x_j^+}}\,
\frac{\displaystyle 1-\frac{g^2}{2x_k^+ x_j^-}}{\displaystyle 1-\frac{g^2}{2x_k^- x_j^-}}}^{i(u_k-u_j)}.
\]
The stringy scattering term modifies the spectrum of the gauge
theory spin chain at three-loop order $\order{g^6}$.
The lowest-lying three-loop energies of the two similar but distinct
spin chains are found in \tabref{tab:TwoEx,tab:SU2}, and may be
obtained either by direct diagonalization or by solving the Bethe
equations \eqref{eq:afsb}, where $\sigma(x_k,x_j)=1$ for gauge theory and 
$\sigma(x_k,x_j)$ as in \eqref{eq:One.StringsAnalytic} for the string chain.

What about the remaining two-component sectors 
$\alg{su}(1|1)$ and $\alg{sl}(2)$? In \cite{Staudacher:2004tk}
the approach of \cite{Arutyunov:2004vx} was extended to these
cases. An approximate stringy S-matrix was obtained for
$\alg{sl}(2)$ from a discretization of the finite gap equation 
describing the classical sigma model in this sector \cite{Kazakov:2004nh}.
Furthermore, it was argued that the spectrum of strings in the
near-plane wave geometry \cite{Parnachev:2002kk}, 
which had recently be obtained in 
\cite{Callan:2003xr,Callan:2004uv,Callan:2004ev,Callan:2004dt,McLoughlin:2004dh}, 
is consistent with a factorized S-matrix.
The latter was extracted for the three two-component sectors,
yielded a stringy S-matrix for $\alg{su}(1|1)$, and
agreed with the $\alg{sl}(2)$ discretization. A comparison
of the three obtained S-matrices then led to the relation
\eqref{eq:grouptheory}.

However, the obtained S-matrices were only designed, by construction, to
reproduce the string results for the near-BMN and Frolov-Tseytlin
limits, i.e.~the $\order{1/L}$ terms in the S-matrix.
In contradistinction to the $\alg{su}(2)$ case, the 
$\alg{su}(1|1)$ and $\alg{sl}(2)$ S-matrices lacked periodicity in the
momenta and could therefore not exactly correspond to a quantum spin chain.
The existence of such a spin chain encompassing all sectors was 
nevertheless conjectured, along with the proposal \eqref{eq:dressing},
which says that the full S-matrix of the string and the gauge
chain should differ by an overall multiplicative, flavor-independent
dressing factor. {}From the above $\alg{su}(2)$ results we then find the 
factor to be
\[\label{eq:dressingguess}
\hat{S}\upper{dressing}(x_k,x_j)=\sigma^2(x_k,x_j)\, ,
\]
with $\sigma(x_k,x_j)$ as in \eqref{eq:One.StringsAnalytic}.
Given its rather complicated structure we are not sure whether
we have already found the final, analytically exact form of this 
dressing factor. In fact, one would hope that this is not
the case; we would prefer an interpolating function which
smoothly changes from $\sigma=1$ for the weakly coupled
gauge theory%
\footnote{In fact, 
we currently cannot exclude a dressing factor for gauge theory 
which sets in at higher loop orders or non-perturbatively.
For example, as in \cite{Fischbacher:2004iu}, 
a violation of proper scaling beyond three loops,
which has not yet been completely ruled out,
might be induced by a non-trivial $\sigma$.}
to $\sigma$ as given in \eqref{eq:One.StringsAnalytic}
at strong coupling \cite{Arutyunov:2004vx}.

Given the conjecture \eqref{eq:One.BetheAll} in \secref{sec:One.Bethe}
we may then write a Bethe ansatz for the string chain in all
three sectors:
\[\label{eq:One.StringsBethe}
\lrbrk{\frac{x^+_k}{x^-_k}}^L
=
\prod_{\textstyle\atopfrac{j=1}{j\neq k}}^K
\lrbrk{\frac{x_k^+-x_j^-}{x_k^--x_j^+}}^\sgrad\,
\frac{1-g^2/2x_k^+x_j^-}{1-g^2/2x_k^-x_j^+}\,
\sigma^2(x_k,x_j)\, .
\]
A very important test of this proposal is that to three
loop order \eqref{eq:One.StringsBethe} agrees with 
direct diagonalization of the $\alg{su}(1|1)$
string chain Hamiltonian as discussed in 
\cite{Staudacher:2004tk}. The latter is known from 
\cite{Beisert:2003ys,Beisert:2004jw}.

It would be very interesting to compare the $1/L$ corrections to
the Frolov-Tseytlin limit \cite{Frolov:2003tu,Frolov:2004bh,Park:2005ji} 
on the gauge and string side using our Bethe equations \eqref{eq:One.StringsBethe},
cf.~\cite{Beisert:2005mq,Hernandez:2005nf}.

For the remainder of this paper we will always 
include a dressing factor $\sigma(x_k,x_j)$ in the Bethe equations
and assume $\sigma(x_k,x_j)=1$ for gauge theory 
or the expression in \eqref{eq:One.StringsAnalytic} 
for the string chain.

\subsection{Two-Excitation States}
\label{sec:One.Two}

One non-trivial property the Bethe ans\"atze for all
three sectors should satisfy is related to two-excitation states. 
These states form multiplets 
of the superconformal algebra \cite{Beisert:2002tn}
which have the unique property of 
having representatives in all three sectors. 
For these states, the different Bethe equations should therefore 
reproduce the same values for the energies (and higher charges). 
The two Bethe roots are given by $x_1=x$ and $x_2=-x$ due to 
the momentum constraint \eqref{eq:One.MomConX}. 
The Bethe equation now reads
\[\label{eq:One.TwoBethe}
\lrbrk{\frac{x^+}{x^-}}^{L}
=
\lrbrk{\frac{x^++x^+}{x^-+x^-}}^\sgrad\,
\frac{1+g^2/2x^+x^+}{1+g^2/2x^-x^-}\,
\sigma^2(x,-x).
\]
where we have made use of $x_2^\pm=-x^\mp$.
It is now clear that a state of length $L$ in the 
$\alg{su}(2)$ sector with $\sgrad=+1$ has a corresponding state
of length $L-1$ in the 
$\alg{su}(1|1)$ sector with $\sgrad=0$ and  
a corresponding state
of length $L-2$ in the 
$\alg{su}(1,1)$ sector with $\sgrad=-1$.
This is true for the Bethe ans\"atze for spins as well as for strings,
in generalization of the results obtained in \cite{Staudacher:2004tk}.

\subsection{Thermodynamic Limit}
\label{sec:One.Limit}

Let us consider the thermodynamic limit of 
very long spin chains $L\to\infty$
with $g=\order{L}$ 
while keeping the energy $E$ small.
Here we distinguish between two cases: 
\begin{bulletlist}
\item
The near-BMN/plane-wave limit \cite{Parnachev:2002kk,Callan:2003xr} with a 
fixed number of excitations
and $E=\order{1/L^2}$.
The spectrum is described by a
scattering problem.

\item
The Frolov-Tseytlin limit \cite{Frolov:2003qc} with $\order{L}$ excitations
and $E=\order{1/L}$.
The spectrum is described by 
the spectral curve or equivalently by a
Riemann-Hilbert problem.

\end{bulletlist}

Using the expressions in 
\appref{sec:Thermo.Scatter}
it is straightforward to read off the
phases for the pairwise scattering 
from \eqref{eq:One.StringsBethe}.
For gauge theory with $\sigma(x_k,x_j)=1$
we find  
\[\label{eq:One.LimitScatterGauge}
\Phi=
2\sgrad\scatsym+(1-\sgrad)(2\scataux+\scataux_{1,2}-\scataux_{2,1}).
\]
Here $\scatsym$ is the main scattering phase
\[\label{eq:One.LimitScatterMain}
\scatsym(x_k,x_j)=
-i\log\frac{u_k-u_j-i/2}{u_k-u_j+i/2}
=\frac{1}{u_j-u_k}+\order{1/L^2}
\]
and the auxiliary phases, 
cf.~\eqref{eq:One.StringsAux,eq:One.StringsAuxTerm}, 
are given by
\<\label{eq:One.LimitScatterAux}
\scataux(x_k,x_j)\eq
\frac{g^2/2x_k^2}{1-g^2/2x_k^2}\,
\frac{g^2/2x_j^2}{1-g^2/2x_j^2}\,
\frac{1/x_j-1/x_k}{1-g^2/2x_jx_k}
+\order{1/L^2},
\nln
\scataux_{1,2}(x_k,x_j)\eq
\frac{1/x_k}{1-g^2/2x_k^2}\,
\frac{g^2/2x_j^2}{1-g^2/2x_j^2}
+\order{1/L^2},
\nln
\scataux_{2,1}(x_k,x_j)\eq
\frac{g^2/2x_k^2}{1-g^2/2x_k^2}\,
\frac{1/x_j}{1-g^2/2x_j^2}
+\order{1/L^2}.
\>
The corresponding scattering term for the string chain is
\<\label{eq:One.LimitScatterString}
\Phi\eq2\sgrad\scatsym-2\sgrad\scataux+(1-\sgrad)(\scataux_{1,2}-\scataux_{2,1}).
\nln\eq2\sgrad\scatmain+(1+\sgrad)\scataux_{1,2}-(1-\sgrad)\scataux_{2,1}.
\>
with $\scatmain$ the main scattering phase in the $x$-plane
\[\label{eq:One.LimitScatterSec}
\scatmain(x_k,x_j)=
-i\log\frac{x_k-x^+_j}{x_k-x^-_j}
=\frac{1}{1-g^2/2x_j^2}\,\frac{1}{x_j-x_k}+\order{1/L^2}.
\]
The total phase $\Phi$ agrees precisely with the asymptotic form 
for near plane-wave strings derived in \cite{Staudacher:2004tk}.

For many, $K=\order{L}$, excitations, the Bethe equations 
turn into integral equations
\[\label{eq:One.LimitBethe}
2\sgrad \resolvHsl(x)
+F(x)
=-2\pi n_a,
\qquad
\mbox{for } x\in \contour_a.
\]
The functions $H(x),G(x)$ are two different types of resolvents
\[\label{eq:One.LimitResolv}
H(x)=\int_{\contour}\frac{dy\,\rho(y)}{u(y)-u(x)}\,,\qquad
G(x)=\int_{\contour}\frac{dy\,\rho(y)}{1-g^2/2y^2}\,\frac{1}{y-x}\,,
\]
and a slash implies a principal value prescription,
$\resolvHsl(x)=\half H(x+\varepsilon)+\half H(x-\varepsilon)$.
The potential $F(x)$ for gauge theory reads
\[\label{eq:One.LimitPotentialGauge}
F\indup{g}(x)=
\frac{L/x}{1-g^2/2x^2}
+(1-\sgrad)\lrbrk{
2 G(g^2/2x)
-\frac{G'(0)\,g^2/2x}{1-g^2/2x^2}
-\frac{G(0)\,(2-g^2/2x^2)}{1-g^2/2x^2}
}
\]
and for string theory we obtain 
\[\label{eq:One.LimitPotentialString}
F\indup{s}(x)=
\frac{L/x}{1-g^2/2x^2}
-2\sgrad G(g^2/2x)
+2\sgrad G(0)
+\frac{(1+\sgrad)G'(0) g^2/2x}{1-g^2/2x^2}
-\frac{(1-\sgrad)G(0) g^2/2x^2}{1-g^2/2x^2}.
\]
Note that we can write the resulting integral equation for string chains as 
\[\label{eq:One.LimitBetheString}
2\sgrad \resolvsl(x)
+\frac{(1+\sgrad) g^2/2x}{1-g^2/2x^2}\,G'(0)
-\frac{(1-\sgrad)g^2/2x^2}{1-g^2/2x^2} \,G(0)
+\frac{L/x}{1-g^2/2x^2}
=-2\pi n_a,
\quad
\mbox{for } x\in \contour_a
\]
This expression agrees precisely with 
the classical string sigma model in 
\cite{Kazakov:2004qf,Kazakov:2004nh,Beisert:2005bm}.

The potentials $F\indup{g}$ and $F\indup{s}$ are similar
for small values of $g$. This leads to an agreement 
between gauge theory and string theory up to two loops.
Let us try to explain the agreement using 
general features of the Bethe equations. 
The expansion of the Bethe equation for Bethe root $x$
around $x=\infty$ usually yields
the Noether charges.
Here, $2\sgrad H(x)+F(x)$ yields the single global charge
of the symmetry group of the the model in both cases
\[\label{eq:One.LimitPotentialExpand}
2\eta H(x)+F(x)=
\frac{1}{x}
\lrbrk{
L+2\eta K+\half (1-\sgrad)\,\delta D
}+\order{1/x^2}.
\]
When we combine this feature with proper scaling of $x$
we see that an expansion in $1/x$ is equivalent to 
an expansion in $g^2$. 
The agreement of $F\indup{g}$ and $F\indup{s}$ 
up to $\order{1/x}$ leads to an agreement
of the Bethe equations at
$\order{g^2}$. For the energy shifts it means that
gauge theory and string theory agree up to $\order{g^4}$,
i.e.~two loops. 
Starting at three loops the Bethe equations differ
due to $F\indup{s}-F\indup{g}\sim 1/x^2 \sim g^4$.

\subsection{Strong-Coupling Limit}
\label{sec:One.Strong}

One of the fascinating features of the $\alg{su}(2)$ Bethe ansatz for
quantum strings \eqref{eq:afsb} is that it quantitatively reproduces 
the expected strong coupling behavior of anomalous dimensions
\cite{Arutyunov:2004vx}. It was found that this behavior is, for
two-excitation states with mode number $n$
\[\label{eq:strong}
D=D_0+g^2 E=2\,\sqrt[4]{\lambda n^2}\, .
\]
For multi-excitation states the formula is identical, with 
$n=\sum_{k=1}^{K_+} n_k$ where $K_+$ is the number of modes with positive
mode number $n_k>0$. It turned out that to this leading 
order in strong coupling, the result was entirely due to the
scattering effects resulting from the dressing factor, 
i.e.~the relevant Bethe equation,
cf.~\eqref{eq:dressingguess,eq:One.StringsAnalytic}, 
leading to \eqref{eq:strong} is
\[
\prod_{\textstyle\atopfrac{j=1}{j\neq k}}^K
\hat{S}\upper{dressing}(x_k,x_j)=
\prod_{\textstyle\atopfrac{j=1}{j\neq k}}^K
\sigma^2(x_k,x_j)=1\, .
\]
Since our Bethe ansatz for the string chain in the different
sectors differs from gauge theory as in \eqref{eq:dressing}, 
and since it is easy to see that $S\upper{gauge}$ is always
subleading%
\footnote{Strictly speaking this is not true
for the scattering among left-movers and the scattering among
right-movers, i.e.~when $p_k\cdot p_j>0$. It is however easy to see,
cf.~\cite{Arutyunov:2004vx}, that the resulting phase shifts sum to
zero upon substitution into the dispersion relation.}
we see that \eqref{eq:strong}  immediately generalizes 
to those sectors. In fact, getting ahead of ourselves, in light of
\eqref{eq:dressing} it will
turn out to be true for the entire $\alg{psu}(2,2|4)$ string chain,
in line with general expectations \cite{Gubser:1998bc}.

\section{The Factorized S-Matrix of the $\alg{su}(1|2)$ Sector}
\label{sec:SU12}

In the $\superN=4$ gauge theory the planar $\alg{su}(1|2)$ sector 
consists of operators of the type
\[\label{eq:su12ops}
\Tr \,\fldF^{K_1} \fldS^{K_2-K_1} \fldZ^{L-K_2}+ \ldots ,
\]
where $\fldZ$ and $\fldS$ are two out of the
three complex adjoint scalars of the $\superN=4$ model, and
$\fldF$ is an adjoint gaugino (in $\superN=1$ connotation).
The dots indicate that we need to consider all possible
orderings of the fields inside the trace, and
diagonalize the set of such operators with respect to dilatation.
As in the previous chapter, this is most easily done 
when interpreting the dilatation operator as a Hamiltonian
acting on a spin chain of length $L$. This requires opening up
the trace and replacing it by a quantum mechanical state on a 
one-dimensional lattice of $L$ sites:
\[\label{eq:replace}
\makebox[0cm][l]{\huge \hspace{-0.05em}\raisebox{-0.1em}{$\times$}} 
\Tr \left( \fldZ \fldZ \fldS \fldF \ldots \fldS \fldZ \right)
\longrightarrow
| \fldZ \fldZ \fldS \fldF \ldots \fldS \fldZ \rangle.
\]
The fact that the original operators \eqref{eq:su12ops} are single-trace
has two consequences for the spin chain interpretation, to be
distinguished: ($i$) The trace links the matrix indices of the first and
the last constituent field. The boundary conditions of the spin chain are
therefore periodic. ($ii$) The trace has the property of cyclicity. This means
that only the subset of translationally invariant states
of the spin chain are relevant to gauge theory.%
\footnote{In the presence of fermions, as in \eqref{eq:replace},
we have to use a shift operator which is properly graded.}

A new feature compared to \Secref{sec:One}
is that we now have a three-component system.
If we choose the fields $\fldZ$ as our reference (=vacuum) configuration,
the fields $\fldS$, $\fldF$ may be regarded as two distinct
types of excitations of the chain.
The total number of excitations is $K_2$. 

As in \Secref{sec:One} we will assume that a modified version of the
spin chain describing weakly coupled gauge theory applies to strongly
coupled quantum string theory. This is very natural, as we can think of
the $\alg{su}(1|2)$ sector as a ``unification{}'' of the $\alg{su}(2)$ and
$\alg{su}(1|1)$ sectors of the previous sectors. The deformation which
takes us from gauge to string theory in these (compact) two-component sectors 
immediately ``lifts{}'' to their three-component unification. 
It is nevertheless important to keep in mind that we do not currently know 
how to find the deformed spin chain, i.e.~the analog of \eqref{eq:su12ops},  
directly from the quantum string sigma model. 

\subsection{One-Loop Scattering in the $\alg{su}(1|2)$ Sector}
\label{sec:su12.1loop}

The planar one-loop Hamiltonian in the closed $\alg{su}(1|2)$ sector reads
\[\label{eq:su12.hamiltonian}
\ham_0=\sum_{\ell=1}^L \,(1-\Pi_{\ell,\ell+1})\, .
\]
It may be extracted from the complete one-loop $\superN=4$ 
dilatation operator \cite{Beisert:2003jj} and rewritten
with the help of the graded permutation operator $\Pi_{\ell,\ell+1}$
which exchanges the partons at sites $\ell$ and $\ell+1$, picking
up a minus sign if the exchange involves two fermions~$\fldF$.

This sector unifies the $\alg{su}(2)$ and the $\alg{su}(1|1)$ subsectors
of the previous \Secref{sec:One}. At 
one loop it again corresponds to an important nearest-neighbor 
integrable spin chain of condensed matter theory, the famous 
supersymmetric $t$-$J$ model
believed to be relevant to the understanding of high $T\indup{c}$ 
superconductivity \cite{Anderson:1987gf,Zhang:1988aa}. 
Its supersymmetry was apparently first 
noticed in \cite{Wiegmann:1988qn,Forster:1989nn}.
It was first solved by the coordinate-space Bethe ansatz in
\cite{Schlottmann:1987aa} and by the algebraic Bethe ansatz in
\cite{Essler:1992nk,Foerster:1992uk}. An equivalent lattice gas model
was solved much earlier by a coordinate-space Bethe ansatz
\cite{Lai:1974aa,Sutherland:1975vr}. As a warm-up for the long-range case,
let us briefly review the latter method, extending the
explanations of \cite{Staudacher:2004tk} to the three-component case. 

Since the Hamiltonian \eqref{eq:su12.hamiltonian} is integrable, 
it again suffices to consider two-body states 
in order to derive the many-body S-matrix. 
The cases of two $\fldS$'s or two $\fldF$'s have been considered
in \cite{Staudacher:2004tk} and we are left to consider
the mixed case of one $\fldS$ and $\fldF$ each.
These states are, in an obvious notation 
\[\label{eq:su12.state}
|\Psi \rangle=
\left(\begin{array}{c}
 \\
|\fldF \fldS\rangle \\
 \\
 \\
|\fldS \fldF\rangle \\
 \\ 
\end{array}\right)=
\left(\begin{array}{c}
\\
\sum_{1\leq \ell_1 < \ell_2\leq L} \Psi_{\fldF \fldS}(\ell_1,\ell_2)~
| ... \fldZ \smash{\MMM{\fldF}{\ell_1}} \fldZ ...\fldZ \smash{\MMM{\fldS}{\ell_2}} \fldZ...\rangle \\
\\
\\
\sum_{1\leq \ell_1 < \ell_2\leq L} \Psi_{\fldS \fldF}(\ell_1,\ell_2)~
| ... \fldZ \smash{\MMM{\fldS}{\ell_1}} \fldZ ...\fldZ \smash{\MMM{\fldF}{\ell_2}} \fldZ...\rangle \\
 \\
\end{array}\right)\, ,
\]
%
where $\ell_{1,2}$ denotes the positions (with $\ell_1<\ell_2$)
of the two ``particles{}''
$\fldF$ and $\fldS$ inside the opened trace. In this lattice gas picture
the fields $\fldZ$ are ``vacancies{}'', i.e.~unoccupied lattice sites.
In contrast to the two-component case we now need to take into
account that the two excitations are distinguishable. The
number of components of the above wave function \eqref{eq:su12.state}
corresponds to their possible orderings (two, for the moment).

Acting with the Hamiltonian \eqref{eq:su12.hamiltonian} on the
state \eqref{eq:su12.state} we find the Schr\"odinger
equations $\ham_0 \cdot |\Psi_{..}\rangle=E_0\,|\Psi_{..}\rangle$
for the position space wave functions
\<\label{eq:su12.schroedingerO}
\mbox{for} \quad \ell_2> \ell_1+1 : \earel{} \\ 
E_0\,\Psi_{\fldF \fldS}(\ell_1,\ell_2)\eq 
2\,\Psi_{\fldF \fldS}(\ell_1,\ell_2)-\Psi_{\fldF \fldS}(\ell_1-1,\ell_2)-
\Psi_{\fldF \fldS}(\ell_1+1,\ell_2)+
\nl
+2\,\Psi_{\fldF \fldS}(\ell_1,\ell_2)-\Psi_{\fldF \fldS}(\ell_1,\ell_2-1)-
\Psi_{\fldF \fldS}(\ell_1,\ell_2+1)\, ,
\nonumber  \\ \label{eq:su12.schroedingerII}
\mbox{for} \quad  \ell_2= \ell_1+1:\earel{}
 \\
E_0\,\Psi_{\fldF \fldS}(\ell_1,\ell_2)\eq 
3\,\Psi_{\fldF \fldS}(\ell_1,\ell_2)-\Psi_{\fldF \fldS}(\ell_1-1,\ell_2)-
\Psi_{\fldF \fldS}(\ell_1,\ell_2+1)-
\nl
-\Psi_{\fldS \fldF}(\ell_1,\ell_2)\, ,
\nonumber 
\>
with a second set of equations for the wave functions 
$\Psi_{\fldS \fldF}(\ell_1,\ell_2)$ which are identical to  
\eqref{eq:su12.schroedingerO,eq:su12.schroedingerII} 
after exchanging $\fldF \leftrightarrow \fldS$ everywhere.

These difference equations are solved by an appropriate Bethe ansatz:
\<\label{eq:su12.1loopBA}
\Psi_{\fldF \fldS}(\ell_1,\ell_2)
\eq
A_{\fldF \fldS}\,e^{i p_1 \ell_1+i p_2 \ell_2}+
A'_{\fldF \fldS}\,e^{i p_2 \ell_1+i p_1 \ell_2}
\nln
\Psi_{\fldS \fldF}(\ell_1,\ell_2)
\eq
A_{\fldS \fldF}\,e^{i p_1 \ell_1+i p_2 \ell_2}+
A'_{\fldS \fldF}\,e^{i p_2 \ell_1+i p_1 \ell_2}\, .
\>
The idea is that the partons, coming in as an arbitrary mixed state
with initial amplitudes 
$A_{\fldF \fldS}$, $A_{\fldS \fldF}$ propagate freely
down the trace with fixed momenta $p_1$,$p_2$ until they scatter
at $\ell_2=\ell_1+1$. Before they do, i.e. when
$\ell_2>\ell_1+1$, the Schr\"odinger equation 
\eqref{eq:su12.schroedingerO} is satisfied for arbitrary 
amplitudes as long as the dispersion law
\[\label{eq:1Ldispersion}
E_0=\sum_{k=1}^{K_2}\, 4\, \sin^2(\half p_k)\, 
\]
(with $K_2=2$, for the moment) holds. However, when the particles hit
each other they may exchange momenta,
and in addition, unlike in the two-component case, exchange their flavors.
This scattering process is \emph{non-diffractive} if the individual
momenta $p_k$ are separately conserved.
The Bethe ansatz \eqref{eq:su12.1loopBA} assumes this to be true
in that the outgoing configuration, with amplitudes
$A'_{\fldF \fldS}$, $A'_{\fldS \fldF}$, is simply added to the wave function.
This is clearly still consistent with the generic case $\ell_2>\ell_1+1$
if the energy expression \eqref{eq:1Ldispersion} holds. 
But now we must check whether the 
amplitudes may be adjusted such that
the Schr\"odinger equation \eqref{eq:su12.schroedingerII} at
$\ell_2=\ell_1+1$ is satisfied as well. This may be achieved by the ansatz
\[\label{eq:su12.trans-ref}
\left(
\begin{array}{c} A'_{\fldS \fldF} \\ \\ A'_{\fldF \fldS} \end{array}
\right)
=
\left(\begin{array}{cc}
T^{\fldF \fldS}_{\fldF \fldS}(p_2,p_1)&R^{\fldF \fldS}_{\fldS \fldF}(p_2,p_1)
\\
\\
R^{\fldS \fldF}_{\fldF \fldS}(p_2,p_1)&T^{\fldS \fldF}_{\fldS \fldF}(p_2,p_1) 
\end{array}\right)
\left(
\begin{array}{c} A_{\fldF \fldS} \\ \\ A_{\fldS \fldF} \end{array}
\right)\, .
\]
and one finds, after substituting the Bethe ansatz
\eqref{eq:su12.1loopBA} into \eqref{eq:su12.schroedingerII}
\<\label{eq:ttrr}
T^{\fldF \fldS}_{\fldF \fldS}(p_1,p_2)=T^{\fldS \fldF}_{\fldS \fldF}(p_1,p_2)\eq
\frac{e^{i p_1}-e^{i p_2}}{e^{i p_1+i p_2}-2 e^{i p_2}+1}\, ,
\nln
R^{\fldF \fldS}_{\fldS \fldF}(p_1,p_2)=R^{\fldS \fldF}_{\fldF \fldS}(p_1,p_2)\eq
-\frac{(1-e^{i p_1})(1-e^{i p_2})}{e^{i p_1+i p_2}-2 e^{i p_2}+1}\, .
\>
Here $T$ denotes the transmission amplitudes (the particles pass
through each other) and $R$ the reflection amplitudes (the particles
backscatter). Note that the order of the particles in the
outgoing amplitude $A'$ on the l.h.s.~of \eqref{eq:su12.trans-ref}
is reversed: Indeed, by our conventions, if the particles transmit 
their order will change. 
Together with the well-known 
expressions for the two-body one-loop S-matrices 
for $\fldS$-$\fldS$ scattering and $\fldF$-$\fldF$ scattering 
in the $\fldZ$ vacuum 
(see \cite{Staudacher:2004tk} for a derivation in the same elementary
fashion as just presented)%
\footnote{As opposed to \cite{Staudacher:2004tk} 
we use the convention that the wave function has implicit 
factors of $-1$ from the crossing of fermionic fields. 
Therefore the element of the S-matrix 
$S^{\fldF \fldF}_{\fldF \fldF}=+1$
merely contributes the scattering phase on top of statistics
(none here).}
\[\label{eq:su2.u11}
S^{\fldS \fldS}_{\fldS \fldS}(p_1,p_2)=
-\frac{e^{i p_1+i p_2}-2 e^{i p_1}+1}{e^{i p_1+i p_2}-2 e^{i p_2}+1}\, ,
\qquad 
S^{\fldF \fldF}_{\fldF \fldF}(p_1,p_2)=1\, ,
\]
we now have the full two-body S-matrix.
We will write it down in the basis 
$|\fldS \fldS \rangle$,
$|\fldF \fldS \rangle$,
$|\fldS \fldF \rangle$,
$|\fldF \fldF \rangle$,
using a vector notation 
$(1,0,0,0)$, $(0,1,0,0)$, $(0,0,1,0)$, $(0,0,0,1)$
for these two-body states.
We can think of these as the possible configurations of
a very short ``auxiliary{}'' two-component spin chain.
The two-body S-matrix is then an operator acting on this short
spin chain and we thus find, 
in the so-called transmission-diagonal convention,
\[\label{eq:su12.1loop.Sp}
S_{k,j}(p_k,p_j)=
\left(\begin{array}{cccc}
S^{\fldS \fldS}_{\fldS \fldS}(p_k,p_j)& & & 
\\
&T^{\fldF \fldS}_{\fldF \fldS}(p_k,p_j)&R^{\fldF \fldS}_{\fldS \fldF}(p_k,p_j)&
\\
\\
&R^{\fldS \fldF}_{\fldF \fldS}(p_k,p_j)&T^{\fldS \fldF}_{\fldS \fldF}(p_k,p_j)& 
\\
& & & S^{\fldF \fldF}_{\fldF \fldF}(p_k,p_j)
\end{array}\right)\, ,
\]
where matrix elements which are zero were left empty.

Integrability now means that, when we consider an arbitrary
number of particles $\fldS,\fldF$, the many-body S-matrix 
factorizes into a product of the two-body S-matrices
\eqref{eq:su12.1loop.Sp}.
A necessary consistency condition
for this to be true is the Yang-Baxter equation
\[\label{eq:ybe}
S_{3,2} S_{3,1} S_{2,1}=S_{2,1} S_{3,1} S_{3,2}\, ,
\]
along with unitarity $S_{2,1} S_{1,2}=1$,  
where we have abbreviated $S_{k,j}=S_{k,j}(p_k,p_j)$.
These properties may be checked explicitly for
our case \eqref{eq:su12.1loop.Sp} (best done
with a symbolic manipulation program)
by acting with both sides of \eqref{eq:ybe} on 
the eight-dimensional vector space spanned by three particles of
two possible flavors, i.e.~the basis of this state space is 
$|\fldS \fldS \fldS\rangle$,
$|\fldF \fldS \fldS\rangle$,
$|\fldS \fldF \fldS\rangle$,
$|\fldS \fldS \fldF\rangle$,
$|\fldF \fldF \fldS\rangle$,
$|\fldF \fldS \fldF\rangle$,
$|\fldS \fldF \fldF\rangle$,
$|\fldF \fldF \fldF\rangle$.
Again, it is useful to visualize this state space as the one
of a short two-component spin chain, this time of length three.
An important consequence is that \eqref{eq:ybe} allows to extend the
Bethe ansatz \eqref{eq:su12.1loopBA} to an arbitrary number
of particles. One may check that that the following
Bethe wave function is an eigenfunction of the Hamiltonian
with eigenvalue \eqref{eq:1Ldispersion}
\[\label{eq:multiBA}
\Psi_{\ldots\, \fldS \fldS \fldF \fldS\, \ldots}(\ell_1,\ldots,\ell_{K_2})
=
\sum_{\sigma}
A_{\ldots\, \fldS \fldS \fldF \fldS\, \ldots}^{(\sigma)}\,
\exp \left(i \sum_{k=1}^{K_2} p_{\sigma(k)} \ell_k\right)\, ,
\]
where $\ell_1<\ell_2<\ldots<\ell_{K_2}$.
We now need to distinguish all possible orderings of the
$K_2-K_1$ particles of type $\fldS$ and the $K_1$ particles
of type $\fldF$. As before we may identify the
various configurations with the states 
$|\ldots\, \fldS \fldS \fldF \fldS\, \ldots\rangle$
of a shorter (length=$K_2$)
spin chain. The sum in \eqref{eq:multiBA} runs over all 
$K_2!$ permutations $\sigma$, caused by the scattering, of the particle 
momenta $p_k$.

We found the S-matrix by studying the scattering problem on an
infinite lattice. The Schr\"odinger equation is satisfied
for any values of the $K_2$ momenta. They become quantized upon
imposing periodic boundary conditions on Bethe's wave function
\eqref{eq:multiBA}. If we push any one, say the $k$'th, of the $K_2$ particles once
around the circle of the chain (i.e.~transforming $\ell_k \rightarrow \ell_k+L$
in \eqref{eq:multiBA}) it acquires a free particle phase factor 
$\exp(i p_k L)$, which gets shifted by the collision with
the $K_2-1$ fellow particles.
This procedure leads to the set of $K_2$ Bethe equations
\[\label{eq:mbethe}
e^{i p_k L}\, |\Psi \rangle
=S_{k,k+1} \ldots S_{k,K_2} S_{k,1} 
\ldots S_{k,k-1} \,  \cdot |\Psi \rangle \, ,
\]
where we have again abbreviated $S_{k,j}=S_{k,j}(p_k,p_j)$.
However, note that, unlike in the two-component case, these
are {\it matrix} Bethe equations. We still need to find the
states $ |\Psi \rangle$ such that \eqref{eq:mbethe} is
simultaneously satisfied for {\it all} $k=1,\ldots,K_2$, i.e. the eigenvector
$|\Psi\rangle$ should {\it not} depend on $k$. 
We may consider the vector $ |\Psi \rangle$ as a state in an 
\emph{inhomogeneous} 
nearest-neighbor spin chain made from two components.
The inhomogeneity is due to the fact that the particles
in the short chain carry labels, namely their momenta in the long chain.
The (somewhat tedious) diagonalization of this ``smaller{}'' spin chain 
hidden in the original chain will be postponed to \secref{sec:nba}. 
The reason is that we shall find that the all-loop case is not a bit harder
than the one-loop case! 

For simplicity, we change variables from momenta $p_k$ to rapidities
$u_k=\half \cot(\half p_k)$. 
The S-matrix \eqref{eq:su12.1loop.Sp} then becomes,
using \eqref{eq:ttrr},
\[\label{eq:su12.Su}
S_{k,j}(u_k,u_j)=\frac{1}{u_k-u_j-i}
\left(\begin{array}{cccc}
u_k-u_j+i& & & 
\\
&u_k-u_j&i&
\\
\\
&i&u_k-u_j& 
\\
& & & u_k-u_j-i
\end{array}\right)\, .
\]

Given the insights and findings of \Secref{sec:One} we could
go on and try to immediately write down, using e.g.~the rules
\eqref{eq:One.SpecIdent}, a conjecture for the all-loop 
(asymptotic) generalization of the one-loop S-matrix \eqref{eq:su12.Su}. 
Focusing on the numerator $u_k-u_j$ of the transmission amplitudes, we 
however face the problem that there are three combinations which
reduce at weak coupling to this expression:
$x_k^+-x_j^+$, $x_k^--x_j^-$, as well as $x_k-x_j$.
Likewise, it is not clear how to ``deform{}'' the off-diagonal
reflection factors $i$.
Let us therefore go back to the three-loop $\alg{su}(1|2)$
dilatation operator \cite{Beisert:2003ys} and apply the
perturbative asymptotic Bethe ansatz \cite{Staudacher:2004tk} 
in order to find the proper higher-loop generalization of
the S-matrix \eqref{eq:su12.Su}.

\subsection{Three-Loop S-Matrix Extraction for $\alg{su}(1|2)$ }
\label{sec:PABA}

The $\alg{su}(1|2)$ sector is manifestly embedded into the
$\alg{su}(2|3)$ sector studied in \cite{Beisert:2003ys}.
If we act with the three-loop Hamiltonian derived in \cite{Beisert:2003ys}
on the Bethe wave function \eqref{eq:su12.1loopBA} we find that it
is still a solution if the excitations are ``sufficiently far{}''
apart, i.e.~iff $\ell_2 > \ell_1+3$. 
It is therefore reasonable to expect that \emph{asymptotically} we still have%
\footnote{More generally the
asymptotic region is  $\ell_2 > \ell_1+\ell$, where $\ell$ is
the order of perturbation theory.}
\<\label{eq:su12.ABA}
\Psi_{\fldF \fldS}(\ell_1,\ell_2)
\sim
A_{\fldF \fldS}\,e^{i p_1 \ell_1+i p_2 \ell_2}+
A'_{\fldF \fldS}\,e^{i p_2 \ell_1+i p_1 \ell_2}\, ,
\nln
\Psi_{\fldS \fldF}(\ell_1,\ell_2)
\sim
A_{\fldS \fldF}\,e^{i p_1 \ell_1+i p_2 \ell_2}+
A'_{\fldS \fldF}\,e^{i p_2 \ell_1+i p_1 \ell_2}\,.
\>
The idea to extract the 
S-matrix of a long-range system from the asymptotic wave function
is apparently due to Sutherland \cite{Sutherland:1978aa}.
Accordingly, we expect to still be able to infer, using
\eqref{eq:su12.trans-ref}, the correct transmission and reflection 
coefficients from \eqref{eq:su12.ABA}. However, in order
to find the asymptotic phase shifts we need to
multiply the various exponentials in  \eqref{eq:su12.ABA} by
appropriate ``correction factors{}''
$C(\ell_2-\ell_1,p_1,p_2,g)$ with the property that
$C = 1+\order{g^6}$ for $\ell_2 > \ell_1+3$:
\<\label{eq:su12.3loopPABA}
\Psi_{\fldF \fldS}(\ell_1,\ell_2)
\eq
A_{\fldF \fldS}\,C_{\fldF \fldS}(\ell_2-\ell_1,p_1,p_2,g)\,
e^{i p_1 \ell_1+i p_2 \ell_2}+
\nl
+A'_{\fldF \fldS}\,C'_{\fldF \fldS}(\ell_2-\ell_1,p_1,p_2,g)\,
e^{i p_2 \ell_1+i p_1 \ell_2}\, ,
\nln
\Psi_{\fldS \fldF}(\ell_1,\ell_2)
\eq
A_{\fldS \fldF}\,C_{\fldS \fldF}(\ell_2-\ell_1,p_1,p_2,g)\,
e^{i p_1 \ell_1+i p_2 \ell_2}+
\nl
+A'_{\fldS \fldF}\,C'_{\fldS \fldF}(\ell_2-\ell_1,p_1,p_2,g)\,
e^{i p_2 \ell_1+i p_1 \ell_2}\, .
\>
This is the perturbative asymptotic Bethe ansatz (PABA) proposed
in \cite{Staudacher:2004tk}. A choice%
\footnote{As in the case of the $\alg{su}(2)$ sector \cite{Fischbacher:2004iu}
we need a slightly more general three-loop ansatz compared to the
$\alg{su}(1|1)$ sector initially treated in \cite{Staudacher:2004tk},
where it turned out that $C^{(2)}(1,p1,p2)=C^{(2)}(2,p1,p2)$.}
which allows to satisfy the necessary consistency conditions 
in the interaction region $\ell_1+4> \ell_2 > \ell_1$ is
\<\label{eq:fudge}
C_{\fldF \fldS}(\ell_2-\ell_1,p_1,p_2,g)
\eq
1+C_{\fldF \fldS}^{(2)}(\ell_2-\ell_1,p_1,p_2)\, g^{2(\ell_2-\ell_1)}+
\nl\quad+C_{\fldF \fldS}^{(4)}(\ell_2-\ell_1,p_1,p_2)\, g^{2+2(\ell_2-\ell_1)}
+{\cal O}(g^6)\, .
\>
%
%
with analogous expressions for the remaining three factors (with either
$\fldF \leftrightarrow \fldS$, or $C \leftrightarrow C'$, or both).
The only role of the correction factors is to correctly disentangle
short-range effects from the actual long-range phase shifts determining
\eqref{eq:su12.ABA}.

After a significant amount of algebra one finds 
from \cite{Beisert:2003ys} for the two-loop correction factors
\<
C^{(2)}_{\fldF \fldS}(1,p_1,p_2)
\eq
\quarter\,(e^{-i p_2}-1)(1+4\,i\,\gamma_1-2\,e^{i p_1})\, ,
\nln
{C'}^{(2)}_{\fldF \fldS}(1,p_1,p_2)
\eq
C^{(2)}_{\fldF \fldS}(1,p_2,p_1)\, ,
\nln
C^{(2)}_{\fldS \fldF}(1,p_1,p_2)
\eq
C^{(2)}_{\fldF \fldS}(1,-p_2,-p_1)\, ,
\nln
{C'}^{(2)}_{\fldS \fldF}(1,p_1,p_2)
\eq
C^{(2)}_{\fldF \fldS}(1,-p_1,-p_2)\, .
\>
Here $\gamma_1$ is one of the three gauge parameters
appearing in the two-loop vertex.
The two-loop transmission amplitudes are found to be
\<\label{eq:tt}
T^{\fldF \fldS}_{\fldF \fldS}(p_1,p_2)\eq
\frac{e^{i p_1}-e^{i p_2}}{e^{i p_1+i p_2}-2 e^{i p_2}+1}
\nl
+
\frac{(1-e^{i p_1})(1-e^{i p_2})^2
(3-e^{2 i p_1}-2 e^{-i p_1+i p_2}-e^{i p_1-i p_2}+e^{i p_1+i p_2})}
{2 (e^{i p_1+i p_2}-2 e^{i p_2}+1)^2}\, g^2
\nl
+
\order{g^4}\, ,
\nln
T^{\fldS \fldF}_{\fldS \fldF}(p_1,p_2)\eq
\frac{e^{i p_1}-e^{i p_2}}{e^{i p_1+i p_2}-2 e^{i p_2}+1}
\nl+
\frac{(1-e^{i p_1})^2(1-e^{i p_2})
(3 e^{i p_2}-e^{i p_1}+e^{-i p_1}-e^{-i p_2}-2 e^{-i p_1+2 i p_2})}
{2 (e^{i p_1+i p_2}-2 e^{i p_2}+1)^2}\, g^2
\nl+
\order{g^4}\, ,
\>
while the two-loop reflection amplitudes are
\<\label{eq:rr}
R^{\fldF \fldS}_{\fldS \fldF}(p_1,p_2)\eq
-\frac{(1-e^{i p_1})(1-e^{i p_2})}{e^{i p_1+i p_2}-2 e^{i p_2}+1}\,
\left(1-(i \gamma_1+\quarter) (e^{i p_1}+e^{-i p_1}-e^{i p_2}-e^{-i p_2})\,
g^2 \right)
\nl
-
\frac{(1-e^{i p_1})(1-e^{i p_2})^2}
{2 (e^{i p_1+i p_2}-2 e^{i p_2}+1)^2}
(1+2 e^{- i p_1}-2 e^{i p_1}+e^{2 i p_1}-2 e^{-i p_1+i p_2}+
\nl
\qquad \qquad \qquad \qquad \qquad \qquad \qquad
+e^{i p_1-i p_2}-2 e^{-i p_2}+2 e^{i p_2}-e^{i p_1+i p_2})\, g^2
\nl+
\order{g^4}\, ,
\nln
R^{\fldS \fldF}_{\fldF \fldS}(p_1,p_2)\eq
-\frac{(1-e^{i p_1})(1-e^{i p_2})}{e^{i p_1+i p_2}-2 e^{i p_2}+1}\,
\left(1+(i \gamma_1+\quarter) (e^{i p_1}+e^{-i p_1}-e^{i p_2}-e^{-i p_2})\, 
g^2 \right)
\nl+
\frac{(1-e^{i p_1})^2(1-e^{i p_2})}
{2 (e^{i p_1+i p_2}-2 e^{i p_2}+1)^2} 
(2+e^{-i p_1}-e^{i p_1}-2 e^{-i p_1+i p_2}+2 e^{i p_1+i p_2}+
\nl
\qquad \qquad \qquad \qquad \qquad \qquad \qquad 
+2 e^{-i p_1+2 i p_2}-e^{-i p_2}-e^{i p_2}-2 e^{2 i p_2})\, g^2
\nl+
\order{g^4}\, .
\>

We have also explicitly obtained, using computer algebra, the 
three-loop correction factors
$C^{(2)}(2,p_1,p_2)$, $C^{(4)}(1,p_1,p_2)$
as well as the three-loop modifications of the
S-matrix elements, i.e.~the $\order{g^4}$ corrections to
$T^{\fldF \fldS}_{\fldF \fldS}$,$T^{\fldS \fldF}_{\fldS \fldF}$ in
\eqref{eq:tt} and 
$R^{\fldF \fldS}_{\fldS \fldF}$,$R^{\fldS \fldF}_{\fldF \fldS}$ in
\eqref{eq:rr}. The results are too long to
display here. They are however crucial for checking
our conjectures for the asymptotic all-loop S-matrices in the  
$\alg{su}(1|2)$ sectors of gauge and string theory 
in the following \Secref{sec:su12conjecture}.

\subsection{All-Loop Factorized S-Matrix for $\alg{su}(1|2)$: A Conjecture}
\label{sec:su12conjecture}

Armed with the perturbative three-loop results for the S-matrix
we may now extend the all-loop conjectures of \secref{sec:One}
in an informed fashion to $\alg{su}(1|2)$.
The first thing to notice about the amplitudes  
we found is that neither the two transmission amplitudes \eqref{eq:tt}
nor the two reflection amplitudes \eqref{eq:rr} are, respectively,
identical, in contradistinction to the one-loop case \eqref{eq:su12.Su}. 
Let us first focus on transmission. As already mentioned at
the end of \secref{sec:su12.1loop}, in $x$-space there are three natural
combinations which reduce to the numerator $u_k-u_j$ of the
one-loop transmission amplitude.
Expanding these to three-loop order and changing variables
to momenta $p_k$, $p_j$ one finds, excitingly,
that choosing for the numerators of the amplitudes the
two combinations $x_k^+-x_j^+$ and $x_k^--x_j^-$ exactly reproduces,
respectively, the highly involved three-loop expressions for
$T^{\fldF \fldS}_{\fldF \fldS}$ and $T^{\fldS \fldF}_{\fldS \fldF}$,
as displayed (to two loops) in \eqref{eq:tt}.

Turning our attention to reflection, we see that
the amplitudes are not only asymmetric, but also
depend on the gauge parameters. 
They comprise some of the undetermined 
coefficients of the Hamiltonian which correspond to
similarity transformations.
These ambiguities are inevitable in any renormalization scheme.
To proceed, observe that at one loop 
the eigenvalues of the fermion-boson block
\[
\left(\begin{array}{cc}
T^{\fldF \fldS}_{\fldF \fldS}&R^{\fldF \fldS}_{\fldS \fldF}
\\
\\
R^{\fldS \fldF}_{\fldF \fldS}&T^{\fldS \fldF}_{\fldS \fldF} 
\end{array}\right)
\]
are precisely the S-matrix elements
$S^{\fldS \fldS}_{\fldS \fldS}$ and $S^{\fldF \fldF}_{\fldF \fldF}$. 
One may check that this remains true at two and three loops.
Assuming this to be true at any order, and using our just
presented conjecture for transmission, one finds that
\[\label{eq:zwang}
(R^{\fldS \fldF}_{\fldF \fldS})_{k,j}\,(R^{\fldF \fldS}_{\fldS \fldF})_{k,j}=
\frac{S_0^2(x_k,x_j)}{(x_k^- - x_j^+)^2}\,   
(x_k^+ - x_k^-)(x_j^+ - x_j^-)\, .
\]
We now observe that the combinations $x_k^+ - x_k^-$ and
$x_j^+ - x_j^-$ reduce at one loop precisely to the numerators
$i$ of the reflection amplitudes, cf.~\eqref{eq:su12.Su}.
After some experimenting, and taking into account \eqref{eq:zwang} 
one finds that the combinations 
${(x_j^+ - x_j^-)}\,\omega_{k}/\omega_{j}$ and 
${(x_k^+ - x_k^-)}\,\omega_{j}/\omega_{k}$ reproduce, respectively, 
the two- and three-loop numerators of the reflection amplitudes
$R^{\fldF \fldS}_{\fldS \fldF}$ and $R^{\fldS \fldF}_{\fldF \fldS}$.
The gauge parameters appearing in the reflection amplitudes may be
absorbed in the functions $\omega_k$. 
To two loops, cf.~\eqref{eq:rr}, one has
\[\label{eq:omgam}
\omega_k=(x_k^+-x_k^-)^{1/2+2 i \gamma_1}\,+\order{g^4}.
\]
At three loops, the functional form of the function $\omega_k$ becomes
more involved if one keeps all gauge parameters.

We are then led to the following form for the all-loop asymptotic S-matrix:
\[\label{eq:su12.Sx}
S_{k,j}(x_k,x_j)=\frac{S_0(x_k,x_j)}{x_k^- - x_j^+}
\left(\begin{array}{cccc}
x_k^+ - x_j^-& & & 
\\
&x_k^+ - x_j^+&(x_j^+ - x_j^-)\,\omega_{k}/\omega_{j}&
\\
\\
&(x_k^+ - x_k^-)\,\omega_{j}/\omega_{k}&x_k^- - x_j^-& 
\\
& & & x_k^- - x_j^+
\end{array}\right)\, .
\]
%
%
%
The global dressing factor $S_0(x_{2,k},x_{2,j})$,
differing slightly between gauge and string theory, 
is the same as found in \secref{sec:One}:
%
\[\label{eq:longrangeS}
S_0(x_k,x_j)=
\frac{1-g^2/2x_k^+ x_j^-}{1-g^2/2x_k^- x_j^+}\,
\sigma^2(x_k,x_j)\, .
\]
%
Excitingly, this \emph{all-loop} S-matrix still satisfies the
Yang-Baxter equation (YBE) \eqref{eq:ybe}. Note that, unlike in
most known solutions of the YBE, our S-matrix 
\eqref{eq:su12.Sx} appears (as far as we
can see) to not be expressible in terms of the difference
of some suitable spectral parameter.%
\footnote{Note that \eqref{eq:su12.Sx} actually satisfies the YBE without 
any assumption on the relation between $x^+$ and $x^-$. It would be interesting
to know whether \eqref{eq:su12.Sx} may be transformed to a known
solution of the YBE, or whether it constitutes a hitherto unknown, 
novel solution.}

We expect the spectrum to be independent of the function $\omega_k$, 
as it merely represents a renormalization of the basis of states of the form
\[
\state{\fldS\ldots\MMM{\fldF}{k_1}\ldots \MMM{\fldF}{\ldots}\ldots\MMM{\fldF}{k_{K_1}}\ldots\fldS}
\longrightarrow
\omega_{k_1}\cdots
\omega_{k_{K_1}}
\state{\fldS\ldots\MMM{\fldF}{k_1}\ldots \MMM{\fldF}{\ldots}\ldots\MMM{\fldF}{k_{K_1}}\ldots\fldS}.
\]
Applying a reality condition on the Hamiltonian 
one can actually eliminate such spurious degrees of freedom
from the S-matrix (at least up to three loops).
This should then lead to a symmetric S-matrix. 
{}From \eqref{eq:zwang,eq:su12.Sx} we find the unique result to be
\[
\omega_k=\sqrt{x_k^+-x_k^-}\, .
\]
To second loop order this corresponds to setting the gauge
parameter $\gamma_1=0$, as seen from \eqref{eq:omgam}.
Actual computations, however, are simplified by the
choice $\omega_k=1$ which we adopt in the following.
We leave it as an exercise for the reader to 
confirm the independence of all observables on $\omega_k$.

\subsection{Nested All-Loop Asymptotic Bethe Ansatz for $\alg{su}(1|2)$}
\label{sec:nba}

The nested Bethe ansatz was discovered, along with the Yang-Baxter
equation \eqref{eq:ybe}, in a seminal paper by C.-N.~Yang \cite{Yang:1967bm}.
This article is very concisely written and we found it useful
to consult with the more detailed accounts 
\cite{Fung:1981kj,Andrei:1982cr}, and in particular
\cite{Sutherland:1985aa}.

Bethe equations for the long-range chain are derived as in the
one-loop case by imposing periodic boundary conditions on the
wave function. Ideally this should be done for the exact, all-loop 
wave functions, but these are currently not known.
What we can do is impose periodicity on the asymptotic wave functions 
such as \eqref{eq:su12.ABA}. Correspondingly we may only hope
to find {\it asymptotic} Bethe equations. They are expected to break
down when the region of interaction reaches the
size of the system and the ``asymptotic region{}'' effectively
shrinks to zero. In gauge theory this is expected to happen around 
${\cal O}(g^{2 L})$ of perturbation theory.

The all-loop asymptotic Bethe equations are identical in form to the
one-loop equations \eqref{eq:mbethe} and we merely use the
all-loop S-matrix \eqref{eq:su12.Sx} in place of \eqref{eq:su12.Su}. 
While the latter is conjectural starting from four loops,
it contains the three-loop gauge theory S-matrix calculated in 
\secref{sec:su12conjecture}.
Our Bethe ansatz is thus expected to properly diagonalize
arbitrary $\alg{su}(1|2)$ states of $\superN=4$ gauge theory to
at least third loop order. 

Let us then focus on the matrix Bethe equations
\[\label{eq:mbethe.x}
\lrbrk{\frac{x^+_{2,k}}{x^-_{2,k}}}^L |\Psi \rangle
=S_{k,k+1} \ldots S_{k,K_2} S_{k,1} 
\ldots S_{k,k-1} \,  \cdot |\Psi \rangle \, ,
\]
where we have expressed, as in \secref{sec:One}
momenta $p_k$ via \eqref{eq:One.SpecMom,eq:One.SpecRoots} by rapidities 
$x_{2,k}^{\pm}:=x_k^{\pm}$. For reasons that will become clear shortly we
have added a further index $2$ to all rapidities. 
The two-body S-matrix $S_{k,j}=S_{k,j}(x_{2,k},x_{2,j})$
is given in \eqref{eq:su12.Sx}.
This matrix eigenvalue equation is to be satisfied 
simultaneously for {\it all} $k=1,\ldots,K_2$, i.e. the eigenvector
$|\Psi\rangle$ is not allowed to depend on $k$.

We mentioned in \secref{sec:su12.1loop} that we should think of
$|\Psi\rangle$ as a state in a short spin chain of length $K_2$.
Let us start by picking a vacuum (=reference state) on this chain,
say the bosonic fields $\fldS$. 
We may then say that, cf.~\eqref{eq:su12ops}, 
that we have a length-$K_2$ spin chain 
doped with $K_1$ ``magnons{}'' $\fldF$. 
Let us define the reduced two-body scattering operator
\[\label{eq:Sreduced}
s_{k,j}=
\left(\begin{array}{cccc}
1& & & 
\\
&\left(t^{\fldF \fldS}_{\fldF \fldS}\right)_{k,j}&
(r^{\fldF \fldS}_{\fldS \fldF})_{k,j}&
\\
\\
&(r^{\fldS \fldF}_{\fldF \fldS})_{k,j}&
(t^{\fldS \fldF}_{\fldS \fldF})_{k,j}& 
\\
& & & (s^{\fldF \fldF}_{\fldF \fldF})_{k,j}
\end{array}\right)\, ,
\]
with 
\[\label{eq:Ss}
S_{k,j}(x_{2,k},x_{2,j})=S_0(x_{2,k},x_{2,j})\,
\frac{x_{2,k}^+ - x_{2,j}^-}{x_{2,k}^- - x_{2,j}^+}\,
s_{k,j}(x_{2,k},x_{2,j})\, .
\]
Defining the common eigenvalue of the reduced many-body scattering
operator by
\[\label{eq:lambda}
\lambda_k= 
\lrbrk{\frac{x^+_{2,k}}{x^-_{2,k}}}^L
\prod_{\textstyle\atopfrac{j=1}{j\neq k}}^{K_2}
S_0^{-1}(x_{2,k},x_{2,j})\,
\frac{x_{2,k}^- - x_{2,j}^+}{x_{2,k}^+ - x_{2,j}^-}\, ,
\]
we may rewrite \eqref{eq:mbethe.x} as
\[\label{eq:mbethe.red}
\lambda_k |\Psi \rangle
=s_{k,k+1} \ldots s_{k,K_2} s_{k,1} 
\ldots s_{k,k-1} \,  \cdot |\Psi \rangle \, .
\]
On the vacuum of the short spin chain the reduced operators $s_{k,j}$
act trivially and we find immediately for all values of $k$
\[
\lambda_k |\fldS \fldS ... \fldS \fldS \rangle=
|\fldS \fldS ... \fldS \fldS \rangle\, ,
\]
i.e.~that $\lambda_k=1$, which is nothing but our 
$\alg{su}(2)$ Bethe equation \eqref{eq:One.BetheSU2}
or \eqref{eq:afsb}. 

Let us next solve the one-magnon
problem of the short chain. Unfortunately we cannot solve the
problem by a common Fourier transform, because our short chain
is inhomogeneous and thus not translationally invariant. 
However, introducing a coordinate-space wave function $\psi_k$ through
\[
\label{eq:position}
|\Psi\rangle=\sum_{1\leq k \leq K_2} \psi_k~
| \fldS ... \fldS \MMM{\fldF}{k} \fldS... \fldS \rangle\, ,
\]
one may write down and solve recursion relations for the
amplitudes $\psi_k$. Interestingly the necessary calculations
and considerations, displayed in great detail in \cite{Sutherland:1985aa},
are essentially identical to the one-loop case.
The idea is to consider \eqref{eq:mbethe.red} and to
recursively apply the string of reduced two-body operators 
$s_{k,k-j}$ from ``right to left{}''
to the state $|\Psi \rangle$ and investigate how the latter changes.
One then derives the following recursion relation on
the amplitudes $\psi_k$:
\[\label{eq:master}
\frac{\psi_{k-j-1}}{\psi_{k-j}}=
\frac{(t^{\fldF \fldS}_{\fldF \fldS})_{k,k-j} \lambda_k
-\Delta_{k,k-j}}
{\lambda_k-(t^{\fldS \fldF}_{\fldS \fldF})_{k,k-j-1}}\, 
\frac{(r^{\fldS \fldF}_{\fldF \fldS})_{k,k-j-1}}
{(r^{\fldS \fldF}_{\fldF \fldS})_{k,k-j}}\, ,
\]
where 
\[
\Delta_{k,k-j}=
(t^{\fldF \fldS}_{\fldF \fldS})_{k,k-j} 
(t^{\fldS \fldF}_{\fldS \fldF})_{k,k-j}
-(r^{\fldF \fldS}_{\fldS \fldF})_{k,k-j} 
(r^{\fldS \fldF}_{\fldF \fldS})_{k,k-j}\, .
\]
Now, using \eqref{eq:su12.Sx,eq:Sreduced} for our
transmission amplitudes
\[
(t^{\fldF \fldS}_{\fldF \fldS})_{k,k-j}=
\frac{x_{2,k}^+-x_{2,k-j}^+}{x_{2,k}^+-x_{2,k-j}^-}\, ,
\qquad
(t^{\fldS \fldF}_{\fldS \fldF})_{k,k-j}=
\frac{x_{2,k}^--x_{2,k-j}^-}{x_{2,k}^+-x_{2,k-j}^-}\, ,
\]
and reflection amplitudes 
\[
(r^{\fldS \fldF}_{\fldF \fldS})_{k,k-j}=
\frac{x_{2,k}^+-x_{2,k}^-}{x_{2,k}^+-x_{2,k-j}^-}\, ,
\qquad
(r^{\fldF \fldS}_{\fldS \fldF})_{k,k-j}=
\frac{x_{2,k-j}^+-x_{2,k-j}^-}{x_{2,k}^+-x_{2,k-j}^-}\, , 
\]
we may calculate the determinant $\Delta_{k,k-j}$ to be
\[\label{eq:determinant}
\Delta_{k,k-j}=
\frac{x_{2,k}^--x_{2,k-j}^+}{x_{2,k}^+-x_{2,k-j}^-}=
(s^{\fldF \fldF}_{\fldF \fldF})_{k,k-j}\, ,
\]
i.e.~it happens to coincide with the reduced fermion-fermion
scattering amplitude.
We may then rewrite \eqref{eq:master} as
\[\label{eq:master2}
\frac{\psi_{k-j-1}}{\psi_{k-j}}=
\frac{x_{2,k-j}^+-\frac{x_{2,k}^+\lambda_k-x_{2,k}^-}{\lambda_k-1}}
{x_{2,k-j-1}^--\frac{x_{2,k}^+\lambda_k-x_{2,k}^-}{\lambda_k-1}}\, .
\]
The left hand side of this equation only depends on the
difference $k-j$. Therefore, the equation can only be consistent
if the right hand side does {\it not} depend on the index $k$.
We then conclude that 
\[\label{eq:rapid}
x_1:=\frac{x_{2,k}^+\lambda_k-x_{2,k}^-}{\lambda_k-1}
\]
must be a constant. Iterating \eqref{eq:master2} one finds the
following elegant result for the one-magnon wave functions
\[\label{eq:1magnon}
\psi_k(x_1)=\prod_{j=1}^{k-1}\frac{x_1-x_{2,j}^-}{x_1-x_{2,j+1}^+}\, .
\]
We may interpret $x_1$ as a new rapidity variable, parametrizing
the ``momentum{}'' of our magnon $\fldF$. One may also 
check explicitly, with some work \cite{Fung:1981kj,Andrei:1982cr},
that the wave function \eqref{eq:1magnon} we just derived indeed satisfies
the matrix eigenvalue equation \eqref{eq:mbethe.red}. The
eigenvalue is found from inverting \eqref{eq:rapid}
\[\label{eq:eigenwert}
\lambda_k=\lambda_k(x_1):=
\frac{x_{2,k}^{-}-x_1}{x_{2,k}^{+}-x_1}\, .
\]
Is the value of the magnon rapidity $x_1$ arbitrary? It turns
out that it is not: Demanding periodic boundary conditions
for our short chain leads to the quantization condition
\[\label{eq:periodicity}
\prod_{k=1}^{K_2} \lambda_k =1\, .
\]
This completely solves the one-magnon problem $K_1=1$.

Next we will study the matrix eigenvalue equation \eqref{eq:mbethe.red}
for the two-magnon problem $K_1=2$. 
The states
are now described by 
coordinate-space wave functions $\psi_{k_1,k_2}$ through
\[
\label{eq:position2}
|\Psi\rangle=\sum_{1\leq k_1 < k_2 \leq K_2} \psi_{k_1,k_2}~
| \fldS ... \fldS \MMM{\fldF}{k_1} \fldS...\fldS \MMM{\fldF}{k_2} \fldS... \fldS \rangle\, ,
\]
Denoting the level-two rapidities of
the two magnons by $x_{1,1}$ and $x_{1,2}$, a first guess for the solution 
would be that the one-magnon eigenvalue of \eqref{eq:eigenwert}
is now replaced by the product
\[\label{eq:eigenwert2}
\lambda_k=\lambda_k(x_{1,1})\,\lambda_k(x_{1,2})=
\frac{x_{2,k}^{-}-x_{1,1}}{x_{2,k}^{+}-x_{1,1}}\,
\frac{x_{2,k}^{-}-x_{1,2}}{x_{2,k}^{+}-x_{1,2}}\, ,
\]
and that the two-body wave function is just the product 
$\psi_{k_1}(x_{1,1})\,\psi_{k_2}(x_{1,2})$ of
one-body amplitudes \eqref{eq:1magnon}.
The first part of this guess, \eqref{eq:eigenwert2}, is indeed correct.
However, in order to find the correct wave function we need to take into
account the scattering of the particles.
If the scattering is non-diffractive, the particles may only exchange
their rapidities. It is then reasonable to attempt a (secondary) Bethe Ansatz:%
\footnote{The sign in front of $B'$ is due to the
exchange of two fermions.}
\[\label{eq:Bethe2}
\psi_{k_1,k_2}(x_{1,1},x_{1,2})=
B\, \psi_{k_1}(x_{1,1})\,\psi_{k_2}(x_{1,2})
-B'\, \psi_{k_1}(x_{1,2})\,\psi_{k_2}(x_{1,1})\, .
\]
In similarity to the treatment for the original, long spin chain
we have to check explicitly that this ansatz indeed satisfies 
\eqref{eq:mbethe.red} with \eqref{eq:eigenwert2}.
Abbreviating, respectively, 
$\psi_k(x_{1,1})$,$\psi_k(x_{1,2})$ by $\psi_k$,$\psi'_k$ and
$\lambda_k(x_{1,1})$,$\lambda_k(x_{1,2})$ by $\lambda_k$,$\lambda'_k$
we find the consistency condition for factorized scattering to be
\[\label{eq:BprimeB1}
\frac{B'}{B}=
\frac
{\Delta_{k,k-j}\,\lambda'_{k-j}\,\psi'_k\,\psi_{k-j}\,
\lambda_{k-1}\,\ldots\lambda_{k-j}+
\psi_k\,\psi'_{k-j}\,\lambda'_k 
\ldots\lambda'_{k-j}}
{\Delta_{k,k-j}\,\lambda_{k-j}\,\psi_k\,\psi'_{k-j}\,
\lambda'_{k-1}\ldots\lambda'_{k-j}+
\psi'_k\,\psi_{k-j}\,\lambda_k 
\ldots\lambda_{k-j}}\, ,
\]
where obviously the amplitude $\Delta$ for the $\fldF$-$\fldF$ exchange
enters, see \eqref{eq:determinant}.
Our ansatz requires that the left hand side of \eqref{eq:BprimeB1}
should neither depend on $k$ nor on $j$. 
Now, using \eqref{eq:determinant,eq:1magnon,eq:eigenwert} we find
\[\label{eq:BprimeB2}
\frac{B'}{B}= 
\frac
{(x_{k-j}^+ -x_k^-)(x_{k-j}^- -x_{1,2})(x_k^+-x_{1,1})+
(x_{k-j}^- -x_k^+)(x_k^--x_{1,2})(x_{k-j}^+ -x_{1,1})}
{(x_{k-j}^+ -x_k^-)(x_{k-j}^- -x_{1,1})(x_k^+-x_{1,2})+
(x_{k-j}^- -x_k^+)(x_k^--x_{1,1})(x_{k-j}^+ -x_{1,2})}\, .
\]
After a short computation one ends up with the remarkably simple result
\[\label{eq:BprimeB3}
\frac{B'}{B}=1\, .
\]
We see that the unwanted dependence on the indices $k$ and $j$
has disappeared and that the secondary Bethe ansatz \eqref{eq:Bethe2}
yields indeed the correct two-magnon wave function with
eigenvalue \eqref{eq:eigenwert2}.

In general we do not have to find the wave function of the short
chain explicitly, but merely impose periodic boundary conditions. 
Due to the simplicity of \eqref{eq:BprimeB3}, this is easily done here:
Note that the two-magnon wave function is of the form of a
Slater determinant:
\[\label{eq:2magnon}
\psi_{k_1,k_2}(x_{1,1},x_{1,2})=
\psi_{k_1}(x_{1,1})\,\psi_{k_2}(x_{1,2})
-\psi_{k_1}(x_{1,2})\,\psi_{k_2}(x_{1,1})\, ,
\]
indicating that the fermions $\fldF$ in the short, inhomogeneous chain
are ``free{}'' to all loop orders. And indeed, since scattering is
factorized in the short chain, we may immediately write down
the $K_1$-magnon wave functions as a $K_1 \times K_1$ Slater determinant:
\[\label{eq:multimagnon}
\psi_{k_1,\ldots,k_{K_1}}(x_{1,1},\ldots,x_{1,K_1})=
\det_{\mu,\nu} \psi_{k_\mu}(x_{1,\nu}) \, ,
\]
To prevent confusion let us reiterate that
the indices $\mu,\nu$ are labels for the $K_1$ magnons in the
auxiliary, short spin chain of length $K_2$. The indices $k_\mu$
indicate the position of these magnons in the auxiliary chain and
therefore take values in the set $\{1,\ldots,K_2\}$. Likewise,
the $x_{1,\mu}$ are the rapidities of the $K_1$ magnons describing
their motion in the short chain of length $K_2$. They are
not to be confused with the rapidities $x_{2,k}$ describing
the motion of the original $K_2$ magnons in the long chain of length $L$.  
Note that the one-body wave functions $\psi_{k_\mu}(x_{1,\nu})$
in \eqref{eq:multimagnon,eq:2magnon} depend implicitly also on the
magnon rapidities $x_{2,k}$ of the original chain, as seen from the result
\eqref{eq:1magnon}.  

The eigenvalue associated to the wave function \eqref{eq:multimagnon}
is clearly given by the product
$\lambda_k=\lambda_k(x_{1,1})\ldots\lambda_k(x_{1,K_1})$,
in generalization of the one and two-magnon expressions
\eqref{eq:eigenwert,eq:eigenwert2}.
Using \eqref{eq:lambda}, we thus derived the first set of 
asymptotic Bethe equations for the
$\alg{su}(1|2)$ sector. They read, for $k=1,\ldots,K_2$
\[\label{eq:SU12.Bethe.Bethe1}
\lrbrk{\frac{x^+_{2,k}}{x^-_{2,k}}}^L
=
\prod_{\textstyle\atopfrac{j=1}{j\neq k}}^{K_2}
S_0(x_{2,k},x_{2,j})
\prod_{\textstyle\atopfrac{j=1}{j\neq k}}^{K_2}
\frac{x_{2,k}^{+}-x_{2,j}^{-}}
{x_{2,k}^{-}-x_{2,j}^{+}}
\prod_{j=1}^{K_1}
\frac{x_{2,k}^{-}-x_{1,j}}{x_{2,k}^{+}-x_{1,j}}\, .
\]
Finally we need to impose periodic boundary conditions on the
small chain. Generalizing the one-magnon case \eqref{eq:periodicity},
this yields a second set of Bethe equations. Taking each of the   
$k=1,\ldots,K_1$ particles once around the chain of length $K_2$,
we then find (recall that the fermions are free)
\[\label{eq:SU12.Bethe.Bethe2}
1
=
\prod_{j=1}^{K_2}
\frac{x_{1,k}-x_{2,j}^{+}}{x_{1,k}-x_{2,j}^{-}}\, ,
\]
which is nothing but an inhomogenous version of the free particle
quantization law.

We could alternatively have picked the fermions $\fldF$ as a
reference state in the short spin chain of length $K_2$.
We should then consider the $\tilde{K}_1$ scalars $\fldS$ as excitations
on this vacuum state:
\[\label{eq:su12opsalt}
\Tr \,\fldS^{\tilde{K}_1} \fldF^{K_2-\tilde{K}_1} \fldZ^{L-K_2}+ \ldots\, .
\]
In this case one should replace the reduced two-body scattering
operator \eqref{eq:Sreduced} by
\[\label{eq:Sreduced2}
\tilde{s}_{k,j}=
\left(\begin{array}{cccc}
(\tilde{s}^{\fldS \fldS}_{\fldS \fldS})_{k,j}& & & 
\\
&\left(\tilde{t}^{\fldF \fldS}_{\fldF \fldS}\right)_{k,j}&
(\tilde{r}^{\fldF \fldS}_{\fldS \fldF})_{k,j}&
\\
\\
&(\tilde{r}^{\fldS \fldF}_{\fldF \fldS})_{k,j}&
(\tilde{t}^{\fldS \fldF}_{\fldS \fldF})_{k,j}& 
\\
& & & 1
\end{array}\right)\, ,
\]
and express the S-matrix entering the matrix Bethe equation
\eqref{eq:mbethe.x} as
\[\label{eq:Ss2}
S_{k,j}(x_{2,k},x_{2,j})=S_0(x_{2,k},x_{2,j})\,
\tilde{s}_{k,j}(x_{2,k},x_{2,j})\, .
\]
The further analysis proceeds in exactly the way we just presented
for the bosonic vacuum and it is straightforward to obtain
the expressions for the eigenvalues and wave functions. 
In particular the magnons $\fldS$ again behave like free 
particles in the reduced chain. We will just
state the corresponding Bethe equations which read,
for $k=1,\ldots,K_2$
\[\label{eq:SU12.Bethe.Dual1}
\lrbrk{\frac{x^+_{2,k}}{x^-_{2,k}}}^L
=
\prod_{\textstyle\atopfrac{j=1}{j\neq k}}^{K_2}
S_0(x_{2,k},x_{2,j})
\prod_{j=1}^{\tilde K_1}
\frac{x_{2,k}^{+}-\tilde x_{1,j}}{x_{2,k}^{-}-\tilde x_{1,j}}\, ,
\]
where $S_0(x_{2,k},x_{2,j})$ is the same long-range scattering
matrix \eqref{eq:longrangeS} as in the first form \eqref{eq:SU12.Bethe.Bethe1}
of the Bethe ansatz. The second set of equations is,
with $k=1,\ldots,\tilde K_1$
\[\label{eq:SU12.Bethe.Dual2}
1
=
\prod_{j=1}^{K_2}
\frac{\tilde x_{1,k}-x_{2,j}^{-}}{\tilde x_{1,k}-x_{2,j}^{+}}\, .
\]
Later in \Secref{sec:PSU112.Duality} we will see, in much greater generality, 
that this second form \eqref{eq:SU12.Bethe.Dual1,eq:SU12.Bethe.Dual2} of 
the Bethe equations may also be obtained directly from the first form 
\eqref{eq:SU12.Bethe.Bethe1,eq:SU12.Bethe.Bethe2} by a duality transformation
\cite{Essler:1992nk,Essler:1992uc,Gohmannn:2003aa,Beisert:2005di}.

\subsection{Spectrum}
\label{sec:su12.Spectrum}

\begin{table}\centering
$\begin{array}{|c|cc|l|}\hline
L&K_2&K_1&(E_0,E_2,E\indup{4g}|E\indup{4s})^P\\\hline\hline
7&4&2&(10,-\frac{75}{4},\frac{4315}{64}|\frac{3963}{64})^\pm\,\surd\\
 & & &(6,-\frac{33}{4},\frac{1557}{64}|\frac{1461}{64})^\pm\,\surd\\
 &5&3&(12,-\frac{45}{2},\frac{1281}{16}|\frac{1137}{16})^+\\
 & & &(8,-\frac{25}{2},\frac{687}{16}|\frac{655}{16})^+\\\hline
8&4&2&({\scriptstyle 31E^3-350E^2+1704E-3016},
     {\scriptstyle -50E^3+1111E^2-7971E+18452},\\ &&&\qquad
     {\scriptstyle \frac{337}{2}E^3-\frac{18363}{4}E^2+38740E-102390}|
     {\scriptstyle \frac{621}{4}E^3-\frac{17213}{4}E^2+36730E-97840})^\pm\\
 &5&2&(8,-13,\frac{343}{8}|\frac{311}{8})^\pm\\
 & & &(15E-48,-23E+135,\frac{595}{8}E-\frac{4023}{8}|\frac{541}{8}E-\frac{3735}{8})^\pm\\
 &5&3&(7,-\frac{21}{2},\frac{3241}{94}|\frac{6083}{188})^\pm\\
 & & &({\scriptstyle 33E^2-358E+1279},
       {\scriptstyle -\frac{119}{2}E^2+1283E-\frac{13675}{2}},\\&&&\qquad
       {\scriptstyle \frac{19601}{94}E^2-\frac{1059061}{188}E+\frac{1693257}{47}}|
       {\scriptstyle \frac{8784}{47}E^2-\frac{972991}{188}E+\frac{6323573}{188}})^\pm\\
 &6&4&(12,-22,78|72)^\pm\\\hline
\end{array}$
\caption{Spectrum of lowest-lying states genuinely in the $\alg{su}(1|2)$ sector.}
\label{tab:SU12}
\end{table}

We have computed the spectrum of all states
of the $\alg{su}(1|2)$ spin chain with $L\leq 8$ 
using the three-loop Hamiltonian found in \cite{Beisert:2003ys}.
The results are presented in \tabref{tab:SU12}
excluding those which have been given 
before in \tabref{tab:TwoEx,tab:SU2,tab:U11}.
The notation is explained 
in \secref{sec:One.Spectrum}.
The states marked as ``$\surd$''
have been computed using the Bethe equations
\eqref{eq:SU12.Bethe.Bethe1,eq:SU12.Bethe.Bethe2}.
Note that the state with one-loop energy $E_0=6$ is singular,
i.e.~it has a pair of roots of type $2$ at $u_2=\pm \sfrac{i}{2}+\delta u$,
where $\delta u=\order{g^2}$ is the same for both roots.
We kindly invite the reader to confirm that the energies 
of the remaining cases also agree with our nested Bethe ansatz.

Note that the excitation numbers $K_2,K_1$ correspond to
the Dynkin labels $[q_1,p,q_2]$ of $\alg{su}(4)$ 
\[
q_1=K_2-K_1,\qquad p=L+K_1-2K_2,\qquad q_2=K_2
\]
and the labels $[s_1,r,s_2]$ of $\alg{su}(2,2)$ 
\[
s_1=K_1,\qquad r=-L-K_1-2K_2-\delta D,\qquad s_2=0.
\]
%

\section{The Factorized S-Matrix of the $\alg{su}(1,1|2)$ Sector}
\label{sec:PSU112}

In the previous \Secref{sec:SU12} we have unified the bosonic $\alg{su}(2)$ and
fermionic $\alg{su}(1|1)$ sectors into a supersymmetric, long-range  $t$-$J$
model with  $\alg{su}(1|2)$ symmetry. In turn, in \cite{Staudacher:2004tk} it was
demonstrated that the integrable structure of the non-compact  bosonic 
$\alg{sl}(2)$ sector is also closely related to the  $\alg{su}(2)$
and $\alg{su}(1|1)$ sectors, confer \eqref{eq:grouptheory} and the discussion
in \secref{sec:One}. It is therefore very natural to attempt to find the S-matrix and
the associated Bethe ansatz for the smallest closed sector unifying all three two-component
sectors. This is the $\alg{su}(1,1|2)$ sector (see \cite{Beisert:2004ry}).
In addition to the scalars $\fldZ,\fldS$, 
the derivative $\fldD$ and the gaugino $\fldF$,
it requires a second fermion $\fldG$.
The possible states at a given lattice site are then  
\[
\fldD^k \fldZ,\quad
\fldD^k \fldS,\quad
\fldD^k \fldF,\quad
\fldD^k \fldG
\]
where $k$ may be any non-negative integer.
We consider $\fldZ$ as the vacuum and its
single excitations are 
\[
\fldZ\to\fldS,\quad
\fldZ\to\fldF,\quad
\fldZ\to\fldG,\quad
\fldZ\to \fldD \fldZ.
\]
Note that the scalar $\fldZ\to\fldS$ is a hard-core excitation,
there can be only one such excitation per site.
Conversely, the derivative $\fldZ\to\fldD\fldZ$ is a soft-core excitation,
there can be arbitrarily many excitations per site
and they also exist on sites which are already occupied 
by scalars or fermions.
In this context it is useful to consider the fermions 
as excitation with a mixed type of core:
A fermion cannot coexist with a scalar or another fermion
of the same type. 
A mixture of the two fermions
$\fldZ\to\fldF$ and $\fldZ\to\fldG$ however is possible and
should be considered as the double excitation $\fldZ\to\fldD\fldS$
in agreement with supersymmetry transformation rules.

This is the largest sector where the spin chain remains ``static{}'', i.e.~the 
length does not fluctuate \cite{Beisert:2003ys}. 
We are therefore still on firm grounds and should 
be able to apply the technology established 
in the previous section to this extended non-compact sector.
However, it still exhibits an exciting 
new feature as compared to the previously discussed cases. The interactions allow
for \emph{flavor change}:
\[\label{eq:mutation}
\fldF\,\fldG\leftrightarrow \fldS\,(\fldD\fldZ)\, .
\]
This means that a pair of fermions may annihilate 
and produce a pair of bosons. 
While particle annihilation and production are often believed to
destroy integrability, here the number of excitations is actually
preserved by the flavor change: This is related to the above claim
that a combination of the two excitations $\fldZ\to\fldF$ and $\fldZ\to\fldG$ 
is equivalent to the double excitation $\fldZ\to\fldD\fldS$.
Among the above four single excitations there is one linear
dependence (the sector has rank $3$) which allows for the flavor change.

As in \cite{Staudacher:2004tk} we are currently unable to derive the higher loop
S-matrix by the PABA in this sector since the dilatation operator is not known beyond one loop.
However, we may inspect the one-loop S-matrix in this sector
and generalize it to all loops according to the principles discovered above. 
Excitingly, we shall find that the result still 
satisfies the Yang-Baxter equation! 
In addition, our conjecture is consistent with multiplet splitting, 
dualization and the expected thermodynamics.

\subsection{One-Loop Scattering in the $\alg{su}(1,1|2)$ Sector}
\label{sec:PSU112.One}

Let us consider the vacuum state 
\[
\state{0}=\state{\fldZ\ldots\fldZ}
\]
which has zero energy. 
An eigenstate
with a single excitation 
$\fldA=\fldS,\fldF,\fldG,\fldD\fldZ$
is given by
\[
\state{p_{\fldA}}=
\sum_{\ell}e^{ip\ell}\,\state{\fldZ\ldots\fldZ\MMM{\fldA}{\ell}\fldZ\ldots\fldZ}
\]
Its energy eigenvalue is $e_0(p)=4\sin^2(p/2)$.
We will now represent the states in a slightly different fashion 
using oscillator excitations
\[\label{eq:PSU112.One.Single}
\state{p_{\osc{A},\osc{\dot A}}}
=
\sum_{\ell}e^{ip\ell}
\osc{A}^\dagger_{\ell}
\osc{\dot A}^\dagger_{\ell}
\state{0}.
\]
We have replaced the field $\fldA$ by a pair 
of harmonic oscillators $\osc{A}=\osc{a},\osc{c}$ and
$\osc{\dot A}=\osc{\dot a},\osc{\dot c}$ acting on the vacuum at site $\ell$. 
The oscillators $\osc{a},\osc{\dot a}$ are bosonic
while $\osc{c},\osc{\dot c}$ are fermionic.
The four different elementary excitations $\fldA=\fldS,\fldF,\fldG,\fldD\fldZ$
are represented by a combination of one $\osc{A}^\dagger$ and one 
$\osc{\dot A}^\dagger$
\[
\state{\fldS}=\osc{c}^\dagger\osc{\dot c}^\dagger\state{\fldZ},\qquad
\state{\fldF}=\osc{a}^\dagger\osc{\dot c}^\dagger\state{\fldZ},\qquad
\state{\fldG}=\osc{c}^\dagger\osc{\dot a}^\dagger\state{\fldZ},\qquad
\state{\fldD\fldZ}=\osc{a}^\dagger\osc{\dot a}^\dagger\state{\fldZ}.
\]
In this notation we will be able to write 
the scattering eigenstate of two excitations 
in a very concise form. 
The eigenstate is defined by the equation 
\[
\ham_0
\state{p_{\osc{A},\osc{\dot A}};q_{\osc{B},\osc{\dot B}}}
=
\bigbrk{e_0(p)+e_0(q)}
\state{p_{\osc{A},\osc{\dot A}};q_{\osc{B},\osc{\dot B}}}
\]
and the boundary condition that the 
wave function is a product of two 
instances of \eqref{eq:PSU112.One.Single}
when the excitations are far apart.
We find the following set of independent states
specified by the momenta $p,q$ and the 
oscillators $\osc{A},\osc{\dot A},\osc{B},\osc{\dot B}$
\<\label{eq:PSU112.One.Double}
\state{p_{\osc{A},\osc{\dot A}};q_{\osc{B},\osc{\dot B}}}
\eq
\sum_{\ell_1<\ell_2}e^{ip\ell_1+iq\ell_2} 
\osc{A}^\dagger_{\ell_1}
\osc{\dot A}^\dagger_{\ell_1}
\osc{B}^\dagger_{\ell_2}
\osc{\dot B}^\dagger_{\ell_2}\state{0}
\nl
+\sum_{\ell_1=\ell_2}
\frac{u-v}{u-v-i}\,
e^{ip\ell_1+iq\ell_2} 
\osc{A}^\dagger_{\ell_1}
\osc{\dot A}^\dagger_{\ell_1}
\osc{B}^\dagger_{\ell_1}
\osc{\dot B}^\dagger_{\ell_1}\state{0}
\nl
+\sum_{\ell_1>\ell_2}
\frac{(u-v)^2}{(u-v-i)(u-v+i)}\,
e^{ip\ell_1+iq\ell_2} 
\osc{A}^\dagger_{\ell_1}
\osc{\dot A}^\dagger_{\ell_1}
\osc{B}^\dagger_{\ell_2}
\osc{\dot B}^\dagger_{\ell_2}\state{0}
\nl
+\sum_{\ell_1>\ell_2}
\frac{i(u-v)}{(u-v-i)(u-v+i)}\,
e^{ip\ell_1+iq\ell_2} 
\osc{A}^\dagger_{\ell_2}
\osc{\dot A}^\dagger_{\ell_1}
\osc{B}^\dagger_{\ell_1}
\osc{\dot B}^\dagger_{\ell_2}\state{0}
\nl
+\sum_{\ell_1>\ell_2}
\frac{i(u-v)}{(u-v-i)(u-v+i)}\,
e^{ip\ell_1+iq\ell_2} 
\osc{A}^\dagger_{\ell_1}
\osc{\dot A}^\dagger_{\ell_2}
\osc{B}^\dagger_{\ell_2}
\osc{\dot B}^\dagger_{\ell_1}\state{0}
\nl
+\sum_{\ell_1>\ell_2}
\frac{i^2}{(u-v-i)(u-v+i)}\,
e^{ip\ell_1+iq\ell_2} 
\osc{A}^\dagger_{\ell_2}
\osc{\dot A}^\dagger_{\ell_2}
\osc{B}^\dagger_{\ell_1}
\osc{\dot B}^\dagger_{\ell_1}\state{0}.
\>
The rapidities $u,v$ are defined via
\[
e^{ip}=\frac{u+\sfrac{i}{2}}{u-\sfrac{i}{2}}\,,\qquad
e^{iq}=\frac{v+\sfrac{i}{2}}{v-\sfrac{i}{2}}\,.
\]
The first line represents the incoming excitations.
The second line represents the wave-function 
when the two excitations overlap and is
only present for particles which experience soft-core scattering.
The remaining four lines represent
outgoing excitations and they encode the S-matrix.

Let us explain how to obtain the S-matrix from the last four lines 
in more detail.
The simplest case is when all four oscillators 
$\osc{A},\osc{\dot A},\osc{B},\osc{\dot B}$ are different.
In that case flavor changes are possible because the excitations
$\osc{A},\osc{\dot A}$ and 
$\osc{B},\osc{\dot B}$ can recombine
as $\osc{A},\osc{\dot B}$ and 
$\osc{B},\osc{\dot A}$ (middle two lines).
If no flavor change takes place, there can either be
transmission (first line) or reflection (last line).
Note that despite the recombinations we have not changed the order of oscillators 
in \eqref{eq:PSU112.One.Double} because some of the oscillators 
can be fermionic. This would lead to various additional signs
which are not necessary in the form \eqref{eq:PSU112.One.Double}.
When two of the oscillators are equal, say $\osc{A}=\osc{B}$,
there can only be transmission and reflection. The contributions
to the S-matrix elements now come from adding two lines,
the first two (transmission) or the latter two (reflection).
Here the statistics of oscillators determines the outcome of the sum:
It can either yield a factor of $u-v+i$ or $u-v-i$, 
both of which will cancel against one of the denominators 
and yield elements of the sort \eqref{eq:su12.Su}.
If in addition $\osc{\dot A}=\osc{\dot B}$, both excitations
are equal and all four lines combine according to statistics.

In fact, the scattering state
\eqref{eq:PSU112.One.Double} is completely
general for any unitary algebra $\alg{sl}(m|n)$ with
the spin sites in oscillator representations.
In particular it applies to 
the complete $\superN=4$ one-loop spin chain \cite{Beisert:2003jj,Beisert:2003yb}.
There the excitations are made out of two sets of four (instead of two) 
oscillators $\osc{A}=\osc{a},\osc{b},\osc{c},\osc{d}$
and $\osc{\dot A}=\osc{\dot a},\osc{\dot b},\osc{\dot c},\osc{\dot d}$.
The oscillators $\osc{a},\osc{b},\osc{\dot a},\osc{\dot b}$
are bosonic whereas $\osc{c},\osc{d},\osc{\dot c},\osc{\dot d}$
as fermionic. The excitations are made up from a pair of oscillators
$\osc{A},\osc{\dot A}$ and consequently there 
are $8$ bosonic and $8$ fermionic excitations in total,
cf.~\cite{Berenstein:2002jq,Beisert:2005di}.

\subsection{The Asymptotic S-Matrix for $\alg{su}(1,1|2)$ }
\label{sec:PSU112.All}

Based on the above complete one-loop S-matrix
we make an educated guess of how to extend the all-loop S-matrix 
from the $\alg{su}(1|2)$ sector to the $\alg{su}(1,1|2)$ sector
which also includes the derivative sector $\alg{su}(1,1)=\alg{sl}(2)$
of \secref{sec:One}.
As before, the S-matrix should have an overall prefactor $S_0$
which contains auxiliary scattering terms and which distinguishes 
between gauge theory and the string chain.
The remaining terms have a common denominator which,
in agreement with \eqref{eq:PSU112.One.Double}, 
is composed from two factors $(x_k^+-x_j^-)(x_k^--x_j^+)$. 
The individual elements of the S-matrix differ only by 
the numerator $N$ which is also a product of two terms ``$x^\pm_{k,j}-x^\pm_{k,j}$''
\[\label{eq:PSU112.S.Pre}
S_{ab}^{cd}(x_k,x_j)=S_0(x_k,x_j)\,
\frac{N_{ab}^{cd}(x_k,x_j)}{(x_k^+-x_j^-)(x_k^--x_j^+)}\,.
\]
{}From \eqref{eq:One.BetheAll} we can read off 
the single-flavor scattering terms of 
scalars, fermions and derivatives
(we shall abbreviate the excitation 
$\fldD\fldZ$ by just $\fldD$)
\[\label{eq:PSU112.S.Diag}
\begin{array}[b]{rclccrcl}
N_{\fldS\fldS}^{\fldS\fldS}\eq (x_k^+-x_j^-)(x_k^+-x_j^-),
&\quad&
N_{\fldG\fldG}^{\fldG\fldG}\eq (x_k^+-x_j^-)(x_k^--x_j^+),
\\[0.3em]
N_{\fldF\fldF}^{\fldF\fldF}\eq (x_k^+-x_j^-)(x_k^--x_j^+),
&\quad&
N_{\fldD\fldD}^{\fldD\fldD}\eq (x_k^--x_j^+)(x_k^--x_j^+).
\end{array}\]
The scattering of scalars $\fldS$ and fermions $\fldF$ we know
from \eqref{eq:su12.Sx}
\[\label{eq:PSU112.S.XU}
\begin{array}[b]{rclccrcl}
N_{\fldS\fldF}^{\fldS\fldF}\eq (x_k^+-x_j^-)(x_k^--x_j^-),
&\quad&
N_{\fldF\fldS}^{\fldS\fldF}\eq (x_k^+-x_j^-)(x_k^+-x_k^-),
\\[0.3em]
N_{\fldS\fldF}^{\fldF\fldS}\eq (x_k^+-x_j^-)(x_j^+-x_j^-),
&\quad&
N_{\fldF\fldS}^{\fldF\fldS}\eq (x_k^+-x_j^-)(x_k^+-x_j^+).
\end{array}\]
Note that we have omitted the factors $\omega$ because
we have seen in \secref{sec:su12conjecture} 
that they are irrelevant for the spectrum.
To obtain a symmetric S-matrix one would have 
to replace both $x_k^+-x_k^-$ and $x_j^+-x_j^-$
by their geometric mean 
$\sqrt{x_k^+-x_k^-}\sqrt{x_{\smash{j}}^+-x_{\smash{j}}^-}$
at all places.
As the fermion $\fldG$ should have the same interactions
with the scalars as $\fldF$ we expect the S-matrix elements
of $\fldS$ with $\fldG$ to be given by the same expressions
\eqref{eq:PSU112.S.XU}.

The scattering between fermions and derivatives most likely
takes a very similar form but with some changes in the signs
(same for $\fldF$ instead of $\fldG$)
\[\label{eq:PSU112.S.DU}
\begin{array}[b]{rclccrcl}
N_{\fldG\fldD}^{\fldG\fldD}\eq (x_k^--x_j^+)(x_k^--x_j^-),
&\quad&
N_{\fldD\fldG}^{\fldG\fldD}\eq (x_k^--x_j^+)(x_k^--x_k^+),
\\[0.3em]
N_{\fldG\fldD}^{\fldD\fldG}\eq (x_k^--x_j^+)(x_j^--x_j^+),
&\quad&
N_{\fldD\fldG}^{\fldD\fldG}\eq (x_k^--x_j^+)(x_k^+-x_j^+).
\end{array}\]
The largest sector of the S-matrix is the scattering 
between the scalar $\fldS$ and the derivative $\fldD$ which can mix with
the scattering of both fermions $\fldF$ and $\fldG$.
Our proposal for the remaining numerators is%
\footnote{In our convention there are no signs if
the two involved excitations for $S_{k,j}$ are adjacent, 
i.e.~when $j=k+1$. Otherwise various signs arise from 
permuting with intermediate fields.}
\[\label{eq:PSU112.S.Mix}
\begin{array}[b]{rclccrcl}
N_{\fldS\fldD}^{\fldS\fldD}\eq  (x_k^--x_j^-)(x_k^--x_j^-),
&\quad&N_{\fldF\fldG}^{\fldS\fldD}\eq  (x_k^--x_j^-)(x_k^+-x_k^-),
\\[0.3em]
N_{\fldS\fldD}^{\fldF\fldG}\eq  (x_k^--x_j^-)(x_j^+-x_j^-),
&&N_{\fldF\fldG}^{\fldF\fldG}\eq  (x_k^--x_j^-)(x_k^+-x_j^+),
\\[0.3em]
N_{\fldS\fldD}^{\fldG\fldF}\eq  (x_k^--x_j^-)(x_j^--x_j^+),
&&N_{\fldF\fldG}^{\fldG\fldF}\eq  (x_k^--x_k^+)(x_j^+-x_j^-),
\\[0.3em]
N_{\fldS\fldD}^{\fldD\fldS}\eq  (x_j^--x_j^+)(x_j^--x_j^+),
&&N_{\fldF\fldG}^{\fldD\fldS}\eq  (x_k^+-x_j^+)(x_j^+-x_j^-),
\\\\
N_{\fldG\fldF}^{\fldS\fldD}\eq  (x_k^--x_j^-)(x_k^--x_k^+),
&&N_{\fldD\fldS}^{\fldS\fldD}\eq  (x_k^--x_k^+)(x_k^--x_k^+),
\\[0.3em]
N_{\fldG\fldF}^{\fldF\fldG}\eq  (x_k^--x_k^+)(x_j^+-x_j^-),
&&N_{\fldD\fldS}^{\fldF\fldG}\eq  (x_k^+-x_j^+)(x_k^+-x_k^-),
\\[0.3em]
N_{\fldG\fldF}^{\fldG\fldF}\eq  (x_k^+-x_j^+)(x_k^--x_j^-),
&&N_{\fldD\fldS}^{\fldG\fldF}\eq  (x_k^+-x_j^+)(x_k^--x_k^+),
\\[0.3em]
N_{\fldG\fldF}^{\fldD\fldS}\eq  (x_k^+-x_j^+)(x_j^--x_j^+),
&&N_{\fldD\fldS}^{\fldD\fldS}\eq  (x_k^+-x_j^+)(x_k^+-x_j^+),
\end{array}\]
All of these expressions have the 
correct one-loop limit \eqref{eq:PSU112.One.Double}.

As a first and important test of this S-matrix 
we have checked the validity of three features:
Parity invariance
\[
S_{a_ka_j}^{b_kb_j}(x_k,x_j)=
(-1)^{[a_j][a_k]+[b_j][b_k]}
S_{a_ja_k}^{b_jb_k}(-x_j,-x_k)
\]
and the unitarity condition 
\[
(-1)^{[b_j][b_k]+[c_j][c_k]}
S_{a_ka_j}^{b_kb_j}(x_k,x_j)
S_{b_jb_k}^{c_jc_k}(x_j,x_k)
=
\delta_{a_k}^{c_k}\delta_{a_j}^{c_j}
\]
are rather easy to confirm.
Here $[a]$ is the grading of the particle labelled by $a$;
it is even for bosons and odd for fermions.
We have verified the Yang-Baxter equation 
\<
\earel{}
(-1)^{[b_j][a_\ell]+[b_j][b_\ell]}
 S_{a_k a_j}^{b_k b_j}(p_k,p_j)
S_{b_k a_\ell}^{c_k b_\ell}(p_k,p_\ell)
S_{b_j b_\ell}^{c_j c_\ell}(p_j,p_\ell)
\nln\eq
(-1)^{[b_j][b_\ell]+[b_j][c_\ell]}
S_{a_j a_\ell}^{b_j b_\ell}(p_j,p_\ell)
S_{a_k b_\ell}^{b_k c_\ell}(p_k,p_\ell)
S_{b_k b_j}^{c_k c_j}(p_k,p_j)
\>
in \texttt{Mathematica}.
The signs arise from the application of the second S-matrix $S_{1,3}$
which requires to bring the excitations with momenta $p_k$ and $p_\ell$
next to each other, i.e.~we must permute $p_j$ with $p_\ell$.

\subsection{Nested Asymptotic Bethe Ansatz for $\alg{su}(1,1|2)$}
\label{sec:PSU112.NBA}

For the second level of the nested Bethe ansatz we have to specify
a new vacuum. As before we shall pick the scalar $\fldS$. 
The elementary excitations of this vacuum 
are given by the two fermions $\fldF$ and $\fldG$. 
Although $\fldD\fldZ$ used to be an elementary excitation of the
first level $\fldZ$-vacuum, it is no longer elementary in a sea of
$\fldS$'s. This can be inferred from the above S-matrix:
While the fermions $\fldF,\fldG$ are stable, 
the derivative $\fldD\fldZ$ is not. 
In other words, according to \eqref{eq:PSU112.S.XU}, 
the fermions can only flip positions with $\fldS$'s
while according to \eqref{eq:PSU112.S.Mix} $\fldD\fldZ$ can decay into 
two fermions $\fldF,\fldG$ using one of vacuum fields $\fldS$.
We should therefore consider $\fldD\fldZ$ as a double
excitation in great similarity to the multiple
excitations $\fldD^k\fldZ,\fldD^k\fldS,\fldD^k\fldF,\fldD^k\fldG$ 
of the vacuum $\fldZ$.

To perform the nested Bethe ansatz for the system of the two 
fermions $\fldF$ and $\fldG$ we can largely rely on the results
of \secref{sec:nba}:
The propagation of a single excitation 
and the scattering of two excitations of
type $\fldF$ have been solved there.
As the two flavors of fermions are very similar,
the propagation and scattering among $\fldG$'s works
precisely the same way. 
The only point left to be investigated is the
scattering between a $\fldF$ and a $\fldG$. 

Before we consider scattering, 
let us briefly review propagation.
A single-excitation eigenstate $\state{x_{\fldF}}$ 
with some fixed value for $x$ 
is described by a wave function $\psi_k(x)$
\[
\state{x_{\fldF}}=\sum_{k=1}^{K_2} \psi_{k}(x)\,\state{k_{\fldF}},
\qquad
\state{k_{\fldF}}=\state{\fldS\ldots\fldS\MMM{\fldF}{k}\fldS\ldots\fldS}
\]
We are interested in the wave function $\psi_k(x)$ and the eigenvalue
of the total scattering operator
\[
s_{k,k-K_2+1}\cdots s_{k,k-1}\state{x_{\fldF}}
=\lambda_k\state{x_{\fldF}}.
\]
Note that we assume the indices $k$ specifying
the position on the reduced lattice to be periodic. 
The solution can be found by induction on the number
of $s_{k,k-j}$'s applied to $\state{x_{\fldF}}$. We shall
denote the state after $j$ steps by 
\[
\state{x^{(k,j)}_{\fldF}}=
s_{k,k-j}\cdots s_{k,k-1}\state{x_{\fldF}}=
\sum_{l=1}^{K_2} \psi^{(k,j)}_{l}(x)\state{l_{\fldF}}.
\]
The induction is based on the assumption
\[\label{eq:PSU112.NBA.Assumption}
\frac{\psi^{(k,j-1)}_{k}(x)}{\psi^{(k,j-1)}_{k-j}(x)}
=\frac{x-x_{2,k-j}^-}{x-x_{2,k}^+}\,.
\]
which ensures that 
in the induction step
$s_{k,k-j}\state{x_{\fldF}^{(k,j-1)}}=\state{x_{\fldF}^{(k,j)}}$
we can make use of the following two identities
of the S-matrix 
\<\label{eq:PSU112.NBA.Ident}
s^{\fldF\fldS}_{\fldF\fldS}(x_k,x_j)
+\frac{x-x_{k}^+}{x-x_{j}^-}\,
s^{\fldF\fldS}_{\fldS\fldF}(x_k,x_j)\eq
\frac{x-x_{j}^+}{x-x_{j}^-}\,,
\nln
\frac{x-x^-_{j}}{x-x^+_{k}}\,
s^{\fldS\fldF}_{\fldF\fldS}(x_k,x_j)+s^{\fldS\fldF}_{\fldS\fldF}(x_k,x_j)
\eq
\frac{x_{k}^--x}{x_{k}^+-x}\,,
\>
which hold for any $x,x_k,x_j$.
We then find that only two elements of the wave function change
\[
\psi^{(k,j)}_{k}=\psi^{(k,j-1)}_{k}\cdot\frac{x-x_{2,k-j}^+}{x-x_{2,k-j}^-}\,,
\qquad
\psi^{(k,j)}_{k-j}=\psi^{(k,j-1)}_{k-j}\cdot\frac{x_{2,k}^--x}{x_{2,k}^+-x}\,.
\]
It is easy to see that the induction condition of the next step is satisfied and
that the wave function after $j$ steps is given by 
\[\label{eq:PSU.NBA.WaveSteps}
\psi^{(k,j)}_{k-m}
=
\psi_{k-m}
\cdot
\begin{cases}
\displaystyle\prod_{l=1}^j \frac{x-x_{2,k-l}^+}{x-x_{2,k-l}^-}&\mbox{if }m=0,
\\
\displaystyle\frac{x_{2,k}^--x}{x_{2,k}^+-x}&\mbox{if }0 < m\leq j,
\\
1&\mbox{if }m>j.
\end{cases}
\]
Note that the assumption \eqref{eq:PSU112.NBA.Assumption}
for $j=1$ immediately determines the wave-function up
to an overall constant 
in agreement with 
\eqref{eq:1magnon}
\[
\psi_k(x)=\psi_1(x)\prod_{j=1}^{k-1}\frac{x-x_{2,j}^-}{x-x_{2,j+1}^+}\,.
\]
In particular when we impose the periodicity condition 
$\psi_k(x)=\psi_{k+K_2}(x)$ we find the Bethe equation 
\eqref{eq:SU12.Bethe.Bethe2}
\[\label{eq:PSU.NBA.AuxBethe}
\prod_{j=1}^{K_2}\frac{x-x_{2,j}^-}{x-x_{2,j}^+}=1\,,
\]
which determines the admissible values of $x$. 
It is then not difficult to see that after $j=K_2-1$ steps
all elements of the wave function \eqref{eq:PSU.NBA.WaveSteps}
have been multiplied by
\eqref{eq:eigenwert}
\[
\lambda_k=\frac{x_{2,k}^--x}{x_{2,k}^+-x}
\]
and even the element $k$ itself by virtue of 
\eqref{eq:PSU.NBA.AuxBethe}.

Let us now proceed to the two-excitation problem
composed from the basis states
\<
\state{k_1,k_3}\eq +\state{\fldS\ldots\fldS
\MMM{\vphantom{\hat{\fldG}}\fldF}{k_1}\fldS\ldots\fldS
\MMM{\vphantom{\hat{\fldG}}\fldG}{k_3}\fldS\ldots\fldS}
\quad\mbox{when }k_1<k_3,
\nln
\state{k_1,k_1}\eq +\state{\fldS\ldots\fldS
\MMM{(\fldD\fldZ)}{k_1}\fldS\ldots\fldS},
\nln
\state{k_1,k_3}\eq -\state{\fldS\ldots\fldS
\MMM{\vphantom{\hat{\fldG}}\fldG}{k_3}\fldS\ldots\fldS
\MMM{\vphantom{\hat{\fldG}}\fldF}{k_1}\fldS\ldots\fldS}
\quad\mbox{when }k_1>k_3,
\>
We propose that the periodic scattering eigenstate is simply given by
\[
\state{x_1,x_3}=\sum_{k_1,k_3=1}^{K_2}\psi_{k_1}(x_1)\,\psi_{k_3}(x_3)\,\state{k_1,k_3}\,.
\]
As before, this can be proven by applying the partial chain 
of pairwise scatterings and showing that the following
expression satisfies a recursion relation
\[
\state{x_1,x_3{}^{(k,j)}}=s_{k,k-j}\cdots s_{k,k-1}\state{x_1,x_3}
=\sum_{k_1,k_3=1}^{K_2}\psi^{(k,j)}_{k_1}(x_1)\,\psi^{(k,j)}_{k_3}(x_3)\,\state{k_1,k_3}\,.
\]
The recursion is based on the same identities as before, but
we need to separately consider the case
when $s_{k,k-j}$ acts on both $\fldF$ and $\fldG$ at the same time
or when it acts on $\fldD\fldZ$. 
Luckily, this is guaranteed by the identity
\[\label{eq:PSU112.NBA.Ident2}
s^{\fldD\fldS}_{\fldD\fldS}
+\frac{x_3-x_{k}^+}{x_3-x_{j}^-}\,s^{\fldD\fldS}_{\fldF\fldG}
-\frac{x_1-x_{k}^+}{x_1-x_{j}^-}\,s^{\fldD\fldS}_{\fldG\fldF}
+\frac{x_1-x_{k}^+}{x_1-x_{j}^-}\,\frac{x_3-x_{k}^+}{x_3-x_{j}^-}\,s^{\fldD\fldS}_{\fldS\fldD}
=\frac{x_1-x_{j}^+}{x_1-x_{j}^-}\,\frac{x_3-x_{j}^+}{x_3-x_{j}^-}\,.
\]
%
and three similar ones.

As the two-particle wave function is merely the product 
of two one-particle wave functions, there is no phase shift and
periodicity is ensured by two instances of \eqref{eq:PSU.NBA.AuxBethe}, 
i.e.
\[
\prod_{j=1}^{K_2}\frac{x_1-x_{2,j}^-}{x_1-x_{2,j}^+}=1\,,\qquad
\prod_{j=1}^{K_2}\frac{x_3-x_{2,j}^-}{x_3-x_{2,j}^+}=1\,,
\]
and the eigenvalue of the total scatting operator is
\[
\lambda_k=\frac{x_{2,k}^--x_1}{x_{2,k}^+-x_1}\,\frac{x_{2,k}^--x_3}{x_{2,k}^+-x_3}\,.
\]
In conclusion, this means that $\fldF$ and $\fldG$ 
do not feel each other's presence.
The two excitations completely factorize,
there is no further diagonalization required.
In fact, the factorization can be traced back to the
S-matrix which also factorizes in two parts.
Each part governs independently the behavior of 
one type of constituent oscillator
$\osc{A}$ and $\osc{\dot A}$ introduced in \secref{sec:PSU112.One}.
Also the identities of the sort \eqref{eq:PSU112.NBA.Ident2}
are essentially the product
of two identities from \eqref{eq:PSU112.NBA.Ident}.

Note that the factorization is slightly
different from the one for two alike
fermionic excitations
which are subject to the Pauli principle
$\state{x_{\fldF},x_{\fldF}}=0$.
Here the state $\state{x_1,x_3}$ with $x_1=x_3$ 
does exist and the exclusion principle is 
avoided by having two flavors of fermions at our disposal. 
The exclusion principle applies only to excitations of
the same kind.

\subsection{Asymptotic Bethe Equations for $\alg{su}(1,1|2)$}
\label{sec:PSU112.Bethe}


We have now obtained the pairwise scattering 
of excitations $x_{1,2,3}$. As the S-matrix obeys the
YBE, the factorized scattering of more than two excitations
is self-consistent. We can therefore write down the Bethe equations
from the results on scattering phases from above.
The asymptotic Bethe equations for the
$\alg{su}(1,1|2)$ sector read
\<\label{eq:PSU112.Bethe.Bethe}
1
\eq
\prod_{j=1}^{K_2}
\frac{x_{1,k}-x_{2,j}^{+\sgrada}}{x_{1,k}-x_{2,j}^{-\sgrada}}\,,
\nln
1
\eq
\lrbrk{\frac{x^-_{2,k}}{x^+_{2,k}}}^L
\prod_{\textstyle\atopfrac{j=1}{j\neq k}}^{K_2}
\lrbrk{
\frac{1-g^2/2x_{2,k}^+ x_{2,j}^-}{1-g^2/2x_{2,k}^- x_{2,j}^+}\,
\sigma^2(x_{2,k},x_{2,j})}
\nl\qquad\times
\prod_{\textstyle\atopfrac{j=1}{j\neq k}}^{K_2}
\frac{x_{2,k}^{+\sgrada}-x_{2,j}^{-\sgrada}}{x_{2,k}^{-\sgradb}-x_{2,j}^{+\sgradb}}
\prod_{j=1}^{K_1}
\frac{x_{2,k}^{-\sgrada}-x_{1,j}}{x_{2,k}^{+\sgrada}-x_{1,j}}
\prod_{j=1}^{K_3}
\frac{x_{2,k}^{-\sgradb}-x_{3,j}}{x_{2,k}^{+\sgradb}-x_{3,j}}
,
\nln
1
\eq
\prod_{j=1}^{K_2}
\frac{x_{3,k}-x_{2,j}^{+\sgradb}}{x_{3,k}-x_{2,j}^{-\sgradb}}\,.
\>
Here we have presented four equivalent forms
labelled by the grading constants $\sgrada=\pm 1$, $\sgradb=\pm 1$
related to four different Dynkin diagrams and representation vectors.
In the nested Bethe ansatz, these come about by choosing,
at the second level, 
one of the four primary excitations $\fldS,\fldF,\fldG,\fldD\fldZ$
as the new vacuum. Above we have restricted ourselves to $\fldS$ 
which corresponds to $\sgrada=\sgradb=+1$. 
In \secref{sec:PSU112.Duality} we 
shall show how to dualize between the different forms
and thus show their equivalence independently of the nested Bethe ansatz.

As above, $\sigma(x_k,x_j)$ determines 
the model: We set $\sigma(x_k,x_j)=1$
for gauge theory and
$\sigma(x_k,x_j)\neq 1$ 
as in \eqref{eq:One.StringsAnalytic}
for the string chain.
The local charges are obtained 
as before and determined only through the middle, momentum-carrying roots
\[
Q_r=\sum_{j=1}^{K_2}q_r(x_{2,j})
\]
with $q_r(x)$ given in \eqref{eq:One.SpecCharges}.
The anomalous dimension is given by 
\eqref{eq:One.EngX} with $K=K_2$
and for (cyclic) gauge theory states,
the momentum constraint 
\eqref{eq:One.MomConX} must be obeyed.
Note that for $K_1=K_3=0$, the equations reduce to
either one of the three rank-one sectors in 
\eqref{eq:One.BetheAll,eq:One.StringsBethe}
specified by $\sgrad=(\sgrada+\sgradb)/2$.

The Dynkin labels $[q_1,p,q_2]$ of $\alg{su}(4)$ are related to the
excitation numbers by
\<
q_1 \eq -\sgrada K_1+\half(1+\sgrada)K_2,\nln
p \eq +L+\sgrada K_1-\half(2+\sgrada+\sgradb)K_2+\sgradb K_3,\nln
q_2 \eq -\sgradb K_3+\half(1+\sgradb)K_2
\>
while the labels $[s_1,r,s_2]$ of $\alg{su}(2,2)$ are given by
\<
s_1\eq +\sgrada K_1+\half(1-\sgrada)K_2,\nln
r\eq -L-\sgrada K_1-\half(2-\sgrada-\sgradb)K_2-\sgradb K_3-\delta D,\nln
s_2\eq +\sgradb K_3+\half(1-\sgradb)K_2.
\>
%



\subsection{Multiplet Splitting}
\label{sec:PSU112.PSU11}

The $\alg{su}(1,1|2)$ sector has 
a hidden $\alg{psu}(1|1)\times\alg{psu}(1|1)$ symmetry
\cite{Beisert:2004ry}.
Furthermore, all these factors have a common central charge, 
the anomalous dimension $\delta D$, and they
share two abelian external automorphisms,
the length $L$ and the hypercharge $B$.
The symmetry factors originate from the full
$\alg{psu}(2,2|4)$ symmetry. 
The $\alg{psu}(1|1)$ algebras are hidden symmetries
because they act trivially at leading order, i.e.~at $g=0$.

There are two types of multiplets of 
$\alg{psu}(1|1)\ltimes \alg{u}(1)$,
a singlet $\rep{1}$ and a doublet $\rep{1|1}$.
The singlet has zero central charge, in the fully 
interacting theory it is realized only by the vacuum. 
The other states have a non-zero central charge $\delta D$
and therefore come in doublets. However, at
the classical level, the anomalous dimension vanishes
and the doublets split into two singlets. 

Let us see how the Bethe ansatz realizes the 
hidden $\alg{psu}(1|1)$ symmetry.
For the one-loop Bethe ansatz, the multiplet splitting 
implies that each state should have a partner with precisely
the same $\alg{su}(1,1|2)$ quantum numbers,
energy and local charges.
Moreover these two multiplets should
join in the higher-loop Bethe ansatz.

Consider a spin chain eigenstate of length $L$
with $(K_1,K_2,K_3)$ excitations. 
Assume that one of the $x_1$-roots is at $x=0$, 
i.e.~$x_{1,K_1}=0$.
The corresponding Bethe equation \eqref{eq:PSU112.Bethe.Bethe} reads
\[\label{eq:PSU112.PSU11.Cyclic}
1
=\prod_{j=1}^{K_2}
\frac{x_{2,j}^+}{x_{2,j}^-}\,.
\]
Curiously, this is just the cyclicity constraint \eqref{eq:One.MomConX}. 
In other words, $x=0$ can be a $x_1$-root if and only if
the momentum constraint is satisfied. 
Let us see what the effects of 
$x_{1,K_1}=0$ on the other roots are. 
After substitution, the middle equation of \eqref{eq:PSU112.Bethe.Bethe}
turns out to be precisely the equation
for a spin chain of length $L+\sgrada$ with 
$(K_1-1,K_2,K_3)$ excitations.
The outer two equations are not modified as there are
no self-interactions of $x_1$'s and no interactions
between $x_1$'s and $x_3$'s.

We conclude that for every eigenstate with a root $x_1=0$, there 
exist an eigenstate with $L'=L+\sgrada$ and 
$(K_1',K_2',K_3')=(K_1-1,K_2,K_3)$ with the
same set of Bethe roots (except for $x_1=0$). 
Conversely, for every state
in the zero-momentum sector 
without the root $x_1=0$ there exist a 
state with $L'=L-\sgrada$ and $x_1=0$ as
the only additional root. 
As the energy and local charges are determined through the
$x_2$'s alone, they coincide for both states.
They consequently form a doublet of $\alg{psu}(1|1)$.
A similar construction applies for roots $x_3=0$ which
explains the appearance of the second $\alg{psu}(1|1)$ factor. 

Note that an essential requirement of this construction 
is the restriction to cyclic states. 
The hidden $\alg{psu}(1|1)$ symmetry in fact does not exist for generic states. 
This parallels the observations in the construction of 
a similar spin chain in \cite{Beisert:2004ry}, but now
at the level of the Bethe equations.
It does not mean, however, that the spin chain, Bethe ansatz
or $\alg{su}(1,1|2)$ symmetry should be restricted to cyclic states,
they are valid for states with arbitrary total momentum as well.

We can also explain this mechanism at a somewhat deeper level 
of the integrable structure. 
In the standard Bethe ans\"atze, multiplets are
realized as follows: The highest-weight
state has no Bethe root at $u=\infty$. 
Descendants are obtained from the primary 
state by adding roots at $u=\infty$. 
These neither change the Bethe equations, nor do they
modify the energy or higher charge eigenvalues. 
In fact, the point $u=\infty$ is related to 
the generators of the symmetry algebra. 

The situation for the higher-loop Bethe ans\"atze is somewhat different.
Our spectral parameter $x$ resembles the one of string sigma models,
see \cite{Kazakov:2004qf,Beisert:2004hm}. 
There, not only the point $x=\infty$ relates to the symmetry generators,
but, not surprisingly, also the point $x=0$.
In the classical limit, $g=0$, 
the $x$-plane limits to the $u$-plane. 
The points at infinity are identified, but the
limit at the point zero is singular \cite{Beisert:2004ag,Beisert:2005bm}.
This feature spoils the relation between $u=0$ and the symmetry generators,
whereas it is apparent for $x=0$. 
This explains the multiplet splitting/joining mechanism 
in the Bethe ansatz.

\subsection{Exact Degeneracies}
\label{sec:PSU112.Degen}

There is an even larger set of hidden symmetries
that can boost the degeneracy of states by several factors of two.
Let us for simplicity set $\sgrada=\sgradb=+1$.
Consider an eigenstate which has a Bethe root $x$ of flavor $1$ which
is \emph{not} also a root of flavor $3$. Then we can construct
a state with the same set of Bethe roots, but now $x$ being
of flavor $3$ instead of $1$. This state obeys the Bethe equations
\eqref{eq:PSU112.Bethe.Bethe}
because the first and third equation 
coincide and the second equation does
not distinguish between $x_1$ and $x_3$. 
The states have the same higher charges, because 
they share all the roots of flavor $2$.
Nevertheless, the representation of both 
states is different due to different occupation numbers
for the individual flavors.

Degenerate states within the same representation 
can also be constructed. 
Suppose there are at least two roots to which the
above argument applies. For a pair of roots,
we can associate one to flavor $1$ and the other
to flavor $3$ or vice versa. In both cases
the excitation numbers are the same and thus
we get two completely degenerate multiplets of states.
These even have the same parity and therefore this effect
is not related to the pairing described in \cite{Beisert:2003tq}.

We see that (except for the global charges) 
we do not have to distinguish between 
roots of type $1$ and $3$. Nevertheless,
one has to keep in mind that no two Bethe roots 
can occupy the same position.
For these mixed roots the exclusion principle is
circumvented and we can have two
roots at the same position, c.f.~\secref{sec:PSU112.NBA}, 
they merely have to be associated to different flavors.
The situation is somewhat reminiscent of $\superN=2$ supersymmetry
with two flavors of fermions.

It would be interesting to investigate this degeneracy further. 
What is its origin and is it restricted to planar gauge theory only?
Are these degeneracies restricted to the $\alg{su}(1,1|2)$ sector or 
are there even larger degeneracies in the full theory
which are not directly related to the symmetry
group?

\subsection{Duality Transformation}
\label{sec:PSU112.Duality}

Here we shall show the equivalence of the Bethe equations
for different gradings $\sgrada=\pm 1$, $\sgradb=\pm 1$. 
This is an essential step in demonstrating the 
self-consistency of the Bethe ansatz. 
The proof parallels the one for the one-loop level in  
\cite{Essler:1992nk,Essler:1992uc,Gohmannn:2003aa,Beisert:2005di}. 

We rewrite the
Bethe equation for fermionic roots of type $x_1$ 
as the algebraic equation $P(x_{1,k})=0$ with the polynomial
\[\label{eq:PSU112.Duality.PolyDef}
P(x)=
\prod_{j=1}^{K_2}\bigbrk{x-x_{2,j}^+}
-\prod_{j=1}^{K_2}\bigbrk{x-x_{2,j}^-}
.\]
Clearly, all the roots $x_{1,k}$ solve the same equation, but
there are $\tilde K_1=K_2-K_1-1$ further solutions $\tilde x_{1,k}$. 
We can thus write the polynomial as the product of monomials 
\[\label{eq:PSU112.Duality.PolyFactor}
P(x)\sim \prod_{j=1}^{K_1}(x-x_{1,j})\prod_{j=1}^{\tilde K_1}(x-\tilde x_{1,j})
\]
with some fixed factor of proportionality.
Using this identity, 
we can now write some combination of terms which appear in the 
Bethe equations 
\eqref{eq:PSU112.Bethe.Bethe}
using the polynomial
\[\label{eq:PSU112.Duality.Comb1}
\prod_{j=1}^{K_1}
\frac{x_{2,k}^{+\sgrada}-x_{1,j}}{x_{2,k}^{-\sgrada}-x_{1,j}}
\prod_{j=1}^{\tilde K_1}
\frac{x_{2,k}^{+\sgrada}-\tilde x_{1,j}}{x_{2,k}^{-\sgrada}-\tilde x_{1,j}}
=
\frac{P(x_{2,k}^{+\sgrada})}{P(x_{2,k}^{-\sgrada})}\,.
\]
When we substitute the original 
definition of the polynomials 
\eqref{eq:PSU112.Duality.PolyDef}
we obtain a different expression
\[\label{eq:PSU112.Duality.Comb2}
\frac{P(x_{2,k}^{+\sgrada})}{P(x_{2,k}^{-\sgrada})}
=
\frac{\prod_{j=1}^{K_2}\bigbrk{x_{2,k}^{+\sgrada}-x_{2,j}^{+}}
      -\prod_{j=1}^{K_2}\bigbrk{x_{2,k}^{+\sgrada}-x_{2,j}^{-}}}
{\prod_{j=1}^{K_2}\bigbrk{x_{2,k}^{-\sgrada}-x_{2,j}^{+}}
      -\prod_{j=1}^{K_2}\bigbrk{x_{2,k}^{-\sgrada}-x_{2,j}^{-}}}
=
\prod_{\textstyle\atopfrac{j=1}{j\neq k}}^{K_2}
\frac{x_{2,k}^{+\sgrada}-x_{2,j}^{-\sgrada}}{x_{2,k}^{-\sgrada}-x_{2,j}^{+\sgrada}}\,.
\]
For $\sgrada=+1$, the first term in the numerator and the second term
in the denominator are trivially zero and vice versa for $\sgrada=-1$. 
The sign is cancelled by removing the factor with $j=k$ which is always $-1$. 
Combining the two equations we find an identity which relates 
terms of the Bethe equations using the roots $x_1$ and their duals
$\tilde x_1$
\[\label{eq:PSU112.Duality.Identity}
\prod_{\textstyle\atopfrac{j=1}{j\neq k}}^{K_2}
\bigbrk{x_{2,k}^{+\sgrada}-x_{2,j}^{-\sgrada}}
\prod_{j=1}^{K_1}
\frac{x_{2,k}^{-\sgrada}-x_{1,j}}{x_{2,k}^{+\sgrada}-x_{1,j}}
=
\prod_{\textstyle\atopfrac{j=1}{j\neq k}}^{K_2}
\bigbrk{x_{2,k}^{-\sgrada}-x_{2,j}^{+\sgrada}}
\prod_{j=1}^{\tilde K_1}
\frac{x_{2,k}^{+\sgrada}-\tilde x_{1,j}}{x_{2,k}^{-\sgrada}-\tilde x_{1,j}}\,.
\]
We apply this identity to the middle equation of
\eqref{eq:PSU112.Bethe.Bethe} and 
invert the first equation to obtain the dualized equations
\<\label{eq:PSU112.Duality.Bethe}
1
\eq
\prod_{j=1}^{K_2}
\frac{\tilde x_{1,k}-x_{2,j}^{-\sgrada}}{\tilde x_{1,k}-x_{2,j}^{+\sgrada}}\,,
\nln
1
\eq
\lrbrk{\frac{x^-_{2,k}}{x^+_{2,k}}}^L
\prod_{\textstyle\atopfrac{j=1}{j\neq k}}^{K_2}
S_0(x_{2,k},x_{2,j})
\prod_{\textstyle\atopfrac{j=1}{j\neq k}}^{K_2}
\frac{x_{2,k}^{-\sgrada}-x_{2,j}^{+\sgrada}}{x_{2,k}^{-\sgradb}-x_{2,j}^{+\sgradb}}
\prod_{j=1}^{\tilde K_1}
\frac{x_{2,k}^{+\sgrada}-\tilde x_{1,j}}{x_{2,k}^{-\sgrada}-\tilde x_{1,j}}
\prod_{j=1}^{K_3}
\frac{x_{2,k}^{-\sgradb}-x_{3,j}}{x_{2,k}^{+\sgradb}-x_{3,j}}\,,
\nln
1
\eq
\prod_{j=1}^{K_2}
\frac{x_{3,k}-x_{2,j}^{+\sgradb}}{x_{3,k}-x_{2,j}^{-\sgradb}}\,.
\>
They agree with the original equations after substituting 
$\sgrada\mapsto -\sgrada$ and $x_{1,k}\to \tilde x_{1,k}$.
As the $x_{2}$-roots remain unchanged, the energy and local charges 
remain invariant under dualization.
For $K_3=0$, this proves the equivalence of the 
two sets of equations in \secref{sec:nba} on an independent basis.
The argument for roots of type $3$ is the same.

Note that we will not try to dualize the middle 
node of the Dynkin diagram. This would take the protected ground state of 
scalar fields $\fldZ$ into a highly interacting pseudo-vacuum of 
fermions $\fldF$. Consequently, and in contrast to the above transformation 
as well as the one-loop approximation, 
the dualization of the middle node appears to be substantially more involved.

\subsection{Frolov-Tseytlin Limit}

Here we present the Frolov-Tseytlin limit
of the Bethe equations. 
In this limit, the duality transformation is a mere permutation
Riemann sheets \cite{Beisert:2005di}, so without loss of generality
we can set $\sgrada=\sgradb=+1$.
The limit of the Bethe equations 
for gauge theory with $\sigma(x_k,x_j)=1$ reads%
\footnote{For many states, Bethe roots form strings of stacks,
not merely strings of roots \cite{Beisert:2005di}. 
The corresponding cuts stretch between 
several resolvents $G_j$.
To handle this situation carefully, 
additional resolvents and equations should be introduced.
Nevertheless, the presented equations contain
all the relevant information
and it is not necessary to specify 
the extended equations.}
\<
-2\pi n_{1,a} \eq -G_2(x)
\qquad\qquad\qquad\qquad\qquad\qquad\qquad\qquad\qquad\quad\,\,\,\,
\mbox{for }x\in \contour_{1,a},
\nln
-2\pi n_{2,a} \eq
+2\resolvHsl_2(x)
+\frac{L/x}{1-g^2/2x^2}
\nl
-G_1(x)-G'_1(0)\,\frac{g^2/2x}{1-g^2/2x^2}-G_1(0)\,\frac{g^2/2x^2}{1-g^2/2x^2}
\nl
-G_3(x)-G'_3(0)\,\frac{g^2/2x}{1-g^2/2x^2}-G_3(0)\,\frac{g^2/2x^2}{1-g^2/2x^2}
\qquad
\mbox{for }x\in \contour_{2,a},
\nln
-2\pi n_{3,a} \eq -G_2(x)
\qquad\qquad\qquad\qquad\qquad\qquad\qquad\qquad\qquad\quad\,\,\,\,
\mbox{for }x\in \contour_{3,a}.
\>
The limits of the equations for the string chain 
with $\sigma(x_k,x_j)\neq 1$ are
\<
-2\pi n_{1,a} \eq -G_2(x)\qquad\qquad\qquad\qquad\qquad\qquad\qquad\qquad\qquad\quad\,\,\,\,
\mbox{for }x\in \contour_{1,a},
\nln
-2\pi n_{2,a}\eq
+2\resolvsl_2(x)
+2G'_2(0)\,\frac{g^2/2x}{1-g^2/2x^2}
+\frac{L/x}{1-g^2/2x^2}
\nl
-G_1(x)-G'_1(0)\,\frac{g^2/2x}{1-g^2/2x^2}-G_1(0)\,\frac{g^2/2x^2}{1-g^2/2x^2}
\nl
-G_3(x)-G'_3(0)\,\frac{g^2/2x}{1-g^2/2x^2}-G_3(0)\,\frac{g^2/2x^2}{1-g^2/2x^2}
\qquad
\mbox{for }x\in \contour_{2,a},
\nln
-2\pi n_{3,a} \eq -G_2(x)\qquad\qquad\qquad\qquad\qquad\qquad\qquad\qquad\qquad\quad\,\,\,\,
\mbox{for }x\in \contour_{3,a}.
\>
They perfectly agree with the expressions
derived from the classical superstring sigma model in 
\cite{Beisert:2005bm}.
We have thus found a possible quantization for
superstrings on $AdS_5\times S^5$ restricted to
the subspace $AdS_3\times S^3$
in the spirit of \cite{Arutyunov:2004vx,Beisert:2004jw}.

\section{Complete Asymptotic Bethe Equations}
\label{sec:Odyssee2005}

The $\alg{su}(1,1|2)$ sector is the largest sector 
for which mixing of states of different lengths is suppressed
at all orders in perturbation theory.
Nevertheless, the dynamic nature of the higher-loop spin chain 
for $\superN=4$ gauge theory is not an obvious obstacle for integrability.
Indeed some signs of higher-loop integrability beyond
$\alg{su}(1,1|2)$ were found in \cite{Beisert:2003ys}.
We might therefore hope that the spectrum of the full model 
with $\alg{psu}(2,2|4)$ symmetry can also be described 
by a suitable Bethe ansatz.
In this section we will assemble various pieces 
of the puzzle available in the literature
and construct candidate Bethe equations 
for the complete higher-loop 
spin chain of $\superN=4$ SYM
and the complete string chain.

\subsection{Bethe Equations}
\label{sec:points}

Before we make a proposal for the equations, let us 
present a list of constraints and expected features
(see also \cite{Beisert:2004ry}):
\begin{enumlist}
\item\label{constr1}
The higher-loop equations should turn into 
the one-loop equations of \cite{Beisert:2003yb} when setting $g=0$.

\item\label{constr2}
The equations should turn into 
the equations of the previous section
when restricting to the $\alg{su}(1,1|2)$ sector.

\item\label{constr2b}
The thermodynamic limit of the string chain equations 
($\sigma\neq 1$ as in \eqref{eq:One.StringsAnalytic}) 
should agree with \cite{Beisert:2005bm}.

\item\label{constr3}
The length $L$ and the hypercharge $B$ are
not conserved quantities at higher loops. 
This fact should be reflected by the equations. 

\item\label{constr4}
In the Bethe ansatz, the highest-weight state of a multiplet
is singled out by the absence of Bethe roots at $u=\infty$.
In the $x$-plane this point corresponds to $x=\infty$ or $x=0$.
Bethe roots at $u=\infty$ should indicate descendants. 

\item\label{constr5}
All short multiplets in the free theory
(except the vacuum) 
must join into long multiplets in the interacting theory. 
For the Bethe ansatz this implies that some states 
should appear as primaries at $g=0$, 
but become descendants when $g\neq 0$.

\item\label{constr6}
The spin chains from gauge theory are defined only modulo 
cyclic permutations. 
While at one loop this feature merely led to 
the restriction to the zero-momentum sector
of a general periodic spin chain,
the higher-loop spin chain apparently is 
self-consistent only in the zero-momentum sector 
\cite{Beisert:2003ys,Beisert:2004ry}.
The two main reasons are: 
Firstly, length-changing interactions destroy the
identification of individual sites and 
allow only for relative positions. 
Secondly, multiplet joining cannot
work due to a mismatch of states 
(unless there are many more protected states than expected).

\item\label{constr7}
One should be able to read the Dynkin labels of a state 
from the set of Bethe roots:
When expanding the left hand side of the Bethe equation for 
a root of flavor $j$ around $x=\infty$
while keeping all other roots fixed, 
one should obtain the $j$-th Dynkin label $r_j$
via $1-ir_j/x_{j,k}+\order{1/x_{j,k}^2}$,
see, e.g., \eqref{eq:One.LimitPotentialExpand}.
In particular, for the non-compact algebra
$\alg{psu}(2,2|4)$ some Dynkin labels
contain the anomalous dimension $\delta D$. 

\item\label{constr8}
Somehow the very nature of the algebra
$\alg{psu}(2,2|4)$ should play a role
because the $\superN=4$ superconformal
field theory is a very special model. 
Unlike the standard spin chains which can be 
constructed for an arbitrary symmetry algebra,
the higher-loop spin chain is expected to make 
use of special features of $\alg{psu}(2,2|4)$.

\item\label{constr9}
There should be several equivalent formulations
of the Bethe equations for various forms of the 
Dynkin diagram of $\alg{psu}(2,2|4)$. 

\item\label{constr10}
The spectrum should agree with $\superN=4$ SYM.

\end{enumlist}

\begin{figure}\centering
$\sgrada=+1$
\begin{minipage}{260pt}
\setlength{\unitlength}{1pt}%
\small\thicklines%
\begin{picture}(260,20)(-10,-10)
\put(  0,00){\circle{15}}%
\put(  7,00){\line(1,0){26}}%
\put( 40,00){\circle{15}}%
\put( 47,00){\line(1,0){26}}%
\put( 80,00){\circle{15}}%
\put( 87,00){\line(1,0){26}}%
\put(120,00){\circle{15}}%
\put(127,00){\line(1,0){26}}%
\put(160,00){\circle{15}}%
\put(167,00){\line(1,0){26}}%
\put(200,00){\circle{15}}%
\put(207,00){\line(1,0){26}}%
\put(240,00){\circle{15}}%
\put( -5,-5){\line(1, 1){10}}%
\put( -5, 5){\line(1,-1){10}}%
\put( 75,-5){\line(1, 1){10}}%
\put( 75, 5){\line(1,-1){10}}%
\put(155,-5){\line(1, 1){10}}%
\put(155, 5){\line(1,-1){10}}%
\put(235,-5){\line(1, 1){10}}%
\put(235, 5){\line(1,-1){10}}%
\put( 40,00){\makebox(0,0){$-$}}%
\put(120,00){\makebox(0,0){$+$}}%
\put(200,00){\makebox(0,0){$-$}}%
\end{picture}
\end{minipage}
$\sgradb=+1$%
\medskip\par
$\sgrada=+1$
\begin{minipage}{260pt}
\setlength{\unitlength}{1pt}%
\small\thicklines%
\begin{picture}(260,20)(-10,-10)
\put(  0,00){\circle{15}}%
\put(  7,00){\line(1,0){26}}%
\put( 40,00){\circle{15}}%
\put( 47,00){\line(1,0){26}}%
\put( 80,00){\circle{15}}%
\put( 87,00){\line(1,0){26}}%
\put(120,00){\circle{15}}%
\put(127,00){\line(1,0){26}}%
\put(160,00){\circle{15}}%
\put(167,00){\line(1,0){26}}%
\put(200,00){\circle{15}}%
\put(207,00){\line(1,0){26}}%
\put(240,00){\circle{15}}%
\put( -5,-5){\line(1, 1){10}}%
\put( -5, 5){\line(1,-1){10}}%
\put( 75,-5){\line(1, 1){10}}%
\put( 75, 5){\line(1,-1){10}}%
\put(115,-5){\line(1, 1){10}}%
\put(115, 5){\line(1,-1){10}}%
\put(155,-5){\line(1, 1){10}}%
\put(155, 5){\line(1,-1){10}}%
\put(235,-5){\line(1, 1){10}}%
\put(235, 5){\line(1,-1){10}}%
\put( 40,00){\makebox(0,0){$-$}}%
\put(200,00){\makebox(0,0){$+$}}%
\end{picture}
\end{minipage}
$\sgradb=-1$%
\medskip\par
$\sgrada=-1$
\begin{minipage}{260pt}
\setlength{\unitlength}{1pt}%
\small\thicklines%
\begin{picture}(260,20)(-10,-10)
\put(  0,00){\circle{15}}%
\put(  7,00){\line(1,0){26}}%
\put( 40,00){\circle{15}}%
\put( 47,00){\line(1,0){26}}%
\put( 80,00){\circle{15}}%
\put( 87,00){\line(1,0){26}}%
\put(120,00){\circle{15}}%
\put(127,00){\line(1,0){26}}%
\put(160,00){\circle{15}}%
\put(167,00){\line(1,0){26}}%
\put(200,00){\circle{15}}%
\put(207,00){\line(1,0){26}}%
\put(240,00){\circle{15}}%
\put( -5,-5){\line(1, 1){10}}%
\put( -5, 5){\line(1,-1){10}}%
\put( 75,-5){\line(1, 1){10}}%
\put( 75, 5){\line(1,-1){10}}%
\put(115,-5){\line(1, 1){10}}%
\put(115, 5){\line(1,-1){10}}%
\put(155,-5){\line(1, 1){10}}%
\put(155, 5){\line(1,-1){10}}%
\put(235,-5){\line(1, 1){10}}%
\put(235, 5){\line(1,-1){10}}%
\put( 40,00){\makebox(0,0){$+$}}%
\put(200,00){\makebox(0,0){$-$}}%
\end{picture}
\end{minipage}
$\sgradb=+1$%
\medskip\par
$\sgrada=-1$
\begin{minipage}{260pt}
\setlength{\unitlength}{1pt}%
\small\thicklines%
\begin{picture}(260,20)(-10,-10)
\put(  0,00){\circle{15}}%
\put(  7,00){\line(1,0){26}}%
\put( 40,00){\circle{15}}%
\put( 47,00){\line(1,0){26}}%
\put( 80,00){\circle{15}}%
\put( 87,00){\line(1,0){26}}%
\put(120,00){\circle{15}}%
\put(127,00){\line(1,0){26}}%
\put(160,00){\circle{15}}%
\put(167,00){\line(1,0){26}}%
\put(200,00){\circle{15}}%
\put(207,00){\line(1,0){26}}%
\put(240,00){\circle{15}}%
\put( -5,-5){\line(1, 1){10}}%
\put( -5, 5){\line(1,-1){10}}%
\put( 75,-5){\line(1, 1){10}}%
\put( 75, 5){\line(1,-1){10}}%
\put(155,-5){\line(1, 1){10}}%
\put(155, 5){\line(1,-1){10}}%
\put(235,-5){\line(1, 1){10}}%
\put(235, 5){\line(1,-1){10}}%
\put( 40,00){\makebox(0,0){$+$}}%
\put(120,00){\makebox(0,0){$-$}}%
\put(200,00){\makebox(0,0){$+$}}%
\end{picture}
\end{minipage}
$\sgradb=-1$
\caption{Dynkin diagrams of $\alg{su}(2,2|4)$ for the gradings
$\sgrada,\sgradb=\pm1$.
The signs in the white nodes indicate
the sign of the diagonal elements of the
Cartan matrix.}
\label{fig:DynkinSU224}
\end{figure}
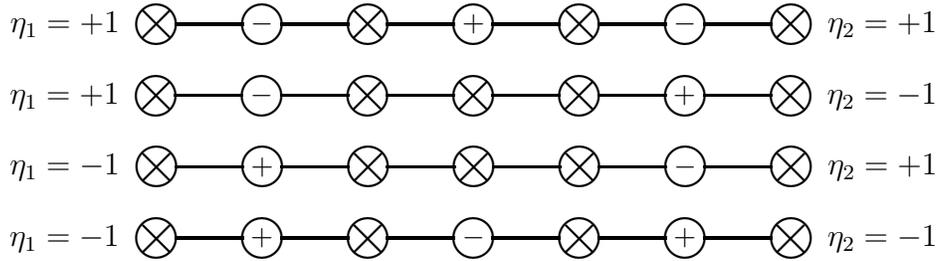

We have found a set of Bethe equations which fulfills 
all of the above conditions or at least does not apparently violate them
(as for point \ref{constr10}). 
There are four forms labelled by the gradings $\sgrada,\sgradb=\pm 1$.
They correspond to different choices for the Cartan matrix
of $\alg{su}(2,2|4)$
\[
M_{j,j'}=\matr{ccccccc}{
&+\sgrada&&&&&\\
+\sgrada&-2\sgrada&+\sgrada&&&&\\
&+\sgrada&&-\sgrada&&&\\
&&-\sgrada&+\sgrada+\sgradb&-\sgradb&&\\
&&&-\sgradb&&+\sgradb&\\
&&&&+\sgradb&-2\sgradb&+\sgradb\\
&&&&&+\sgradb&
}
\]
given by the Dynkin diagrams in \figref{fig:DynkinSU224}. 
The Bethe equations are presented in \tabref{tab:ABE}.

\begin{table}
\<\nn
1\eq
\prod_{j=1}^{K_4}
\frac{x_{4,j}^+}{x_{4,j}^-}\,,
\nonumber\\[1em]
1\eq
\prod_{\textstyle\atopfrac{j=1}{}}^{K_2}
\frac{u_{1,k}-u_{2,j}+\sfrac{i}{2}\sgrada}{u_{1,k}-u_{2,j}-\sfrac{i}{2}\sgrada}
\prod_{j=1}^{K_4}
\frac{1-g^2/2x_{1,k}x_{4,j}^{+\sgrada}}{1-g^2/2x_{1,k}x_{4,j}^{-\sgrada}}\,,
\nln
1\eq
\prod_{\textstyle\atopfrac{j=1}{j\neq k}}^{K_2}
\frac{u_{2,k}-u_{2,j}-i\sgrada}{u_{2,k}-u_{2,j}+i\sgrada}
\prod_{j=1}^{K_3}
\frac{u_{2,k}-u_{3,j}+\sfrac{i}{2}\sgrada}{u_{2,k}-u_{3,j}-\frac{i}{2}\sgrada}
\prod_{j=1}^{K_1}
\frac{u_{2,k}-u_{1,j}+\sfrac{i}{2}\sgrada}{u_{2,k}-u_{1,j}-\frac{i}{2}\sgrada}\,,
\nln
1\eq
\prod_{\textstyle\atopfrac{j=1}{}}^{K_2}
\frac{u_{3,k}-u_{2,j}+\sfrac{i}{2}\sgrada}{u_{3,k}-u_{2,j}-\sfrac{i}{2}\sgrada}
\prod_{j=1}^{K_4}
\frac{x_{3,k}-x_{4,j}^{+\sgrada}}{x_{3,k}-x_{4,j}^{-\sgrada}}\,,
\nln
1\eq
\lrbrk{\frac{x^-_{4,k}}{x^+_{4,k}}}^L
\prod_{\textstyle\atopfrac{j=1}{j\neq k}}^{K_4}
\lrbrk{
\frac{x_{4,k}^{+\sgrada}-x_{4,j}^{-\sgrada}}{x_{4,k}^{-\sgradb}-x_{4,j}^{+\sgradb}}\,
\frac{1-g^2/2x_{4,k}^+ x_{4,j}^-}{1-g^2/2x_{4,k}^- x_{4,j}^+}\,
\sigma^2(x_{4,k},x_{4,j})}
\nl\qquad
\times
\prod_{\textstyle\atopfrac{j=1}{}}^{K_1}
\frac{1-g^2/2x_{4,k}^{-\sgrada} x_{1,j}}{1-g^2/2x_{4,k}^{+\sgrada}x_{1,j}}
\prod_{j=1}^{K_3}
\frac{x_{4,k}^{-\sgrada}-x_{3,j}}{x_{4,k}^{+\sgrada}-x_{3,j}}
\prod_{j=1}^{K_5}
\frac{x_{4,k}^{-\sgradb}-x_{5,j}}{x_{4,k}^{+\sgradb}-x_{5,j}}
\prod_{j=1}^{K_7}
\frac{1-g^2/2x_{4,k}^{-\sgradb}x_{7,j}}{1-g^2/2x_{4,k}^{+\sgradb}x_{7,j}}\,,
\nln
1\eq
\prod_{\textstyle\atopfrac{j=1}{}}^{K_6}
\frac{u_{5,k}-u_{6,j}+\sfrac{i}{2}\sgradb}{u_{5,k}-u_{6,j}-\sfrac{i}{2}\sgradb}
\prod_{j=1}^{K_4}
\frac{x_{5,k}-x_{4,j}^{+\sgradb}}{x_{5,k}-x_{4,j}^{-\sgradb}}\,,
\nln
1\eq
\prod_{\textstyle\atopfrac{j=1}{j\neq k}}^{K_6}
\frac{u_{6,k}-u_{6,j}-i\sgradb}{u_{6,k}-u_{6,j}+i\sgradb}
\prod_{j=1}^{K_5}
\frac{u_{6,k}-u_{5,j}+\sfrac{i}{2}\sgradb}{u_{6,k}-u_{5,j}-\frac{i}{2}\sgradb}
\prod_{j=1}^{K_7}
\frac{u_{6,k}-u_{7,j}+\sfrac{i}{2}\sgradb}{u_{6,k}-u_{7,j}-\frac{i}{2}\sgradb}\,,
\nln
1\eq
\prod_{\textstyle\atopfrac{j=1}{}}^{K_6}
\frac{u_{7,k}-u_{6,j}+\sfrac{i}{2}\sgradb}{u_{7,k}-u_{6,j}-\sfrac{i}{2}\sgradb}
\prod_{j=1}^{K_4}
\frac{1-g^2/2x_{7,k}x_{4,j}^{+\sgradb}}{1-g^2/2x_{7,k}x_{4,j}^{-\sgradb}}\,,
\nonumber\\[1em]
Q_r\eq
\frac{1}{r-1}\sum_{j=1}^{K_4}
\lrbrk{\frac{i}{(x_{4,j}^+)^{r-1}}-\frac{i}{(x_{4,j}^-)^{r-1}}},
\qquad
\delta D=g^2 Q_2=
g^2\sum_{j=1}^{K_4}
\lrbrk{\frac{i}{x_{4,j}^+}-\frac{i}{x_{4,j}^-}}.
\nonumber
\>

\caption{Asymptotic Bethe equations for the complete models.
The first line is the momentum constraint
which is an essential part of the Bethe ansatz.
The following seven equations must hold for all $k=1,\ldots,K_j$. 
The variables $u$ and $x^\pm$ are related to the Bethe roots $x$ by
\protect\eqref{eq:One.SpecRoots,eq:One.SpecRap}.
The last line determines the local charge eigenvalues $Q_r$ and the
anomalous dimension $\delta D$.
The function $\sigma$ specifies the model:
$\sigma=1$ for gauge theory or 
as in \protect\eqref{eq:One.StringsAnalytic} for the string chain.
The gradings $\sgrada,\sgradb=\pm 1$ select one 
of the four Dynkin diagrams in \protect\figref{fig:DynkinSU224}.
}
\label{tab:ABE}
\end{table}

Point \ref{constr2} is easily confirmed by setting $K_1=K_2=K_6=K_7=0$
to restrict to the $\alg{su}(1,1|2)$ sector.
The remaining equations agree with 
\eqref{eq:PSU112.Bethe.Bethe}.
Point \ref{constr1} is almost as
straight-forward: 
One has to set $x=u$, $x^\pm=u\pm \sfrac{i}{2}$ with a finite $u$ and $g=0$.%
\footnote{The limit is subtle:
For the naive limiting equations we assume that all $u=\order{g^0}$.
Special precautions have to be taken when 
some $u=\order{g^1}$ (singular states) 
or $u=\order{g^2}$ (dynamic transformation), see below.}
Then we compare to the Bethe equations in \cite{Beisert:2003yb}
using the appropriate Dynkin diagram 
in \figref{fig:DynkinSU224}.
Concerning point \ref{constr4} we note that
adding a root of any kind at $x=\infty$ 
does not modify the original equations
or the set of higher charge eigenvalues.
Thus a state with roots at $x=\infty$ is a descendant 
and a state without such roots is highest weight.

\subsection{Global Charges}

The Dynkin labels of a state can be read off 
from the Bethe equations as explained at point \ref{constr7}.
We expand the left hand side of the 
Bethe equation for a root of flavor $j$ 
around $x=\infty$ and keep all other roots fixed.
We then obtain the Dynkin labels $r_j$ as the leading 
coefficient%
\footnote{As expected from symmetry arguments, 
the result agrees for both models, i.e.~for gauge and string theory.
Together with the correct scaling behavior of both models,
it implies the exact agreement of spectra 
up to two loops, cf.~\secref{sec:One.Limit}.}
\<
r_1\eq-\sgrada K_2-\half \sgrada \delta D,\nln
r_2\eq-\sgrada K_3+2\sgrada K_2-\sgrada K_1,\nln
r_3\eq+\sgrada K_4-\sgrada K_2+\half \sgrada \delta D,\nln
r_4\eq+L-(\sgrada+\sgradb) K_4+\sgrada K_3+\sgradb K_5+\sfrac{1}{4}(2-\sgrada-\sgradb)\delta D,\nln
r_5\eq+\sgradb K_4-\sgradb K_6+\half \sgradb \delta D,\nln
r_6\eq-\sgradb K_5+2\sgradb K_6-\sgradb K_7,\nln
r_7\eq-\sgradb K_6-\half \sgradb \delta D.
\>
These Dynkin labels depend on the particular choice of Dynkin diagram.
It is more convenient to use the Dynkin labels of the bosonic subalgebras. 
The Dynkin labels $[q_1,p,q_2]$ of $\alg{su}(4)$ are given by
\<
q_1 \eq -\sgrada K_1-(1-\sgrada)K_2-\sgrada K_3+\half(1+\sgrada)K_4,\nln
p \eq +L+\half(1-\sgrada)K_2+\sgrada K_3-\half(2+\sgrada+\sgradb)K_4+\sgradb K_5+\half(1-\sgradb)K_6,\nln
q_2 \eq -\sgradb K_7-(1-\sgradb)K_6-\sgradb K_5+\half(1+\sgradb)K_4
\>
and the labels $[s_1,r,s_2]$ of $\alg{su}(2,2)$ read
\<
s_1\eq +\sgrada K_1-(1+\sgrada)K_2+\sgrada K_3+\half(1-\sgrada)K_4,\nln
r\eq -L+\half(1+\sgrada)K_2-\sgrada K_3-\half(2-\sgrada-\sgradb)K_4-\sgradb K_5+\half(1+\sgradb)K_6-\delta D,\nln
s_2\eq +\sgradb K_7-(1+\sgradb)K_6+\sgradb K_5+\half(1-\sgradb)K_4.
\>
Note that it is sufficient to state just these six charges,
the seventh is determined through
the central charge constraint of $\alg{psu}(2,2|4)$
which reads for the present Cartan matrix
\[
\sgrada r_1-\sgrada r_3+\sgradb r_5-\sgradb r_7=0.
\]
Often it is useful to know 
the scaling dimension $D=-\half s_1-r-\half s_2$, the corresponding charge 
$J=\half q_1+p+\half q_2$ of $\alg{su}(4)$
or the plane-wave light-cone energy $D-J$
\<
J\eq L
+\half \sgrada (K_3-K_1)
-\quarter(2+\sgrada+\sgradb)K_4
+\half \sgradb (K_5-K_7),
\nln
D\eq L
+\half\sgrada (K_3-K_1)
+\quarter(2-\sgrada-\sgradb)K_4
+\half\sgradb (K_5-K_7)
+\delta D,
\nln
D-J\eq K_4+\delta D.
\>
Conversely, the numbers of Bethe roots
in terms of Dynkin labels of 
$\alg{su}(4)$ and $\alg{su}(2,2)$ are%
\footnote{Note that the definition of the highest-weight
state of a multiplet depends on the Dynkin diagram, 
i.e.~on $\sgrada,\sgradb$.}
\<\label{eq:Beyond.ExNo}
K_1\eq 
+\half\sgrada(L-B)
-\sfrac{1}{8}(1-\sgrada)(2r_0+3s_1+s_2)
-\sfrac{1}{8}(1+\sgrada)(2p+3q_1+q_2)
,\nln
K_2\eq 
-\sfrac{1}{4}(2r_0+3s_1+s_2)
-\sfrac{1}{4}(2p+3q_1+q_2)
,\nln
K_3\eq 
-\half\sgrada(L-B)
-\sfrac{1}{8}(3+\sgrada)(2r_0-s_1+s_2)
-s_1
-\sfrac{1}{8}(3-\sgrada)(2p-q_1+q_2)
-q_1
,\nln
K_4\eq 
-r_0-\sfrac{1}{2}s_1-\sfrac{1}{2}s_2
-p-\sfrac{1}{2}q_1-\sfrac{1}{2}q_2
,\nln
K_5\eq 
-\half\sgradb(L+B)
-\sfrac{1}{8}(3+\sgradb)(2r_0+s_1-s_2)
-s_2
-\sfrac{1}{8}(3-\sgradb)(2p+q_1-q_2)
-q_2
,\nln
K_6\eq 
-\sfrac{1}{4}(2r_0+s_1+3s_2)
-\sfrac{1}{4}(2p+q_1+3q_2)
,\nln
K_7\eq 
+\half\sgradb(L+B)
-\sfrac{1}{8}(1-\sgradb)(2r_0+s_1+3s_2)
-\sfrac{1}{8}(1+\sgradb)(2p+q_1+3q_2)
\>
with $r_0=r+\delta D$.
The label $B$ is the $\alg{u}(1)$ hypercharge
of $\alg{pu}(2,2|4)$.

\subsection{Spectrum}

Clearly, the most important constraint on the Bethe equations 
is \ref{constr10}, the equations must give the right answer, 
i.e.~the predicted spectrum must actually agree with $\superN=4$ SYM!
While we are still far from proving this point, 
we can nevertheless compare the energies of several states 
with explicit computations using the three-loop 
$\alg{su}(2|3)$ dynamic spin chain 
Hamiltonian of \cite{Beisert:2003ys}, i.e.~up to third order in $g^2$.

\begin{table}\centering
$\begin{array}{|c|cc|l|}\hline
L&K_1&K_2&(E_0,E_2,E\indup{4g}|E\indup{4s})^P\\\hline\hline
6&4&2&(14E-36,-24E+90,\frac{173}{2}E-315|\frac{151}{2}E-279)^+\,\surd\\\hline
7&4&2&({\scriptstyle 22E^2-144E+248},
       {\scriptstyle -37E^2+460E-1016},
       {\scriptstyle 125E^2-1893E+4438}|
       {\scriptstyle 106E^2-1659E+3942})^-\,\surd\\\hline
8&4&2&(7,-\frac{19}{2},\frac{59}{2}|\frac{115}{4})^\pm\,\surd\\
 & & &({\scriptstyle 44E^5-768E^4+6752E^3-31168E^2+70528E-60224},\\&&& \qquad
       {\scriptstyle -73E^5+2486E^4-31804E^3+188280E^2-506048E+487104},\\&&& \qquad
       {\scriptstyle 251E^5-10452E^4+156202E^3-1041992E^2+3055168E-3125328}|\\&&& \qquad
       {\scriptstyle 223E^5-9500E^4+144054E^3-970456E^2+2864848E-2944656})^+\,\surd\\\hline
\end{array}$
\caption{
Spectrum of lowest-lying $4,2$-excitation
states of the $\alg{su}(3)$ and $\alg{su}(2|1)$ sectors.}
\label{tab:SU3orSU21}
\end{table}

\begin{table}\centering
$\begin{array}{|c|cc|l|}\hline
L&K_1&K_2&(E_0,E_2,E\indup{4g}|E\indup{4s})^P\\\hline\hline
8&5&2&(9,-\frac{31}{2},\frac{103}{2}|\frac{177}{4})^\pm\\
 & & &({\scriptstyle 24E^2-172E+344},
       {\scriptstyle -39E^2+524E-1372},
       {\scriptstyle 138E^2-2209E+6198}|
       {\scriptstyle 127E^2-2069E+5854})^-\,\surd\\\hline
\end{array}$
\caption{Spectrum of lowest-lying states genuinely in the $\alg{su}(3)$ sector.}
\label{tab:SU3}
\end{table}

\begin{table}\centering
$\begin{array}{|c|cc|l|}\hline
L&K_1&K_2&(E_0,E_2,E\indup{4g}|E\indup{4s})^P\\\hline\hline
7&5&2&(9,-15,51|\frac{183}{4})^\pm\,\surd\\\hline
8&5&2&(8,-13,\frac{173}{4}|\frac{157}{4})^\pm\\
 & & &(10,-\frac{67}{4},\frac{3725}{64}|\frac{3437}{64})^\pm\\
 & & &(19E-86,-\frac{133}{4}E+\frac{1169}{4},\frac{7395}{64}E-\frac{79503}{64}|\frac{6707}{64}E-\frac{73359}{64})^\pm\\
8&6&2&({\scriptstyle 30E^2-280E+808},
       {\scriptstyle -52E^2+926E-3800},
       {\scriptstyle \frac{361}{2}E^2-3878E+18246}|
       {\scriptstyle \frac{315}{2}E^2-3476E+16630})^-\\
8&6&3&({\scriptstyle 32E^2-324E+1032},
       {\scriptstyle -54E^2+1038E-4636},
       {\scriptstyle \ast}|
       {\scriptstyle \ast})^+\,\surd\\\hline
\end{array}$
\caption{Spectrum of lowest-lying states genuinely in the 
$\alg{su}(2|1)$ sector.}
\label{tab:SU21}
\end{table}

We have computed the energies of all states for the 
$\alg{su}(2|3)$ spin chain 
of length $L\leq 8$ by direct diagonalization of this Hamiltonian. 
For those multiplets which are part of the
$\alg{su}(1|2)$ sector,
the results have already been presented 
in \tabref{tab:TwoEx,tab:SU2,tab:U11,tab:SU12}.
The multiplets beyond the $\alg{su}(1|2)$ sector 
can be grouped into three classes 
and are given in \tabref{tab:SU3orSU21,tab:SU3,tab:SU21}.
The states which appear up to $L=8$ actually do not saturate the full
$\alg{su}(2|3)$ model, but (at one loop) they are all part of the
$\alg{su}(3)$ or $\alg{su}(2|1)$ subsectors.
We have thus given only the excitation numbers $K'_1,K'_2$ for 
the appropriate subsectors. 
They are related to the excitation numbers for the complete model 
by  $K_1=K_5=K_6=K_7=0$ and $K_4=K'_1$, $K_3=K'_1-2$, $K'_2=K'_2-1$
when we specialize to $\sgradb=-1$ for $\alg{su}(3)$ and 
to $\sgradb=+1$ for $\alg{su}(1|2)$.

We have investigated several states from the tables
using the proposed Bethe equations of \tabref{tab:ABE};
they are marked by ``$\surd$''.
The agreement of the Bethe equations with the tables is perfect. 
This includes one paired state and one state with a relatively large number 
of excitations. There is one point worth mentioning
concerning the state with $(K_1,K_2)=(6,3)$ from \tabref{tab:SU21} 
when the grading is $\sgrada=-1$: 
At one loop it has one root $u=0$ of each of the flavors $1$ and $3$.
Due to its unpaired nature one would expect these two roots to
be exactly $x=0$ even at higher loops. This is however not what
happens, instead they form a pair as $x_3=-g^2/2x_1$ and
are allowed to depart from $x=0$ if $x_1=\order{g}=x_3$. 
In fact, these particular roots turn out to have an expansion 
in odd powers of $g$ unlike all the other roots.

\subsection{Dynamic Bethe Ansatz}

It is well-known that the number of fields in a local operator
is not a conserved quantity at higher loops. 
Similarly, the $\alg{u}(1)$ hypercharge $B$ is anomalous and
broken beyond one loop. 
In the spin chain picture this means that the Hamiltonian 
can add or remove sites of the spin chain. 
This dynamic type of spin chain 
appears to be an altogether novel model 
in the field of integrable spin chains. 
Despite some justified doubts --
the interactions create and destroy particles 
and therefore appear not to be elastic -- 
integrability seems to be an option 
even for dynamic spin chains \cite{Beisert:2003ys}.
A Bethe ansatz for this dynamic chain is expected
to display some novel features. 
Here we will interpret the equations in \tabref{tab:ABE} 
and explain why we believe that they originate from a dynamic chain.

The starting point is the wave function of an eigenstate
of the Hamiltonian. For simplicity let us consider a dilute gas
of excitations on a vacuum state. 
If there exists a Bethe ansatz of more or less familiar kind
then the wave function must be completely described by a set of Bethe roots.
Therefore one set of Bethe roots must be sufficient to describe
states of different lengths or hypercharges.
According to \eqref{eq:Beyond.ExNo}
a change in the length $L$ or the hypercharge $B$, 
while keeping all the conserved Dynkin labels fixed,
leads to modified excitation numbers $K_j$. 
This means that one set of Bethe roots \emph{cannot} 
describe mixed states with different $L$ or $B$.
This apparent dilemma can be solved by supplying 
a rule which maps between two different sets of 
Bethe roots when $L$ or $B$ are changed. 
Let us for definiteness increase both $L$ and $B$ by $\sgradb$
\[
L\mapsto L+\sgradb,\qquad
B\mapsto B+\sgradb.
\]
Then only $K_5$ and $K_7$ change according to 
\[
K_5\mapsto K_5-1,\qquad
K_7\mapsto K_7+1.
\]
The simplest map from one set of Bethe roots to another
changes one root of flavor $5$ into a root of flavor $7$.
In this \emph{dynamic transformation} the value 
of the root need not be invariant,
here we propose the map%
\footnote{
The transformation requires one of the roots
$x_5,x_7$ to be of $\order{g^2}$ if the other one
is finite at $g=0$. 
This is reminiscent of the discussion in 
\cite{Minahan:2005jq}, however, using the
auxiliary roots $x_5,x_7$ instead of the main roots $x_4$.
It would be interesting to see if there can be 
main roots $x_4=\order{g^2}$ and what effect they have for the Bethe ansatz.}
\[\label{eq:Beyond.Trans57}
x_7=g^2/2x_5.
\]

For cyclic spin chains not every set of Bethe roots is admissible,
but the wave-function must be periodic. 
This is ensured by the Bethe equations:
When one takes one excitation once around the trace,
the net phase shift must be zero. The phases are obtained
from permuting the excitation past all other excitations
and all the sites of the vacuum. 
We have to make sure that the change of $L$ or $B$ 
is compatible with the periodicity conditions,
i.e.~\tabref{tab:ABE}. 
The roots $x_5,x_7$ appear in the Bethe equations
for the roots of flavor $4$ through $7$. First of all,
the Bethe equation for $x_{6}$ refers to $x_5,x_7$ only via $u_5,u_7$. 
The transformation \eqref{eq:Beyond.Trans57},
when mapped to the $u$-plane, reads
\[u_5=u_7.\] 
As $u_5$ and $u_7$ appear in precisely the
same term, the Bethe equation for $x_{6}$ is left invariant.
Next, the original Bethe equation for $x_4$ contains the term
\[
\frac{x_4^{-\sgradb}-x_5}{x_4^{+\sgradb}-x_5}
=
\frac{x_4^{-\sgradb}-g^2/2x_7}{x_4^{+\sgradb}-g^2/2x_7}
=
\frac{x_4^{-\sgradb}}{x_4^{+\sgradb}}\,
\frac{1-g^2/2x_4^{+\sgradb}x_7}{1-g^2/2x_4^{-\sgradb}x_7}
=
\lrbrk{
\frac{x_4^{-}}{x_4^{+}}}^{\sgradb}\,
\frac{1-g^2/2x_4^{+\sgradb}x_7}{1-g^2/2x_4^{-\sgradb}x_7}\,.
\]
As we can see, this term is equivalent to the term for $x_7$
when the length $L$ is increased by $\sgradb$.
Finally, the roots of flavor $5$ and $7$ are fermionic and there
are no interactions among different roots of these kinds. 
It is therefore left to confirm that the Bethe equation for
$x_5$ itself is equivalent to the equation for $x_7$.
The coupling to $x_6$ is via $u_5=u_7$ and coincides for both flavors.
The coupling to $x_4$ is
\[
\prod_{j=1}^{K_4}
\frac{x_{5}-x_{4,j}^{+\sgradb}}{x_{5}-x_{4,j}^{-\sgradb}}
=
\lrbrk{
\prod_{j=1}^{K_4}
\frac{x_{4,j}^{+}}{x_{4,j}^{-}}
}^\sgradb
\prod_{j=1}^{K_4}
\frac{1-g^2/2x_{7}x_{4,j}^{+\sgradb}}{1-g^2/2x_{7}x_{4,j}^{-\sgradb}}
=
\prod_{j=1}^{K_4}
\frac{1-g^2/2x_{7}x_{4,j}^{+\sgradb}}{1-g^2/2x_{7}x_{4,j}^{-\sgradb}}\,.
\]
Again we see that the Bethe equation for $x_7$ is automatically satisfied.
We however had to make use of the cyclicity constraint. 
Therefore the transformation \eqref{eq:Beyond.Trans57}
preserves the periodicity of the state and
the Bethe equations are consistent with changes of $L$ and $B$.
Of course, the dynamic transformation between $x_5$ and $x_7$ can be reversed
\[
K_5\mapsto K_5\mp1,\qquad
K_7\mapsto K_7\pm1,\qquad
L\mapsto L\pm\sgradb,\qquad
B\mapsto B\pm\sgradb.
\]
Likewise we can transform on 
the other side of the Dynkin diagram
\[
K_3\mapsto K_3\mp1,\qquad
K_1\mapsto K_1\pm1,\qquad
L\mapsto L\pm\sgrada,\qquad
B\mapsto B\mp\sgrada
\]
using the same map as \eqref{eq:Beyond.Trans57} 
between $x_1$ and $x_3$. 
This proves point \ref{constr3} by relying on 
the cyclic nature of the spin chain,
i.e.~in agreement with point \ref{constr6}.

We can even see from the Bethe equations 
that the length-changing effects 
are a genuinely higher-loop effect:
The map $x\mapsto g^2/2x$ is singular in the
limit $g\to 0$.
A perfectly meaningful finite root $x$ is mapped 
to the point $x=0$ losing all the information of its origin.
Thus there is only one configuration of the Bethe roots 
which can survive in the limit $g\to 0$: 
It is the one where all roots remain finite and do not approach $x=0$.%
\footnote{An exception is a pair of roots $x_{5,7}\sim g$
which cannot be transformed to something finite. 
In this case one of the roots will end up as $x_5=0$ and
one as $x_7=0$.}
The only case where the transformation remains meaningful
at $g=0$ is when $x_5=0$, $x_7=\infty$ and there are no
roots of flavor $6$. 
It is related to multiplet shortening and has already been explained
in \secref{sec:PSU112.PSU11}, as $K_6=0$ restricts to 
that particular subsector. 
Here we understand why a root $x_5=0$ actually corresponds 
to a descendant: 
It is equivalent to $x_7=\infty$ which represents a descendant of the
state without $x_7$. 
For $g\neq 0$, the joining 
transformation of \secref{sec:PSU112.PSU11}
turns into a regular dynamic transformation
which confirms point \ref{constr5}.

Finally, let us consider point \ref{constr8}:
What is special about $\alg{psu}(2,2|4)$?
The equations in \tabref{tab:ABE} are a neat 
but very fragile arrangement of couplings 
of the sort ``$x-x$'' and ``$1-g^2/2xx$''.
The dynamic transformation acts between roots of two flavors. 
Their couplings to other flavors should be alike
in order for the Bethe equations to be preserved
by the transformation.
In particular, they should couple to the 
momentum-carrying flavor in order to be
related to the momentum constraint.
This means that starting from the momentum-carrying node,
the Dynkin diagram can extend only over three further consecutive nodes:
the first node involved in the dynamic transformation,
some other node, and the second node of the dynamic transformation.
This last node couples back to the momentum-carrying one, 
but the link breaks at $g=0$. 
When considering the very limited set of Dynkin diagrams for superalgebras,
we see that we can have no more than two such triples of nodes originating from
the momentum-carrying node. 
This is the case of $\alg{psu}(2,2|4)$ as investigated in this article.
Alternatively, one could have only one triple and potentially
something else, i.e.~$\alg{su}(2|\superN)$ with $\superN\geq 3$.
For $\superN=3$ this would merely be a restriction of a
$\alg{psu}(2,2|4)$ model, see \cite{Beisert:2003ys}. 
For $\superN=4$ we have the $\alg{su}(2|4)$ plane-wave matrix model 
\cite{Berenstein:2002jq,Kim:2003rz,Klose:2003qc,Fischbacher:2004iu},
which has the virtue of a real representation.
Such an algebra with some non-compact signature
could also play a role for integrable subsectors 
of less supersymmetric field theories at higher loops (if they exist).
It would be interesting to see whether dynamic models 
for $\alg{su}(2|\superN)$ with $\superN>4$,
for an orthosymplectic algebra or for exceptional superalgebras exist.

\subsection{Duality Transformation}

To confirm point~\ref{constr9} of \secref{sec:points}
we will now show that the Bethe equations 
for $\sgradb=+1$ are equivalent to the 
ones for $\sgradb=-1$. 
Consider the Bethe equation for roots $x_{5,k}$ of type 5 
\[\label{eq:Beyond.Dual.Bethe}
1=
\prod_{j=1}^{K_6}\frac{u_{5,k}-u_{6,j}+\sfrac{i}{2}}{u_{5,k}-u_{6,j}-\sfrac{i}{2}}
\prod_{j=1}^{K_4}\frac{x_{5,k}-x_{4,j}^+}{x_{5,k}-x_{4,j}^-}\,.
\]
It coincides with the equations for 
roots $x_{7,k}$ of type 7 when $x_{5,k}$ is replaced
by $g^2/2x_{5,k}$
\[
1=
\prod_{j=1}^{K_6}\frac{u_{7,k}-u_{6,j}+\sfrac{i}{2}}{u_{7,k}-u_{6,j}-\sfrac{i}{2}}
\prod_{j=1}^{K_4}\frac{1-g^2/2x_{7,k}x_{4,j}^+}{1-g^2/2x_{7,k}x_{4,j}^-}
\prod_{j=1}^{K_4}\frac{x_{4,j}^+}{x_{4,j}^-}\,
\]
when the cyclicity constraint is imposed.
Using the identities \eqref{eq:One.SpecIdent} we 
can transform the equation to the $x$-plane
\[\label{eq:Beyond.Dual.BetheX}
\prod_{j=1}^{K_4}\frac{x_{5,k}-x_{4,j}^-}{x_{5,k}-x_{4,j}^+}
\prod_{j=1}^{K_6}\frac{x_{5,k}-x_{6,j}^+}{x_{5,k}-x_{6,j}^-}
\prod_{j=1}^{K_6}\frac{x_{5,k}-g^2/2x_{6,j}^+}{x_{5,k}-g^2/2x_{6,j}^+}
=1.
\]
Let us consider a state with $K_5$ roots $x_{5,k}$ and 
$K_7$ roots $x_{7,k}$.
Using the polynomial 
\<\label{eq:Beyond.Dual.Poly}
P(x)\eq
\prod_{j=1}^{K_4}\bigbrk{x-x_{4,j}^+}
\prod_{j=1}^{K_6}\bigbrk{x-x_{6,j}^-}
\prod_{j=1}^{K_6}\bigbrk{x-g^2/2x_{6,j}^-}
\nl-
\prod_{j=1}^{K_4}\bigbrk{x-x_{4,j}^-}
\prod_{j=1}^{K_6}\bigbrk{x-x_{6,j}^+}
\prod_{j=1}^{K_6}\bigbrk{x-g^2/2x_{6,j}^+}
\>
we can write the Bethe equation \eqref{eq:Beyond.Dual.BetheX} as
\[
P(x_{5,k})=0,\qquad
P(g^2/2x_{7,k})=0.
\]
Apart from these roots, there are further solutions:
For counting purposes, we can consider $x=\infty$ to be a solution,
it solves \eqref{eq:Beyond.Dual.BetheX}. 
It is associated to the cancellation of the 
terms $x^{K_4+2K_6}$ in \eqref{eq:Beyond.Dual.Poly}.
Furthermore $x=0$ is a solution if the momentum constraint is
satisfied. This can be viewed as a root of type $7$ at $x=\infty$.
The remaining solutions can be grouped into two classes. 
There are $K_6$ roots of the polynomial \eqref{eq:Beyond.Dual.Poly}
which are of $\order{g^2}$ for small $g$. These are naturally 
associated to Bethe roots of type $7$. 
The remaining $K_4+K_6$ roots are of $\order{1}$ and thus 
correspond to Bethe roots of type $5$.
Therefore there are 
\[\tilde K_5=K_4+K_6-K_5-1,
\qquad
\tilde K_7=K_6-K_7-1
\]
further solutions which we denote by
$\tilde x_{5,k}$ and $g^2/2\tilde x_{7,k}$,
respectively.
We now write the polynomial in factorized form as
\[\label{eq:Beyond.Dual.Poly2}
P(x)\sim x\prod_{j=1}^{K_5}(x-x_{5,j})
\prod_{j=1}^{K_7}(x-g^2/2x_{7,j})
\prod_{j=1}^{\tilde K_5}(x-\tilde x_{5,j})
\prod_{j=1}^{\tilde K_7}(x-g^2/2\tilde x_{7,j})
\]
As before in \secref{sec:PSU112.Duality} we 
use the two equivalent forms 
\eqref{eq:Beyond.Dual.Poly} and \eqref{eq:Beyond.Dual.Poly2}
of $P(x)$ to derive equations which translate between
the two dual forms of the Bethe equations.
The two relevant combinations of $P$ are 
\[
\frac{P(x_{4,k}^+)}{P(x_{4,k}^-)}\,,\qquad
\frac{P(x_{6,k}^-)}{P(x_{6,k}^+)}\,
\frac{P(g^2/2x_{6,k}^-)}{P(g^2/2x_{6,k}^+)}\,.
\]
They lead to 
\<\label{eq:Beyond.Dual.Dualize4}
\earel{}
\prod_{\textstyle\atopfrac{j=1}{j\neq k}}^{K_4}
\bigbrk{x_{4,k}^+-x_{4,j}^-}
\prod_{j=1}^{K_5}
\frac{x_{4,k}^--x_{5,j}}{x_{4,k}^+-x_{5,j}}
\prod_{j=1}^{K_7}
\frac{1-g^2/2x_{4,k}^-x_{7,j}}{1-g^2/2x_{4,k}^+x_{7,j}}
\nln\eq
\prod_{\textstyle\atopfrac{j=1}{j\neq k}}^{K_4}
\bigbrk{x_{4,k}^--x_{4,j}^+}
\prod_{j=1}^{\tilde K_5}
\frac{x_{4,k}^+-\tilde x_{5,j}}{x_{4,k}^--\tilde x_{5,j}}
\prod_{j=1}^{\tilde K_7}
\frac{1-g^2/2x_{4,k}^+\tilde x_{7,j}}{1-g^2/2x_{4,k}^-\tilde x_{7,j}}
\>
and
\<\label{eq:Beyond.Dual.Dualize6}
\earel{}
\prod_{\textstyle\atopfrac{j=1}{j\neq k}}^{K_6}\frac{u_{6,k}-u_{6,j}-i}{u_{6,k}-u_{6,j}+i}
\prod_{j=1}^{K_5}\frac{u_{6,k}-u_{5,j}+\frac{i}{2}}{u_{6,k}-u_{5,j}-\frac{i}{2}}
\prod_{j=1}^{K_7}\frac{u_{6,k}-u_{7,j}+\frac{i}{2}}{u_{6,k}-u_{7,j}-\frac{i}{2}}
\nln
\eq
\prod_{\textstyle\atopfrac{j=1}{j\neq k}}^{K_6}\frac{u_{6,k}-u_{6,j}+i}{u_{6,k}-u_{6,j}-i}
\prod_{j=1}^{\tilde K_5}\frac{u_{6,k}-\tilde u_{5,j}-\frac{i}{2}}{u_{6,k}-\tilde u_{5,j}+\frac{i}{2}}
\prod_{j=1}^{\tilde K_7}\frac{u_{6,k}-\tilde u_{7,j}-\frac{i}{2}}{u_{6,k}-\tilde u_{7,j}+\frac{i}{2}}
\>
which are precisely the equations to turn the Bethe equations for
roots of types $4$ and $6$ for $\sgradb=+1$ into the dual
equations for $\sgradb=-1$ and dual roots $\tilde x_{5,j},\tilde x_{7,j}$.
A similar argument applies
to the duality between $\sgrada=+1$ and $\sgrada=-1$.

Note that these resulting equations are very restrictive for the structure
of the Bethe equations: 
\begin{bulletlist}
\item
If we had assumed there to be no direct coupling between roots
of $4$ and $7$ (or alternatively if we had started out with no roots of type $7$),
the dualization would have generated some new roots $\tilde x_{7,j}$ 
which couple to roots of type $4$. 
It would not be clear how to interpret these additional roots then.

\item
The dualization \eqref{eq:Beyond.Dual.Dualize6}
contains two self-scattering terms for roots of type $6$. 
Contrary to the dualization used in 
\cite{Beisert:2005di} which has only 
one self-scattering term,
this can only work if the roots of type $6$ are bosonic.
In fact, in the framework of 
\cite{Beisert:2005di},
we dualize the (fermionic) roots of type $5$ and $7$ at the same time
and thus generate two self-scattering terms for roots of type $6$.%
\footnote{The dualization does not keep us from 
adding higher-loop self-scattering terms 
to the Bethe equation for roots of flavor $6$.
Whether it is reasonable to do so is a different question.}

\end{bulletlist}
Therefore, once we 
assume \eqref{eq:Beyond.Dual.Bethe} as the Bethe
equation for root of type $5$ 
and we insist on the possibility of dualization,
we can say that much of the structure of the remaining Bethe equations follows.
It is reasonable to assume \eqref{eq:Beyond.Dual.Bethe} 
because it agrees with the string sigma model in \cite{Beisert:2005bm}
in the thermodynamic limit.

For the time being, 
the dualization restricts us to the Dynkin diagrams 
in \figref{fig:DynkinSU224}, but it would be interesting to
see if the other Dynkin diagrams of $\alg{psu}(2,2|4)$ 
can be realized and related by a more general duality
transformation.

\subsection{Frolov-Tseytlin Limit}

In the thermodynamic limit the Bethe equations turn into integral equations. 
We find that the limit of the equations in \tabref{tab:ABE} 
agrees with the generic form (see \appref{sec:Thermo} for a dictionary)
\[\label{eq:Beyond.ThermoBethe}
\sum_{j=1}^7 M_{k,j}\resolvHsl_j(x)+F_k(x)
=-2\pi n_{k,a}
\qquad
\mbox{for }
x\in \contour_{k,a}
\]
where $F_k(x)$ are some functions which specify the details of the model.
The momentum constraint and the anomalous dimension are given by
\[\label{eq:Beyond.ThermoMomEng}
G_4(0)=2\pi n,\qquad
\delta D=g^2 G'_4(0).
\]
The potentials $F_2,F_6$ for the auxiliary bosonic nodes turn out to vanish
\[\label{eq:Beyond.ThermoF26}
F_2(x)=F_6(x)=0.
\]
The remaining auxiliary potentials are all proportional to 
$G_4(g^2/2x)-G_4(0)$
\[\label{eq:Beyond.ThermoF1357}
-\sgrada F_1(x)=+\sgrada F_3(x)
=+\sgradb F_5(x)=-\sgradb F_7(x)
=G_4(g^2/2x)-G_4(0).
\]
The most important function is the potential $F_4$ for the
main excitations for which we find
\<\label{eq:Beyond.ThermoF4}
F_4(x)\eq
-\sgrada\lrbrk{G_1(g^2/2x)-\frac{G_1'(0)\,g^2/2x}{1-g^2/2x^2}-\frac{G_1(0)}{1-g^2/2x^2}}
\nl
+\sgrada\lrbrk{G_3(g^2/2x)-\frac{G_3'(0)\,g^2/2x}{1-g^2/2x^2}-\frac{G_3(0)}{1-g^2/2x^2}}
\nl
+\sgradb\lrbrk{G_5(g^2/2x)-\frac{G_5'(0)\,g^2/2x}{1-g^2/2x^2}-\frac{G_5(0)}{1-g^2/2x^2}}
\nl
-\sgradb\lrbrk{G_7(g^2/2x)-\frac{G_7'(0)\,g^2/2x}{1-g^2/2x^2}-\frac{G_7(0)}{1-g^2/2x^2}}
+F\indup{g,s}(x).
\>
It is the only potential which depends on the particular model,
which is specified by the function $\sigma$, 
through the function $F\indup{g,s}(x)$.
For gauge theory the missing piece is
the same as \eqref{eq:One.LimitPotentialGauge}
\<\label{eq:Beyond.ThermoF4g}
F\indup{g}(x)\eq
\frac{L/x}{1-g^2/2x^2}
\nlnum\nonumber
+\half(2-\sgrada-\sgradb)
\lrbrk{2G_4(g^2/2x)-\frac{G_4'(0)\,g^2/2x}{1-g^2/2x^2}-\frac{G_4(0)(2-g^2/2x^2)}{1-g^2/2x^2}}.
\>
For string theory we recover
the function \eqref{eq:One.LimitPotentialString}
\<\label{eq:Beyond.ThermoF4s}
F\indup{s}(x)\eq
\frac{L/x}{1-g^2/2x^2}
-(\sgrada+\sgradb)\bigbrk{G_4(g^2/2x)-G_4(0)}
\nl
+\half(2+\sgrada+\sgradb)\frac{G_4'(0)\,g^2/2x}{1-g^2/2x^2}
-\half(2-\sgrada-\sgradb)\frac{G_4(0)\,g^2/2x^2}{1-g^2/2x^2}\,.
\>

We can now recast the equations in the
form used in \cite{Beisert:2005bm}. 
This form is useful because it is
closer to the underlying spectral curve
and does not depend on a choice of Dynkin diagram.
The curve is specified by the $4+4$ quasi-momenta
$\tilde p_k(x)$ and $\hat p_k(x)$
corresponding to the $\alg{su}(4)$ and $\alg{su}(2,2)$ parts
of the algebra. The integral equations
\eqref{eq:Beyond.ThermoBethe} become
\<\label{eq:aslkfjasdf}
\sheetsl[\tilde]_l(x)-\sheetsl[\tilde]_k(x)\eq
2\pi \tilde n_{kl,a}
\quad\mbox{for }
x\in \tilde{\contour}_{kl,a},
\nln
\sheetsl[\hat]_l(x)-\sheetsl[\hat]_k(x)\eq
2\pi \hat n_{kl,a}
\quad\mbox{for }
x\in \hat{\contour}_{kl,a},
\nln
\sheetsl[\hat]_l(x)-\sheetsl[\tilde]_k(x)\eq
2\pi n^\ast_{kl,a}
\quad\mbox{for }
x=x^\ast_{kl,a},
\>
where $\tilde{\contour}_{kl,a},\hat{\contour}_{kl,a}$ are branch cuts 
associated to the bosonic subalgebras $\alg{su}(4), \alg{su}(2,2)$,
respectively. The points $x^\ast_{kl,a}$ specify fermionic 
excitations which cannot condense into cuts due to the
Pauli principle \cite{Beisert:2005bm}.
The quasi-momenta are parametrized as follows
\<\label{eq:asdkfjh}
\tilde p_k(x)\eq
\sum_{l=1}^4
\bigbrk{\tilde H_{kl}(x)+H^\ast_{kl}(x)}
+\sheetsign_k \tilde F(x)+F^\ast(x),
\nln
\hat p_k(x)\eq
\sum_{l=1}^4
\bigbrk{\hat H_{lk}(x)+H^\ast_{lk}(x)}
+\sheetsign_k \hat F(x)+F^\ast(x), 
\>
where the functions $G_{kl},H_{kl}$ and $F$ are some combinations
of the functions $G_k,H_k$ and $F_k$.
The coefficients $\sheetsign_k$ equal $(+1,+1,-1,-1)$ 
for $k=(1,2,3,4)$.
By comparing the two different formulations of the integral
equations we obtain 
\<
\tilde F(x)\eq
\half F_4(x)-\quarter(2-\sgrada-\sgradb)\sgrada F_3(x),
\nln
\hat F(x)\eq
\half F_4(x)+\quarter(2+\sgrada+\sgradb)\sgrada F_3(x).
\>
while the resolvents $G_{kl},H_{kl}$ are related to $G_k,H_k$ 
in a canonical way (cf. \cite{Beisert:2005bm} for details).
The fermionic potential $F^\ast$ does not appear in 
the Bethe equations and we cannot determine it here. 
Let us introduce a couple of useful combinations
\<
\tilde G\indup{sum}
\eq
\half \sum_{k,l=1}^4 \sheetsign_k \bigbrk{\tilde G_{kl}+G^\ast_{kl}}
\nln\eq
+\half\sgrada (G_1-G_3)
+\half\sgradb (G_7-G_5)
+\quarter(2+\sgrada+\sgradb)G_4,
\nln
\hat G\indup{sum}
\eq
\half \sum_{k,l=1}^4 \sheetsign_k \bigbrk{\hat G_{lk}+G^\ast_{lk}}
\nln\eq
+\half\sgrada (G_1-G_3)
+\half\sgradb (G_7-G_5)
-\quarter(2-\sgrada-\sgradb)G_4,
\nln
G^\ast\indup{sum}\eq
\half \sum_{k,l=1}^4 G^\ast_{kl}
=
+\half\sgrada (G_1-G_3)
-\half\sgradb (G_7-G_5),
\nln
G\indup{mom}\eq\tilde G\indup{sum}-\hat G\indup{sum}=G_4
\>
and similarly for $H$.
We can then write the potentials for the gauge theory spin chain as
\<
\tilde F\indup{g}(x)\eq
\frac{L/2x}{1-g^2/2x^2}
+\frac{\hat G'\indup{sum}(0)\,g^2/2x}{1-g^2/2x^2}
+\frac{\hat G\indup{sum}(0)}{1-g^2/2x^2}
-\hat G\indup{sum}(g^2/2x),
\\\nonumber
\hat F\indup{g}(x)\eq
\frac{L/2x}{1-g^2/2x^2}
+\frac{\hat G'\indup{sum}(0)\,g^2/2x}{1-g^2/2x^2}
+\frac{\hat G\indup{sum}(0)}{1-g^2/2x^2}
-\hat G\indup{sum}(g^2/2x)
\nl\qquad\qquad\qquad\qquad\qquad\qquad\qquad
+G\indup{mom}(g^2/2x)
-G\indup{mom}(0).
\>
Note that this result, together with
\eqref{eq:aslkfjasdf,eq:asdkfjh},
agrees nicely with the conjectured higher-loop form of the $\alg{so}(6)$ 
Bethe equations in the thermodynamic limit in \cite{Minahan:2004ds}.
The corresponding expressions for the string chain differ slightly
\<
\tilde F\indup{s}(x)\eq
\frac{L/2x}{1-g^2/2x^2}
+\frac{\tilde G'\indup{sum}(0)\,g^2/2x}{1-g^2/2x^2}
+\frac{\hat G\indup{sum}(0)}{1-g^2/2x^2}
-\tilde G\indup{sum}(g^2/2x)
+G\indup{mom}(0),
\nln
\hat F\indup{s}(x)\eq
\frac{L/2x}{1-g^2/2x^2}
+\frac{\tilde G'\indup{sum}(0)\,g^2/2x}{1-g^2/2x^2}
+\frac{\hat G\indup{sum}(0)}{1-g^2/2x^2}
-\hat G\indup{sum}(g^2/2x).
\>
The latter potentials agree with the potentials
of the string sigma model \cite{Beisert:2005bm} 
and confirm point \ref{constr2b}.
The expansion of potentials around $x=\infty$ 
contributes the charges of the vacuum and the anomalous dimension
to the Dynkin labels of a state,
see point \ref{constr7}.
The potentials start out with the same terms for both models
\[
\tilde F(x)=
\frac{L}{2x}
+\order{1/x^2},\qquad
\hat F(x)=
\frac{L+\delta D}{x}
+\order{1/x^2}.
\]

Let us finally investigate the transformation properties
under the map $x\mapsto g^2/2x$.
For string theory the potentials transform according to
\<
\tilde F\indup{s}(g^2/2x)\eq
-\tilde F\indup{s}(x)
-\tilde H\indup{sum}(x)
+G\indup{mom}(0)
,
\nln
\hat F\indup{s}(g^2/2x)\eq
-\hat F\indup{s}(x)
-\hat H\indup{sum}(x).
\>
For definiteness we shall take the missing fermionic
potential $F^\ast$ from the string sigma model \cite{Beisert:2005bm}
\[
F^\ast(x)=
G^{\ast\prime}\indup{sum}(0)\frac{g^2/2x}{1-g^2/2x^2}
+\frac{G^\ast\indup{sum}(0)}{1-g^2/2x^2}
-G^\ast\indup{sum}(g^2/2x).
\]
It transforms according to 
\[
F^\ast\indup{s}(g^2/2x)=
-F^\ast\indup{s}(x)
-H^\ast\indup{sum}(x).
\]
This leads to the following symmetry relations of the quasi-momenta
for the thermodynamic limit of the string chain
\<
\tilde p_k(g^2/2x)\eq
-\tilde p_{k'}(x)
+\sheetsign_k G\indup{mom}(0),
\nln
\hat p_k(g^2/2x)\eq
-\hat p_{k'}(x).
\>
Here the index $k$ is maps to 
the permutation $k'$ by $(1,2,3,4)\mapsto(2,1,4,3)$.
For gauge theory we obtain 
\<
\tilde F\indup{g}(g^2/2x)\eq
-\tilde F\indup{g}(x)
-\tilde H\indup{sum}(x)
+H\indup{mom}(x)
,
\nln
\hat F\indup{g}(g^2/2x)\eq
-\hat F\indup{g}(x)
-\hat H\indup{sum}(x)
+H\indup{mom}(x)
-G\indup{mom}(0)
\>
and the quasi-momenta transform according to 
\<
\tilde p_k(g^2/2x)\eq
-\tilde p_{k'}(x)
+\sheetsign_k H\indup{mom}(x)
,
\nln
\hat p_k(g^2/2x)\eq
-\hat p_{k'}(x)
+\sheetsign_k H\indup{mom}(x)
-\sheetsign_k G\indup{mom}(0).
\>
This inversion appears to be the
only difference between the gauge and string chain 
in the thermodynamic limit.
This is because the Bethe equations
follow from the analyticity properties (which 
are the same for both models) and the symmetry.

\section*{Acknowledgements}

We would like to thank G.~Arutyunov, V.~Bazhanov, V.~Dippel, 
V.~Kazakov, T.~Klose, C.~Kristjansen, T.~M\aa{}nsson, J.~Minahan, 
J.~Plefka, R.~Roiban, K.~Sakai, D.~Serban, 
A.~Tseytlin, H.~Verlinde, M.~Zamaklar and K.~Zarembo
for useful discussions. 
The work of N.~B.~is supported in part by
the U.S.~National Science Foundation Grant No.~PHY02-43680. Any
opinions, findings and conclusions or recommendations expressed in
this material are those of the authors and do not necessarily
reflect the views of the National Science Foundation.

\appendix

\section{Thermodynamic Limit of Terms}
\label{sec:Thermo}

In this appendix we present a dictionary of the various
terms of the discrete Bethe ansatz, 
scattering phases and the resolvents in the thermodynamic limit.
In this limit, the spectral parameter $x$ and the coupling $g$
are both considered to be large and of the same order as the
length of the chain
\[
x=\order{L},\qquad
g=\order{L}.
\]
%

\subsection{Scattering Phases}
\label{sec:Thermo.Scatter}

Let us first state the thermodynamic limit of the
charges
\[\label{eq:Thermo.Charges}
q_r(x_k)
=
\frac{i}{r-1}
\lrbrk{\frac{1}{(x_k^+)^{r-1}}-\frac{1}{(x_k^-)^{r-1}}}
=
\frac{1}{1-g^2/2x_k^2}\,
\frac{1}{x_k^r}+\order{1/L^{r+1}}.
\]
Now we consider two Bethe roots $x_k,x_j$.
Their interactions can be expressed in terms of scattering phases. 
Several useful combinations to write the phases are
given by
\<\label{eq:Thermo.ScatExact}
\scatmain(x_k,x_j)\eq
-i\log\frac{x_k-x_j^+}{x_k-x_j^-}
=
\sum_{r=1}^\infty x_k^{r-1}q_{r}(x_j),
\nln
\scataux_{r,s}(x_k,x_j)\eq(\half g^2)^{(r+s-1)/2}\,q_{r}(x_k)\,q_{s}(x_j)
,
\nln
\scataux(x_k,x_j)\eq\sum_{r=2}^\infty (\scataux_{r,r+1}-\scataux_{r+1,r}),
\nln
\scatsym(x_k,x_j)\eq
-i\log\frac{u_k-u_j-\sfrac{i}{2}}{u_k-u_j+\sfrac{i}{2}}\,.
\>
In the thermodynamic limit, all the scattering phases are
$\order{1/L}$. The phases introduced above are approximated by
\<\label{eq:Thermo.Scat}
\scatmain(x_k,x_j)\eq
\frac{1}{1-g^2/2x_j^2}\,\frac{1}{x_j-x_k}
+\order{1/L^2}
,
\nln
\scataux_{r,s}(x_k,x_j)\eq 
\frac{1}{1-g^2/2x_k^2}\,
\frac{1}{1-g^2/2x_j^2}\,
\frac{(\half g^2)^{(r+s-1)/2}}{x_k^r x_j^s}+\order{1/L^2},
\nln
\scataux(x_k,x_j)\eq
\frac{g^2/2x_k^2}{1-g^2/2x_k^2}\,
\frac{g^2/2x_j^2}{1-g^2/2x_j^2}\,
\frac{1}{1-g^2/2x_kx_j}\,
\frac{x_k-x_j}{x_kx_j}+\order{1/L^2},
\nln
\scatsym(x_k,x_j)\eq
\frac{1}{u_j-u_k}
+\order{1/L^2}
=\sum_{r=1}^\infty u^{r-1}_k u^{-r}_j
+\order{1/L^2}.
\>
The main and auxiliary phases are related by 
\[\label{eq:Thermo.ScatRel}
\scatsym=\scatmain+\scataux+\scataux_{1,2}+\order{1/L^2}.
\]
For the main scattering terms we obtain
\[\label{eq:Thermo.ScatMainX}
-i\log\frac{x_k^a-x_j^c}{x_k^b-x_j^d}
=
\half (b+c-a-d)\scatmain+\half (b-a)\scataux_{1,2}+\half (b-a)\scataux_{2,1}+\order{1/L^2},
\]
where $a,b,c,d=0,\pm 1$ distinguish between $x,x^\pm$.
The limit of the auxiliary terms yields
\[\label{eq:Thermo.ScatAux}
-i\log\frac{1-g^2/2x_k^a x_j^c}{1-g^2/2x_k^b x_j^d}
=
\half(b+c-a-d)\scataux+\half (c-d)\scataux_{1,2}+\half(a-b)\scataux_{2,1}+\order{1/L^2}.
\]
The scattering terms in the $u$-plane limit to 
\[\label{eq:Thermo.ScatMainU}
-i\log\frac{u_k-u_j+\sfrac{i}{2}a}{u_k-u_j+\sfrac{i}{2}b}
=
\half(b-a)\,\scatsym(x_k,x_j)+\order{1/L^2}.
\]
This is compatible with the above expressions using the identity
\eqref{eq:Thermo.ScatRel}.
The combination $\sigma$ for the string chain
yields the auxiliary phase
\[\label{eq:Thermo.ScatString}
-i\log\sigma(x_k,x_j)
=
\scataux(x_k,x_j).
\]
Finally, the limit of the potential term is
\[\label{eq:Thermo.ScatPotential}
-i\log \lrbrk{\frac{x^a_k}{x^b_k}}^L
=
\frac{\half(a-b)L/x_k}{1-g^2/2x_k^2}+\order{1/L}.
\]
%

\subsection{Resolvents}
\label{sec:Thermo.Resolvent}

Let us introduce one resolvent for
the $x$-plane and one for the $u$-plane
\[\label{eq:Thermo.ResolvSum}
G(x)=\sum_{k=1}^{K} \frac{1}{1-g^2/2x_{k}^2}
\,\frac{1}{x_{k}-x}\,,\qquad
H(x)=\sum_{k=1}^{K} 
\,\frac{1}{u_{k}-u(x)}\,.
\]
The two resolvents are related by the identity
\[\label{eq:Thermo.ResolvRel}
H(x)=G(x)+G(g^2/2x)-G(0).
\]
We can also write them in an integral form using a 
density $dx\,\rho(x)=du\,\rho(u)$
\[\label{eq:Thermo.Resolv}
G(x)=\int \frac{dy\,\rho(y)}{1-g^2/2y^2}\,\frac{1}{y-x}\,,\qquad
H(x)=\int \frac{dv\,\rho(v)}{v-u(x)}\,.
\]
The resolvents are related to the summed main scattering phases 
\[\label{eq:Thermo.ResolvMain}
\sum_{j=1}^K\scatmain(x_k,x_j)=G(x_k),\qquad
\sum_{j=1}^K\scatsym(x_k,x_j)=H(x_k).
\]
The partial auxiliary scattering phases yield
derivatives of the resolvent at $x=0$
\[\label{eq:Thermo.ResolvAux}
\sum_{j=1}^K\scataux_{r,s}(x_k,x_j)=\frac{(\half g^2)^{(r+s-1)/2}/x_k^r}{1-g^2/2x_k^2}\,G^{(s-1)}(0).
\]
Finally, the total auxiliary phase translates to
\[\label{eq:Thermo.ResolvString}
\sum_{j=1}^K\scataux(x_k,x_j)=
G(g^2/2x_k)-G(0)-\frac{g^2/2x_k}{1-g^2/2x_k^2}\,G'(0),
\]
so that \eqref{eq:Thermo.ScatRel,eq:Thermo.ResolvMain,eq:Thermo.ResolvRel}
match up.
This dictionary lets us compute the thermodynamic limit
of all expressions straightforwardly.

\section{Transfer Matrices}

\subsection{Rank-One Sectors}
\label{sec:One.Transfer}

Before we test our asymptotic extrapolation,
let us introduce an important object for integrable models,
a transfer matrix. 
Within Bethe ans\"atze there often exist expressions for 
the eigenvalues of the transfer matrices in terms of
the Bethe roots. In fact, the Bethe equations follow from 
these equations by demanding that the transfer matrix has
no poles. For eigenstates with singular 
Bethe roots (usually at $x=0,\pm \sfrac{i}{2}$) 
the cancellation of poles gives the correct prescription
for regularizing the Bethe equations. 
For the three models we heuristically find for 
the eigenvalues of the fundamental transfer matrix
\<\label{eq:One.Transfer}
T\indup{fund}(x)\eq
+\sgrada\lrbrk{\frac{x^+}{x}}^L
\prod_{j=1}^{K}
\lrbrk{
\frac{x^{-\sgrada}_{}-x_{j}^{+\sgrada}}{x-x_{j}}\,
\frac{1-g^2/2x^-x^+_{j}}{1-g^2/2xx_{j}}\,
\sigma^{-1}(x,x_j)}
\nln\earel{}
+\sgradb\lrbrk{\frac{x^-}{x}}^L
\prod_{j=1}^{K}
\lrbrk{
\frac{x^{+\sgradb}_{}-x^{-\sgradb}_{j}}{x-x_{j}}\,
\frac{1-g^2/2x^+x^-_{j}}{1-g^2/2xx_{j}}\,
\sigma^{+1}(x,x_j)}.
\>
Here we use a compact notation with two parameters
$\sgrada,\sgradb=\pm 1$. Together they determine
the model with the total grading $\sgrad=(\sgrada+\sgradb)/2$.
For $\sgrad=\sgrada=\sgradb=+1$ it agrees
with the expression given in \cite{Beisert:2004hm}.
The Bethe equations \eqref{eq:One.BetheAll}
follow from cancelling the poles at $x_{j}$.
The overall factor is ambiguous, we have chosen it 
so that the terms appear in a symmetric way.%
\footnote{One might be tempted to remove the
denominators $1-g^2/2xx_{j}$ 
in order to eliminate poles at $x=g^2/2x_{j}$.
As we do not know how to derive these 
expressions from first principles,
we cannot decide which form is more suitable.}

\paragraph{Thermodynamic Limit.}

In the thermodynamic limit, the transfer matrix 
gives a sum of exponentials
\<
T\indup{fund}(x)=\sgrada\exp\bigbrk{ip_1(x)}+\sgradb\exp\bigbrk{ip_2(x)}.
\>
The exponents
are called quasi-momenta, for gauge theory we obtain 
\<
p_1(x)\eq
+\sgrada H(x)
+\frac{L/2x}{1-g^2/2x^2}
+\half (1-\sgrada) F\indup{g}(x),
\nln
p_2(x)\eq
-\sgradb H(x)
-\frac{L/2x}{1-g^2/2x^2}
-\half (1-\sgradb)F\indup{g}(x)
\>
with the potential
\[
F\indup{g}(x)=
2G(g^2/2x)
-G(0)\,\frac{2-g^2/2x^2}{1-g^2/2x^2}
-G'(0)\,\frac{g^2/2x}{1-g^2/2x^2}\,.
\]
For string theory, the quasi-momenta read 
\<
p_1(x)\eq
+\sgrada G(x)
+\frac{L/2x}{1-g^2/2x^2}
-\half(1-\sgrada)F\indup{s1}(x)
+\half(1+\sgrada)F\indup{s2}(x),
\nln
p_2(x)\eq
-\sgradb G(x)
-\frac{L/2x}{1-g^2/2x^2}
+\half(1-\sgradb)F\indup{s1}(x)
-\half(1+\sgradb)F\indup{s2}(x)
\>
where the potentials are given by
\[
F\indup{s1}(x)=
G(0)\,\frac{g^2/2x^2}{1-g^2/2x^2}
,\qquad
F\indup{s2}(x)=
G'(0)\,\frac{g^2/2x}{1-g^2/2x^2}\,.
\]
These quasi-momenta agree with the expressions 
for classical strings investigated derived in 
\cite{Kazakov:2004qf,Kazakov:2004nh,Beisert:2005bm}
(when fixing $B=0$ in the fermionic case $\sgrada\neq\sgradb$).

\subsection{The $\alg{su}(1,1|2)$ Sector}

The transfer matrix for the $\alg{su}(1,1|2)$ spin chain appears to be
\<\label{eq:Three.Transfer}
T\indup{fund}(x)\eq
-\sgrada\lrbrk{\frac{x^+}{x}}^L
\prod_{j=1}^{K_1}\frac{x^{+\sgrada}-x_{1,j}}{x^{-\sgrada}-x_{1,j}}
\prod_{j=1}^{K_2}\lrbrk{\frac{1-g^2/2x^-x^+_{2,j}}{1-g^2/2x^-x^-_{2,j}}\,
\sigma^{-1}(x,x_{2,j})}
\nln\earel{}
+\sgrada\lrbrk{\frac{x^+}{x}}^L
\prod_{j=1}^{K_1}\frac{x^{+\sgrada}-x_{1,j}}{x^{-\sgrada}-x_{1,j}}
\prod_{j=1}^{K_2}\lrbrk{\frac{1-g^2/2x^-x^+_{2,j}}{1-g^2/2xx_{2,j}}\,
\sigma^{-1}(x,x_{2,j})\,
\frac{x^{-\sgrada}-x^{+\sgrada}_{2,j}}{x-x_{2,j}}}
\nln\earel{}
+\sgradb\lrbrk{\frac{x^-}{x}}^L
\prod_{j=1}^{K_3}\frac{x^{-\sgradb}-x_{3,j}}{x^{+\sgradb}-x_{3,j}}
\prod_{j=1}^{K_2}\lrbrk{
\frac{1-g^2/2x^+x^-_{2,j}}{1-g^2/2xx_{2,j}}\,
\sigma^{+1}(x,x_{2,j})\,
\frac{x^{+\sgradb}-x^{-\sgradb}_{2,j}}{x-x_{2,j}}}
\nln\earel{}
-\sgradb\lrbrk{\frac{x^-}{x}}^L
\prod_{j=1}^{K_3}\frac{x^{-\sgradb}-x_{3,j}}{x^{+\sgradb}-x_{3,j}}
\prod_{j=1}^{K_2}\lrbrk{
\frac{1-g^2/2x^+x^-_{2,j}}{1-g^2/2x^+x^+_{2,j}}\,
\sigma^{+1}(x,x_{2,j})}.
\>
The Bethe equations
\eqref{eq:PSU112.Bethe.Bethe} follow from 
the above expression $T\indup{fund}(x)$ by cancelling the poles 
at $x=x^{+\sgrada}_{1,k},x^{}_{2,k},x^{-\sgradb}_{3,k}$.

\paragraph{Dualization.}

Let us verify that the above 
expression for the transfer matrix 
is valid for all choices of $\sgrada,\sgradb$.
Due to the relation \eqref{eq:PSU112.Duality.PolyFactor} 
the following identity holds
\[
P(x^{-\sgrada})\prod_{j=1}^{K_1}\frac{x^{+\sgrada}-x_{1,j}}{x^{-\sgrada}-x_{1,j}}
=
P(x^{+\sgrada})\prod_{j=1}^{\tilde K_1}\frac{x^{-\sgrada}-\tilde x_{1,j}}{x^{+\sgrada}-\tilde x_{1,j}}
\]
Using the original definition \eqref{eq:PSU112.Duality.PolyDef} 
of the polynomial, the first two lines 
of $T\indup{fund}(x)$ in \eqref{eq:Three.Transfer} 
can be written as
\[
P(x^{-\sgrada})\prod_{j=1}^{K_1}\frac{x^{+\sgrada}-x_{1,j}}{x^{-\sgrada}-x_{1,j}}
\lrbrk{\frac{x^+}{x}}^L
\prod_{j=1}^{K_2}\lrbrk{
\frac{1-g^2/2x^-x^+_{2,j}}{1-g^2/2xx_{2,j}}\,
\sigma^{-1}(x,x_{2,j})\,
\frac{1}{x-x_{2,j}}}
\]
When we flip the sign $\sgrada$ and 
use $\tilde x_{1,k}$ instead of $x_{1,k}$ 
the first two lines 
in $T\indup{fund}(x)$ become equivalent to
\[
P(x^{+\sgrada})\prod_{j=1}^{\tilde K_1}\frac{x^{-\sgrada}-\tilde x_{1,j}}{x^{+\sgrada}-\tilde x_{1,j}}
\lrbrk{\frac{x^+}{x}}^L
\prod_{j=1}^{K_2}\lrbrk{
\frac{1-g^2/2x^-x^+_{2,j}}{1-g^2/2xx_{2,j}}\,
\sigma^{-1}(x,x_{2,j})\,
\frac{1}{x-x_{2,j}}}
\]
The same applies to dualization of $x_3$-roots.
Therefore $T\indup{fund}(x)$ remains valid after the duality transformation.

\paragraph{Thermodynamic Limit.}

The thermodynamic limit of the transfer matrix for $\sgrada=\sgradb=+1$ 
can be written as
\<
p_1(x)\eq -G_1(x)-F_1(x)
\nln
p_2(x)\eq -G_1(x)+G_2(x)-F_1(x)
\nln
p_3(x)\eq +G_3(x)-G_2(x)+F_3(x),
\nln
p_4(x)\eq +G_3(x)+F_3(x).
\>
The potential for gauge theory is
\[
F_{\mathrm{g},j}=
-\frac{L/2x}{1-g^2/2x^2}
-G_2(g^2/2x_k)
+G_2(0)
+G'_j(0)\,\frac{g^2/2x_k}{1-g^2/2x_k^2}+G_j(0)\,\frac{g^2/2x_k^2}{1-g^2/2x_k^2}\,,
\]
whereas the one for the string chain reads
\[
F_{\mathrm{s},j}=
-\frac{L/2x}{1-g^2/2x^2}
-G'_2(0)\,\frac{g^2/2x_k}{1-g^2/2x_k^2}
+G_j(0)\,\frac{g^2/2x_k^2}{1-g^2/2x_k^2}
+G'_j(0)\,\frac{g^2/2x_k}{1-g^2/2x_k^2}\,.
\]
The latter expression agrees with 
classical superstrings on $AdS_3\times S^3$ 
found in \cite{Beisert:2005bm}.

\bibliography{rank1}
\bibliographystyle{nb}

\end{document}